  \documentclass{aa}  
 \usepackage{natbib}
  \bibpunct{(}{)}{;}{a}{}{,}             
 \usepackage{graphicx}
 \usepackage{amsmath}
\usepackage{txfonts}

 \usepackage[breaklinks=true]{hyperref} 
 \hypersetup{colorlinks=true,urlcolor=blue,citecolor=blue,pdfborder= 0 0 0}

\begin{document}

\title{Studying X-ray instruments with galaxy clusters}

\author{J. Nevalainen\inst{1}\fnmsep\thanks{jukka.nevalainen@ut.ee}
        \and
        S. Molendi\inst{2}
       }
\institute{Tartu Observatory, University of Tartu, 61602 T{\~{o}}ravere, Tartumaa, Estonia 
  \and
INAF - IASF Milano, via A.Corti 12 I-20133 Milano, Italy 
 }
   \date{Received ; accepted}
  
  \abstract
   {}  
   {Our aim is to apply a scientific approach to the problem of the effective area cross-calibration of the XMM-Newton EPIC instruments. Using a sample of galaxy clusters observed with the XMM-Newton EPIC, we aim at quantifying the effective area cross-calibration bias between the EPIC instruments as implemented in the public calibration data base on November 2021 in the 0.5-6.1 keV energy band.}
   {We tested two methods for evaluating the effective area cross-calibration bias for CCD-type X-ray instruments. Namely, we compared the evaluation of the cross-calibration bias by modelling it before the convolution of the spectral models with the redistribution matrix or by analysing the convolved products. We applied the methods to a sample of clusters of galaxies observed with XMM-Newton/EPIC instruments. We invested significant efforts in controlling and minimising the systematic uncertainties of the cross-calibration bias below 1\%. The statistical uncertainties are similar and thus we can reliably measure effects at 1\% level.}
    {On average the two methods differ very little; the only difference on the cross-calibration bias is at the highest energies by 3\% at maximum. 
    The effective area cross-calibration in the 0.5-6.1 keV band between MOS and pn is biased at a substantial level.
The MOS/pn bias is systematic suggesting that MOS (pn) effective area may be calibrated too low (high), by $\sim$3-27\%  on average depending on the instrument and energy band. The excellent agreement of the energy dependencies (i.e. shapes) of the effective area of MOS2 and pn suggest that they are correctly calibrated within $\sim$1\% in the 0.5-4.5 keV band. Comparison with an independent data set of point sources (3XMM) confirms this. The cluster sample indicates that the MOS1/pn effective area shape cross-calibration has an approximately linear bias amounting to $\sim$10\% in maximum in the 0.5-4.5 keV band.}
 {The effective area cross-calibration of XMM-Newton/EPIC instruments on November 2021 in the 0.5-4.5 keV band is in a relatively good shape. However, the cluster-to-cluster rms scatter of the bias is  substantial compared to the median bias itself.
Thus, a statistically robust implementation of the cross-calibration uncertainties to a scientific analysis of XMM-Newton/EPIC data should include the propagation of the scatter to the best-fit parameters, instead of a simple average bias correction of the effective area. 
}

   \keywords{calibration -- X-ray -- galaxy clusters
               }
 
\maketitle

\section{Introduction}
In astronomical research instruments are used to investigate astronomical objects. In case of instrument calibration we do the opposite; we use astronomical objects to investigate the instruments. In this work we use clusters of galaxies to investigate the effective area cross-calibration bias between the XMM-Newton/EPIC instruments.

Nearby clusters of galaxies are useful for the X-ray calibration since they provide high numbers of counts with relatively short exposures with modern X-ray telescopes (e.g. \cite{N10}, \cite{Kettula}, \cite{Gerrit}.)
This enables high statistical accuracy of the data which in turn translates into similarly high accuracy of the calibration results. 
Focusing on hotter clusters in particular is preferable, since they provide more counts towards the high energy end ($\sim$10 keV) of the typical CCD-type X-ray detectors (e.g. XMM-Newton/EPIC and Chandra/ACIS). Also, the line emission at $\sim$1 keV energies is reduced when focusing on the hotter clusters and thus 
the complications arising from the calibration accuracy of the energy redistribution are smaller. The number of hot nearby clusters is substantial which enables large samples which is necessary when attempting a statistically meaningful analysis. 

Galaxy clusters are useful for cross-calibration work also because they are stable in human time scales.  Thus, if one desires to compare the effective area cross-calibration between instruments of different missions, one can use observations performed at different epochs. This is different from the case of variable sources often used for cross-mission calibration.
Obtaining a simultaneous cross-mission data set of a single variable target is a huge effort and consequently the samples are typically too small for robust statistical analyses of a given instrument pair (e.g. \cite{Madsen}). 
 
However, the extended nature of the galaxy clusters causes systematic uncertainties not present in the case of point sources.
Our aim in this work is to have a firm control on the systematics and to limit their effect below 1\% of the cross-calibration bias.
We aim at minimising each of the systematic uncertainty component by careful selection of the clusters, observations, spectrum extraction regions and the waveband.

\section{The methods}
\label{methods}

We investigated two methods for evaluating the cross-calibration bias between the pn and MOS instruments onboard XMM-Newton. The method 1 (post-convolution), used in several works (e.g. \cite{Kettula}, \cite{Gerrit} and \cite{Andy}), operates on the data which are obtained after convolution of the spectral models with the energy redistribution matrixes. We also examined the Method 2 (pre-convolution) whereby the cross-calibration bias is modelled before the convolution (e.g. 
Smith et al., 2021\footnote{https://xmmweb.esac.esa.int/docs/documents/CAL-SRN-0382-1-1.pdf}). The latter method is more complex in the sense that the modelling of the reference instrument data has to be very accurate in order to maintain the pre-convolution aspect of the method. 
Otherwise, the inaccuracies of the modelling of the reference instrument data are applied outside the convolution which is the problem we want to avoid with Method 2.

\subsection{Method 1: post-convolution}
Our focus is the total effective area, i.e. essentially the product of the mirror effective area, the filter transmission and the CCD quantum efficiency in the 0.5-6.0 keV band.
We chose EPIC-pn as a reference instrument, whose spectra we fit with absorbed thermal emission models (see Section \ref{spectral-analysis} for the description of the required spectral analysis) obtaining the reference model (${\rm model}_{ref,1}$). 
We use ${\rm model}_{ref,1}$ to form a prediction for the test instrument (MOS1 or MOS2) by multiplying it with the effective area function (${\rm arf}_{test}$) of the test instrument and convolving the product with the energy redistribution matrix ${\rm rmf}_{test}$ of the test instrument (see Eq. \ref{R1.eq}). The parameter R1 (see Eq. \ref{R1.eq}) compares the above prediction with the count rate data (${\rm data}_{test}$) (see Section \ref{Reduction} for the detailed description of the data reduction) in each spectral bin obtained with the test instrument, according to 
\begin{equation}
R1 = \frac{{\rm data}_{test}}{({\rm model}_{ref,1} \times {\rm arf}_{test}) \otimes {\rm rmf}_{test}}~.
\label{R1.eq}
\end{equation}
If the reference model was an accurate description of the data obtained with the reference instrument (${\rm data}_{ref}$), a significant deviation of R1 (Eq. \ref{R1.eq}) from the unity
would indicate significant cross-calibration problems between the reference and test instruments.
However, in the presence of calibration problems, it is difficult to obtain an acceptable match between the data and a physical model using the reference instrument in the first place. On the other hand, the accurate physical modelling is not necessary 
 since we are interested in the $relative$ accuracy between different instruments. Any numerical model which accurately describes the data of the reference instrument would suffice. We attempt to produce such a model by correcting the reference model (${\rm model}_{ref,1}$) in Eq. \ref{R1.eq} by dividing it by the residuals of the reference instrument (R2), according to 
\begin{equation}
R2 =   \frac{{\rm data}_{ref}} {({\rm model}_{ref,1}  \times {\rm arf}_{ref}) \otimes {\rm rmf}_{ref}} ~.
\label{R2.eq}
\end{equation}
Thus the ratio R1/R2 yields the effective area cross-calibration bias parameter J$_{1}$, i.e
\begin{multline}
{\rm J_{1} }  =  \frac{{\rm R}1 } {{\rm R}2} =  \frac{{\rm data}_{test}}{({\rm model}_{ref,1} \times {\rm arf}_{test}) \otimes {\rm rmf}_{test}} \times \\ \frac{({\rm model}_{ref,1}  \times {\rm arf}_{ref}) \otimes {\rm rmf}_{ref}}{{\rm data}_{ref}},
\label{J1.eq}
\end{multline}
which is the focus of this work.

A limitation with correcting the reference model with the residuals as described in Eq. \ref{J1.eq} is that it is done after the reference model is convolved with the test instrument response. By doing so effective area and redistribution issues are mixed with one another and the above approach provides only an approximate correction. We will evaluate the effect of this approximation in Section \ref{comparing-the-methods}.

If the prediction of the residual-corrected reference model on a test instrument deviates significantly from the data observed with the test instrument, then there is a significant cross-calibration bias between the reference instrument and test instrument. This bias manifests itself as a deviation of J$_1$ (see Eq. \ref{J1.eq}) from unity, and thus J$_1$ serves directly as the bias factor.
However, this approach involves another approximation, namely that we evaluate the cross-calibration bias by examining the convolved results while using that to represent the effective area bias which enters the spectral analysis before convolution with the redistribution matrix. We will evaluate the effect of this approximation in Section \ref{comparing-the-methods}.
 
One can also utilise J$_1$ for investigating the accuracy of the cross-calibration of the energy dependence (i.e. the shape) of the effective area.
Namely, in the case that there is no bias in the energy dependence, J$_1$ is constant with energy (E).
Thus, if dJ$_1$/dE deviates significantly from zero, there is a significant bias in the cross-calibration of the energy dependency of the effective area. 
Thus, in the following we will analyse the deviation of J and dJ/dE from unity and zero, respectively. 
 
The statistical accuracy of the above correction to the reference model is set by the statistical uncertainties of the reference instrument data. Additionally the statistics of the test instrument affect the accuracy of J$_1$. Previous works, e.g. \cite{Gerrit} and  \cite{N10}
have indicated $\sim$10\% effects in the effective area cross-calibration between XMM-Newton EPIC instruments. In order to meaningfully investigate the cross-calibration, we set as a requirement to maintain the statistical uncertainties of the sample mean 
cross-calibration bias data below 1\% level at each spectral bin (including the counting statistics of both the reference instrument and the test instrument).

Since we use the numerical model as a description of the data, the physical accuracy of the reference model does not matter (any deviation of the reference model from the data obtained by the reference instrument is corrected with residuals, see Eq. \ref{J1.eq}).
An important consequence is that we can include the spectrally very different cool cores \citep{Hudson10} and the hotter intermediate regions in a single extraction region and thus improve the signal-to-noise ratio.
This approach also helps to minimise the unwanted flux variation due to the point spread function scatter (see Section \ref{psf}).
 
\subsection{Method 2: pre-convolution}
\label{method2}
Given that the above approach includes two potentially significant approximations we decided to compare the results obtained with Method 1 with those obtained with another method which does not include these approximations. 

First, instead of using a purely thermal model (absorbed by the interstellar matter) for describing the reference instrument data, and correcting it with convolved residuals, we now use more complex model which describes the data very accurately.
After experimenting, we settled on a hybrid model which includes an absorbed thermal component modified with cubic splines (piece-wise 3rd order polynomials, see the previous Section for a justification of phenomenological modelling). We included three independent spline components with breaks at 0.5, 2.0, 4.0 and 6.1 keV. This is the reference model 
(${\rm model}_{ref,2}$) when applying Method 2. We used here the optimal spectral bin size of 400 eV (see Section \ref{redistribution}).
The residuals are below 1\% for each cluster and each spectral bin. Thus, the correction term R2 (Eq. \ref{R2.eq}) is very close to unity and can be neglected in further analysis. This way we by-pass the first approximation. We will propagate the statistical uncertainties of the reference instrument data to the following analysis.
 
Second, instead of studying the cross-calibration by analysing the data which has been convolved with the energy redistribution matrix, we now model the effective area bias before convolution. After experimenting with different models we settled on a single 4th order polynomial in the 0.5-6.1 keV band,
\begin{equation}
J_{model} = a_0 + a_1 \times E + a_2 \times E^2  + a_3 \times E^3 + a_4 \times E^4  \,
\label{poly.eq}
\end{equation}
as the model for the cross-calibration bias (E is the photon energy in units of keV). This model is complex enough to describe the broad-band features (see Section \ref{Results} for the details of the fitting) with a relatively small number of free parameters so that the complexity of the analysis is manageable.
In practise we fit the test instrument data in XSPEC with a model consisting of 1) the 
hybrid spline-modified absorbed thermal component (${\rm model}_{ref,2}$) whose 
all parameters are fixed to the best-fit values obtained when fitting the reference instrument data (see above),
multiplied by 2) the model for the cross-calibration bias (Eq. \ref{poly.eq}) whose coefficients are free parameters in the fit.
We thus minimise the deviation of the ratio
\begin{equation}
R3 = \frac{{\rm data}_{test}}{({\rm model}_{ref,2} \times J_{model} \times  {\rm arf}_{test}) \otimes {\rm rmf}_{test}}~
\label{R3.eq}
\end{equation}
from the unity.
We will use the same function (Eq. \ref{poly.eq}) to model the cross-calibration bias values obtained with Method 1 
(Eq. \ref{J1.eq})  for direct comparison in Section \ref{comparing-the-methods}.

\section{Controlling the systematics}
\label{systematics}
Our focus is on the calibration of the effective area. However, there are many aspects of the instrumentation (e.g. the possible bias of other calibration components and the dead detector area due to bad pixels, dominated by the CCD gaps)  
which may produce variation in the measured flux we erroneously attribute to effective area calibration effects.
Our approach is to choose the clusters, observations, extraction regions and the wavebands in a manner that minimises the above effects. 
In particular we investigate energy redistribution, background, point spread function, vignetting and bad pixels.
We discuss these components below and define requirements and procedures for maintaining their effects, i.e. systematic uncertainties, below 1\% of the cluster flux. 
 
\subsection{Energy redistribution}
\label{redistribution}
Due to the limitations of the energy resolution of detectors, a fraction of photons will be detected outside of the channel nominally corresponding to the energy of the incoming photon. 
Also, a given channel will get contribution from photons with true energy outside that channel. While this effect is modelled and taken into account in the data processing and analysis via the energy redistribution matrix, it has some degree of uncertainty.
Thus, there will be additional redistributed flux in a given band which is not accounted for when modelling the data with the nominal energy redistribution matrix file ($rmf$ in Eq. \ref{J1.eq}) and consequently we will interpret the additional flux being due to a bias of the effective area in that band.  

As long as the spectral features are much broader than the width of a symmetric\footnote{This assumption is not valid in case of EPIC and we will investigate the effect below} redistribution function, the scattered flux to and from the neighbouring bins cancels out very accurately, independently of the redistribution calibration. In case of hot thermal plasma, to the first order the continuum  is smooth over scales bigger than the FWHM of the EPIC energy redistribution function ($\sim$ 100 eV) for the hot clusters at energies below 6 keV.
 
The emission lines are problematic in this context. The problem is most severe for the \ion{Fe}{XXV-XXVI} lines of the hottest clusters which dominate the total emission in the $\sim$6--7 keV band. Since most of the useful clusters for our work have redshifts below 0.1 due to photon counting statistics requirements (see below), the iron line flux and thus the possibly related problems are confined above 6 keV energies. Thus, we eliminate the effect by truncating the spectra at 6 keV.

The emission line complex at $\sim$ 1 keV is less strong in hotter clusters which our sample consists of. Yet, for example, it produces $\sim$5\% of the total emission in the 0.8-1.2 keV band 
of the A1795 spectrum. 
Our approach is to minimise the redistribution between neighboring bins by maximising the size of the spectral bins while maintaining useful spectral resolution 
for the cross-calibration study. We studied the effect of different spectral bin sizes for A1795 pn spectrum by comparing the fluxes in such bins obtained 
with the reference model by 1) applying the standard redistribution matrix (rmf-file produced by SAS toot $rmfgen$) and 2) diagonalising the above response using $diagrsp$ - command in XSPEC. The latter approach produces fluxes assuming that the energy resolution is infinite, i.e. there is no redistribution to and from the neighbouring bins.
After experimenting we chose a bin size of 400 eV, which exceeds the pn energy redistribution FWHM by a factor of 2 (4) at 6 keV (1 keV). 

\begin{figure}
\includegraphics[width=9cm,angle=0]{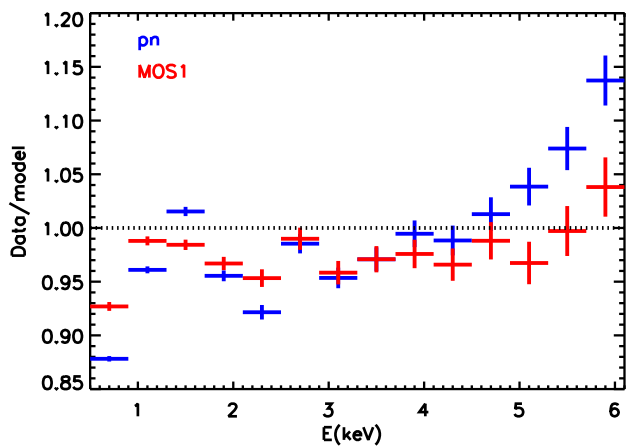}
\caption{The crosses show the ratio of spectral data of A1795 and the prediction of the best-fit hybrid model i.e. the absorbed thermal model modified by the splines (model$_{ref,2}$, see Section \ref{method2}) after diagonalising the redistribution matrix for pn (blue) and MOS1 (red).}
\label{diageff.fig}
\end{figure}

In most of the energy band studied in this work (0.5-6.1 keV) the net flux effect of the pn redistribution is at a few \% level (see Fig. \ref{diageff.fig}).
The larger effect at the low energy end can be understood via the
complex energy-dependent interplay of the effective area, cluster emission and the asymmetric feature (soft tail) of the redistribution function which is wider than 400 eV at 1\% level of the peak flux. 
Namely, the lost fraction of the flux originating from the 0.5-0.9 keV bin due to the soft wing will not be compensated by the redistributed contributions originating from 1) the lower neighbouring energies due to the asymmetry, i.e. there is no hard wing and 2) the neighbouring higher energies since the measured flux decreases with energy which reduces the redistributed wing contribution.
Thus, for A1795 the net redistribution effect in the 0.5-0.9 keV bin leads to a  15\% flux loss for the pn.

At the highest energies the pn redistribution has an opposite effect: 15\% net increase of flux in 5.7-6.1 keV band. While the soft wing is not significant at these energies, another asymmetry arises. Namely, the relatively quickly decreasing cluster emission at these energies will result in larger redistribution contribution originating from lower energy not being compensated by smaller contribution from higher energy. 

The effect on the MOS is different (except for the 0.5-0.9 keV bin);  it is smaller than that in the pn and less strongly energy dependent. The drop of the MOS effective area above $\sim$ 5 keV has an effect of producing an asymmetric situation resulting in redistributed flux excess towards higher energies. Yet, the net effect is smaller than that in pn, this can be ascribed to the better MOS energy resolution at higher energies. The difference of the pn and MOS at the highest energies results in energy-dependent rise of the redistribution effect on the MOS/pn flux ratio. This may be reflected in the differences of the results obtained with different treatments of the redistribution (Methods 1 and 2, see Section \ref{comparing-the-methods}).

Assuming 3\% uncertainty in the calibration of the EPIC redistribution  (Haberl et al., 2010)\footnote{https://xmmweb.esac.esa.int/docs/documents/CAL-SRN-0266-1-3.pdf}, the above estimates for the net redistribution effect with our 400 eV binning (i.e. a few \% of the total flux) yield a systematic redistribution effect which is a fraction of 1\% of the flux in a given bin, satisfying our criteria.
 
\subsection{Point spread function}
\label{psf}
The extended nature of galaxy clusters complicates the analysis. Namely, analogously to the energy redistribution phenomenon due to the limitations of the energy resolution (see Section \ref{redistribution}), the limitations of the spatial resolution cause some of the intended flux to scatter out from the extraction region and some unwanted flux will scatter into our extraction region from outside the studied region. This is modelled as a point spread function (PSF) and SAS software has an option of considering the PSF scatter between the different studied regions of extended sources when computing the effective areas. The modelling of the PSF has some degree of uncertainty and consequently there will be some level of flux variations  which we would erroneously interpret as being due to effective area calibration problems.

Due to its wider PSF we conservatively used pn to estimate the EPIC PSF scatter effect. 90\% of the energy entering the telescope in a single point is spread out within r $\sim$ 1 arcmin circle in the pn detector. The brightness profiles of the non-cool-core clusters are relatively smooth and do not vary much at the 1 arcmin scale. Thus, the scattered flux to and from the neighbouring detector regions cancel out with high accuracy and the net PSF scatter effect for these clusters is negligible. 

A particular problem related to clusters is the central brightness peak in the cool core clusters, analogous to a spectral line in the case of the energy redistribution. The brightness decreases rapidly with increasing radius at scales comparable to the PSF size. Thus, at these scales there will be significant net PSF scatter effect reducing the flux.  
Analogously to the energy redistribution case, our solution is to use extraction regions much bigger than the PSF scale and to include the central regions in our spectrum extraction region. By requiring a minimum of 3 arcmin radius for the extraction region (i.e. $\sim$4 times the 90\% energy encircled radius) we expect to minimise the variation of the extracted flux due to PSF scatter to and from the region outside below $\sim$1\%. In such case, the possible problems in the calibration of the PSF will yield an effect which is a fraction of 1\% of the flux in our extraction region, satisfying our criteria. 

Our expectation is supported by the work of \cite{N10} who examined the PSF scatter issue for a partially overlapping sample of galaxy clusters.
They convolved the surface brightness profiles measured with Chandra with an analytical expression of the XMM-Newton/EPIC PSF obtaining that the PSF-scattered flux from the cool core 
to an annular region at $\sim$1.5-3.0 arcmin distance from the cluster center amounts to less than 1\% of the flux originating from that region. In our case the fraction is much smaller since we include the cool core in the extraction region. Thus an extraction radius of 3 arcmin is our very conservative lower limit for the extraction radius in order to minimise the PSF scatter.  

\subsection{Background}
The data we utilise in our work (data$_{ref}$ and data$_{test}$ in Eqs. \ref{R1.eq}, \ref{R2.eq}, \ref{J1.eq},  \ref{R3.eq} and  \ref{Jmeas.eq}) consist of the background subtracted count rate of the cluster. The systematic uncertainties of the background modelling will introduce unwanted variation in our data. Our choice of the hottest nearby clusters and focusing on their bright central regions is our first order attempt to minimise the effect of the background in our data.
 
 At the low energy end of the spectra the background is dominated by the Galactic emission and the Cosmic X-ray Background. However, the clusters are strong soft X-ray emitters and thus the background-to-source ratio is lowest at the soft X-rays. 

The intrinsic emission of the clusters of galaxies decreases with higher energies in the X-ray band. Since the effective area behaves similarly, the net effect is that the number of source photons per keV decreases strongly with increasing photon energy in the XMM-Newton/EPIC waveband. 
Consequently, due to its much harder X-ray spectrum compared to the hottest clusters, 
the particle-induced background becomes more important at higher energies. We thus conservatively use the 4--6 keV band count rates to set the requirements for the background-to-source ratio.

Since the nearby clusters fill most of the EPIC FOV, we use blank sky and closed cover observations to estimate the background spectra (see Section \ref{Reduction}) following \cite{N10}.
Comparison of the background spectrum estimate composed of the blank sky and closed cover samples with the individual blank sky observations yielded that in the 4-6 keV band the accuracy of such background estimate for EPIC is $\sim$ 5\% at 1 $\sigma$ confidence \citep{N05}. While there are more up-to-date methods for more accurate background treatment
(e.g. \citet{Marelli21,Gastaldello22}) we prefer \cite{N05} method due to its simplicity and because its accuracy is adequate for our purposes. 
Thus, in order to maintain the uncertainty of the count rate due to the background subtraction below 1\% of the source count rate at all wavelengths in the 0.5-6 keV band, we require that the background-to-source count rate ratio be smaller than 20\% in the 4-6 keV band.

\subsection{Vignetting}
The extended nature of the galaxy clusters invokes a problem not present in the case of using on-axis point sources for effective area calibration studies.
Namely, the effective area decreases with the off-axis angle due to partial obscuration of the mirror elements at such angles. 
The SAS tool \texttt{arfgen} calculates the effective area at a given energy by multiplying the on-axis value with the factor obtained by weighting the vignetting function at different positions of the extraction region with the surface brightness of the cluster. In this work we are not interested in the calibration of the vignetting function but rather want to minimise the effect of its calibration problems. 

Our solution is to keep the extraction radius small enough so that the vignetting effect stays small enough.
The effect is stronger at higher energies, leading to $\sim$ 20\% effective area reduction at r = 6 arcmin for 6 keV photons.
Due to radially decreasing surface density of clusters, most of the photons in a r = 6 arcmin circle centered to the cluster center originate from the less vignetted inner parts.   
An example calculation of the emission weighted vignetting factor of A1795 cluster within the central r=6 arcmin region at 6 keV band yields a smaller than 10\% total vignetting effect for the spectrum extracted within a circle of r = 6 arcmin.
Allowing a very generous 10\% uncertainty for the vignetting factor calibration \citep{2003ExA....15...89L}
 results in systematic uncertainties of the effective area by a maximum of 1\% at 6 keV 
when considering a central r = 6 arcmin region, satisfying our criteria. Thus, we set a maximal extraction region radius of 6 arcmin.

\subsection{Obscuration by the bad pixels}
\label{badpix}
On average $\sim$14\% ($\sim$4-5\%) of the full central r = 6 arcmin pn (MOS) region is obscured by the dysfunctional (i.e. bad) pixels, dominated by the CCD gaps \citep{N21}. This complicates the analysis of extended X-ray sources with non-uniform spatial flux distribution such as galaxy clusters. If not accurately accounted for, the bad pixel obscuration may cause significant variation of 
the flux, which is our essential observable (data$_{ref}$ and data$_{test}$ in Eqs. \ref{R1.eq}, \ref{R2.eq}, \ref{J1.eq},  \ref{R3.eq} and \ref{Jmeas.eq}). We summarise here the test of the accuracy of the CCD gap correction we performed in \citep{N21} (see that work for the details of the test).
 
SAS includes an option for recovering the fraction of the flux obtained by a given EPIC instrument which has been lost due to bad pixel obscuration of the intended extraction region.
Since version 17.0.0, SAS has included an option whereby the user can utilise the information of the spatial distribution of the flux (i.e. an image) of the same source observed with another instrument 
or via a synthetic model image. The procedure is implemented to \texttt{arfgen} whereby the effective area associated with the input spectrum is reduced by the lost flux fraction in the full intended extraction region, as informed by the supplementary image. This option is enabled by specifying  \texttt{badpixmaptype=dataset}. 

In \cite{N21} we compared the fluxes of the cluster sample presented in this paper recovered by the above procedure with those obtained with the simulated model images. We found that the accuracy of the recovery method as implemented to SAS18.0.0 is better than 0.1\% on average while in individual cases the recovered flux may be uncertain by $\sim$1\%. Thus, the accuracy of the procedure satisfies our requirement of maintaining the systematics at 1\% level.  

In reality it is the flux, rather than the effective area which is reduced and thus we need to make sure that our study of the effective area is done in a correct way. We use pn (MOS2) images for applying the correction to MOS (pn) effective area. In 
Eqs. \ref{R2.eq} and \ref{J1.eq}, $arf_{ref}$ corresponds to the pn effective area reduced by the above procedure. Thus, when we fit the pn spectra, the best-fit model ($model_{ref}$) will be scaled up corresponding to the full intended extraction region.
The prediction for the reference instrument using the up-scaled model with the down-scaled effective area ($model_{ref} \times arf_{test}$) corresponds to the partially obscured extraction region where the data is extracted from. Thus, the application of of Eq. \ref{R2.eq} with the CCD gap correction is valid. 

We then use the up-scaled pn model to produce the prediction ($model_{ref} \times arf_{test}$) for the test instrument (Eq. \ref{R1.eq}). The effective area of the test instrument $arf_{test}$ is down-scaled due to CCD gap correction procedure and thus the model prediction for the test instrument corresponds to the same partially obscured region of the test instrument, where the test instrument data are extracted from. Thus, the application of Eq. \ref{R1.eq} is also valid.
 
\subsection{Summary of the criteria}
We summarise here the above discussed requirements and procedures we employ in order to maintain the systematic effects of different calibration and instrumentation components 
on the cross-calibration bias below 1\% level in the studied energy band.

\begin{itemize}

\item
Energy redistribution: Minimal spectral bin size of 400 eV.\\

\item
Energy redistribution: Maximal spectral bin energy of 6 keV.\\

\item
Background: Maximal background-to-cluster count rate ratio at 4--6 keV of 20\%.\\

\item
PSF:  Minimal extraction region radius of 3 arcmin.\\

\item
Vignetting: Maximal extraction region radius of 6 arcmin.\\

\end{itemize}

\section{X-ray data}
\label{xraydata}
\subsection{The sample}
The above requirements translate into selection criteria of the suitable clusters, observations and spectrum extraction regions.
Since we want to minimise the statistical uncertainties of the spectra, we will focus on relatively nearby clusters. In order to minimise the background-to-source count rate ratio, we prefer the hottest clusters since they provide intrinsically more counts at the critical higher energies.

In order to carry out a robust statistical analysis we need a substantial sample of cluster observations (we treat repeated observations of the same cluster as independent data sets).
From the XMM-Newton observation archive we constructed a sample of 27 EPIC observations of galaxy clusters hotter than $\sim$ 5 keV (see Table \ref{basic.tab}). These observations comply to an additional requirement that the pointing off-axis angle from the cluster X-ray peak must be smaller than 3 arcmin so that we do not fold in significant asymmetric vignetting features. The combined exposure time of the flare-filtered (see next section) sample is $\sim$600 ks which renders the number of photons in the spectra in our adopted extraction region and bin size (400 eV)
above 50000 at each spectral bin. This renders the statistical uncertainty of the sample flux below 1\% at each bin, i.e. below the systematic uncertainty level.

\begin{table}
 \centering
  \caption{The basic info of the EPIC data sample 
  \label{basic.tab}}
    \begin{tabular}{lccccc}
  \hline\hline
           &            &            &            &                             \\ 
cluster    & Obs. ID    & Rev.       & redshift   & bkg/src\tablefootmark{a}   \\
           &            &            &            &     (\%)                      \\
\hline 
A85        & 0723802101 &  2476      & 0.0551       & 7 \\   
A85        & 0723802201 &  2477      & 0.0551       & 6  \\ 
A401       & 0112260301 &  394       & 0.0737       & 8  \\ 
A478       & 0109880101 &  401       & 0.0881       & 5  \\
A644       & 0744412201 &  2624      & 0.0704       & 8  \\
A754       & 0136740101 &  262       & 0.0542       & 8  \\
A1650      & 0093200101 &  377       & 0.0838       & 18  \\
A1651      & 0203020101 &  835       & 0.0850       & 12  \\
A1795      & 0097820101 &  100       & 0.0625       & 6   \\
A2029      & 0551780301 &  1577      & 0.0773       & 4   \\
A2142      & 0674560201 &  2123      & 0.0909       & 6   \\
A2244      & 0740900101 &  2678      & 0.0968       & 19  \\
A2319      & 0302150101 &  1069      & 0.0557       & 4   \\
A2319      & 0302150201 &  1087      & 0.0557       & 4   \\
A2319      & 0600040101 &  1827      & 0.0557       & 4   \\
A3112      & 0603050101 &  1761      & 0.0753       & 14  \\
A3391      & 0505210401 &  1504      & 0.0514       & 26  \\
A3558      & 0107260101 &  388       & 0.0480       & 10  \\
A3571      & 0086950201 &  483       & 0.0391       &  4   \\
A3827      & 0149670101 &  538       & 0.0984       & 16  \\
Coma       & 0300530101 &  1012      & 0.0231       & 5   \\
Coma       & 0153750101 &  364       & 0.0231       & 4   \\
CygA       & 0302800101 &  1071      & 0.0561       & 5   \\
PKS0745    & 0105870101 &  164       & 0.1028       & 5   \\
Ophiuchus  & 0505150101 &  1416      & 0.028        & 1   \\
Triangulum & 0093620101 &  219       & 0.051        & 4   \\
ZwCl1215   & 0300211401 &  1198      & 0.0913       & 18  \\
 \hline 
\end{tabular}
\tablefoot{\\
\tablefoottext{a}{EPIC-pn background-to-source count rate ratio in the 4--6 keV band for spectra extracted within the central r = 6 arcmin region.} 
}
\end{table}

\subsection{Reduction}
\label{Reduction}
We used SAS19.1.0 for processing the archival raw event files and producing the spectra and responses in Nov 2021, using the latest calibration files (CCF:s) available at that date.
We applied the SAS tools \texttt{epchain} and \texttt{emchain} with default parameters to the raw data in order to produce the event files for the observations and the simulated out-of-time\footnote{http://xmmtools.cosmos.esa.int/external/xmm\_user\_support/documentation- /sas\_usg/USG/epicOoT.html}
(OOT) file for pn. We filtered the event files using patterns 0-4 for pn and 0-12 for MOS
and applying expressions \texttt{flag==0} for pn and \texttt{\#XMMEA\_EM} for MOS.
We further filtered the event files to minimise the flares by accepting only such periods when the E $>$ 10 keV band flux was within $\pm$20\% of the quiescent level.
 
We extracted the spectra at circular regions centered at the location of the X-ray peak. 
While a larger extraction radius improves the statistical precision of the spectrum, it also increases the background to source ratio.
We adopted the extraction radius of 6 arcmin since this value satisfies the requirements due to PSF minimisation (not too small radius) and vignetting (not too large radius) and yields background-to-source ratio (see below) smaller than 20\% in the 4--6 keV band in vast majority of the spectra (see Table \ref{basic.tab}).
  
We extracted the spectra in 5 eV bins and corrected the observational pn spectrum with the OOT spectrum.
We produced the energy redistribution matrices (rmf) using rmfgen tool, using the image of the cluster for weighting.
We produced the effective area files (arf) as described in Section \ref{badpix}. 
Importantly, we did not apply the average EPIC-MOS effective area correction described in the XMM-Newton calibration note XMM-CCF-REL-382 
(see Section \ref{discussion} for discussion on this approach in comparison to ours).
 
\begin{figure*}
\hbox{
\includegraphics[width=8cm,angle=0]{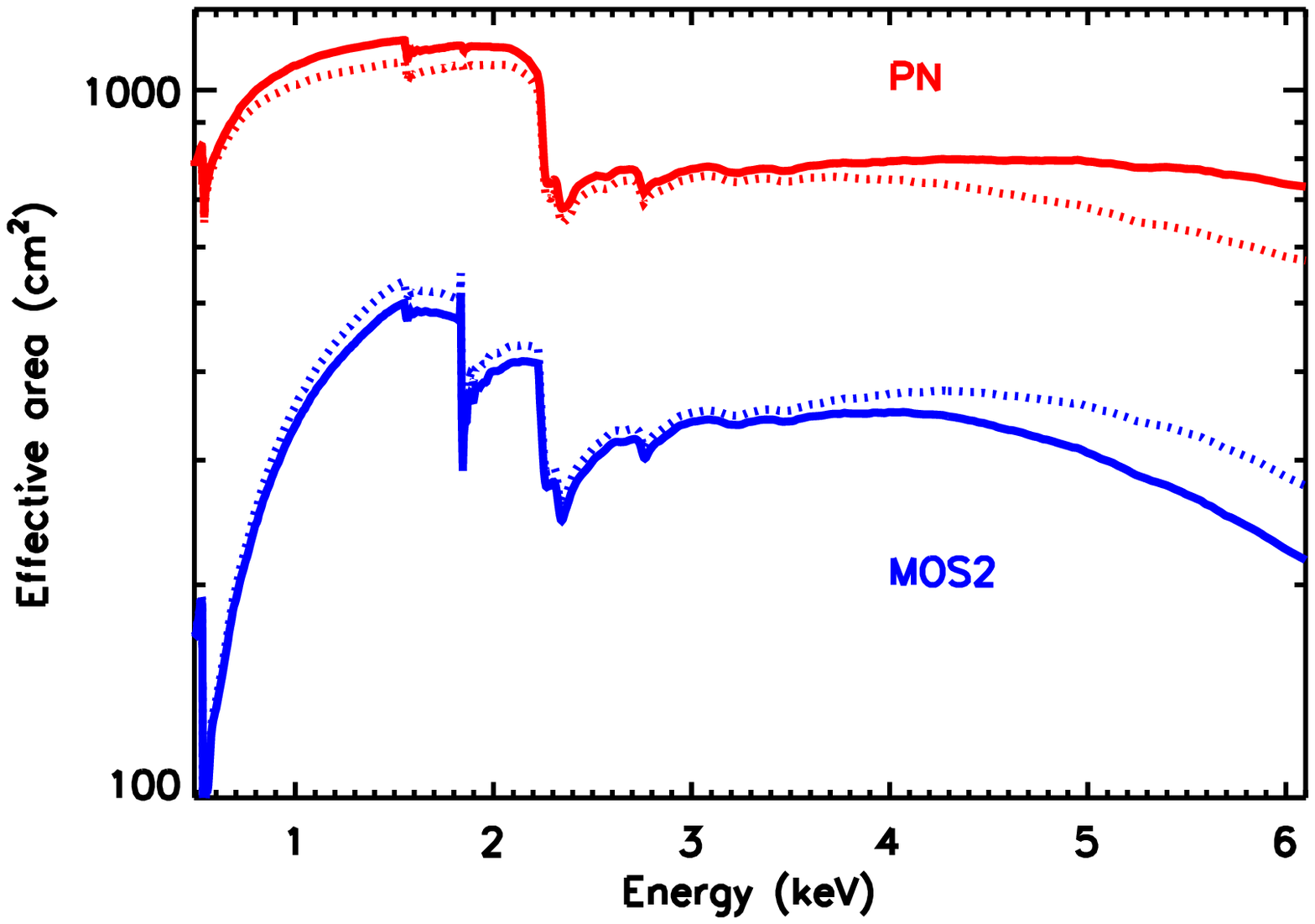}
\includegraphics[width=8cm,angle=0]{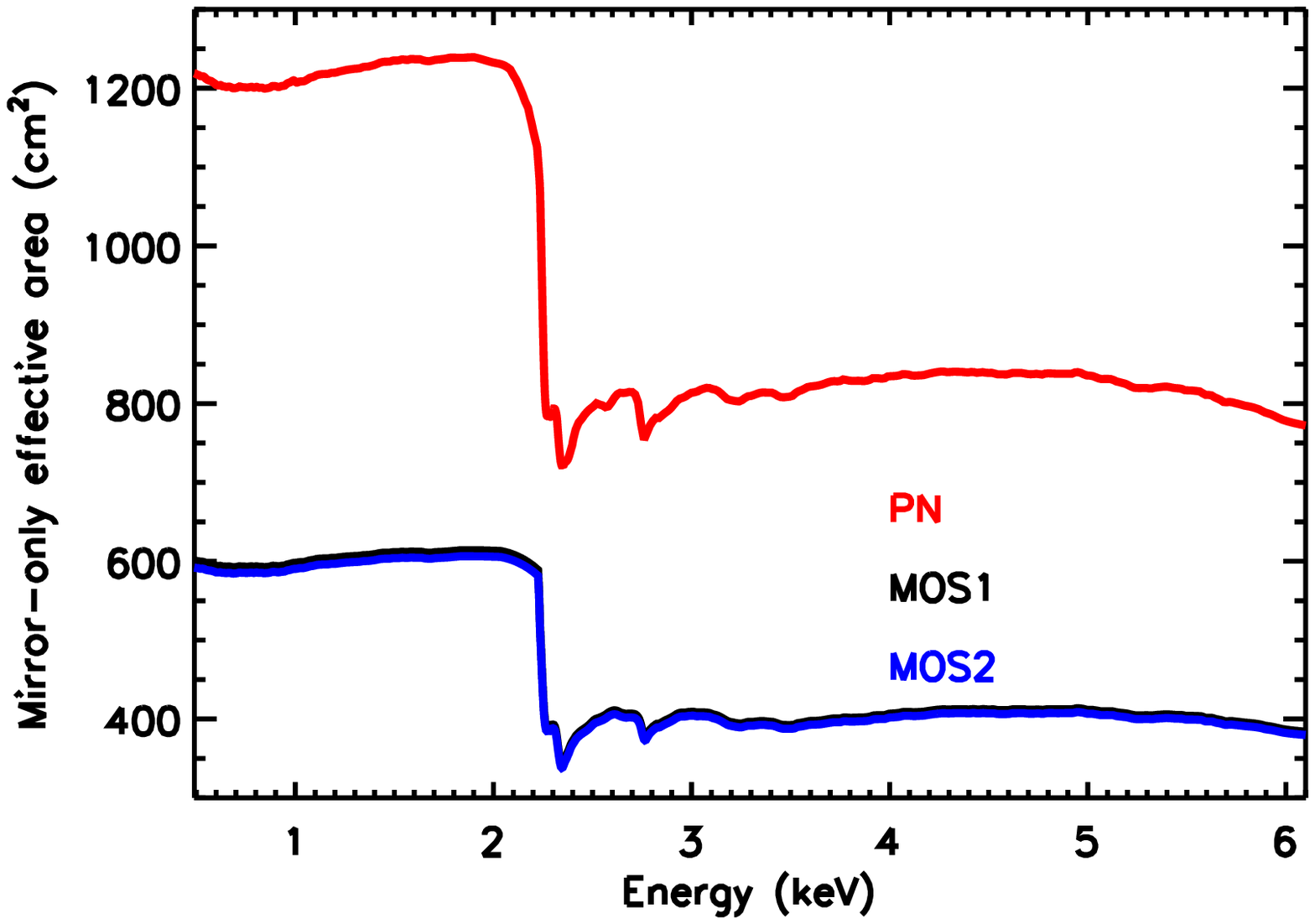}
}
\hbox{
\includegraphics[width=8cm,angle=0]{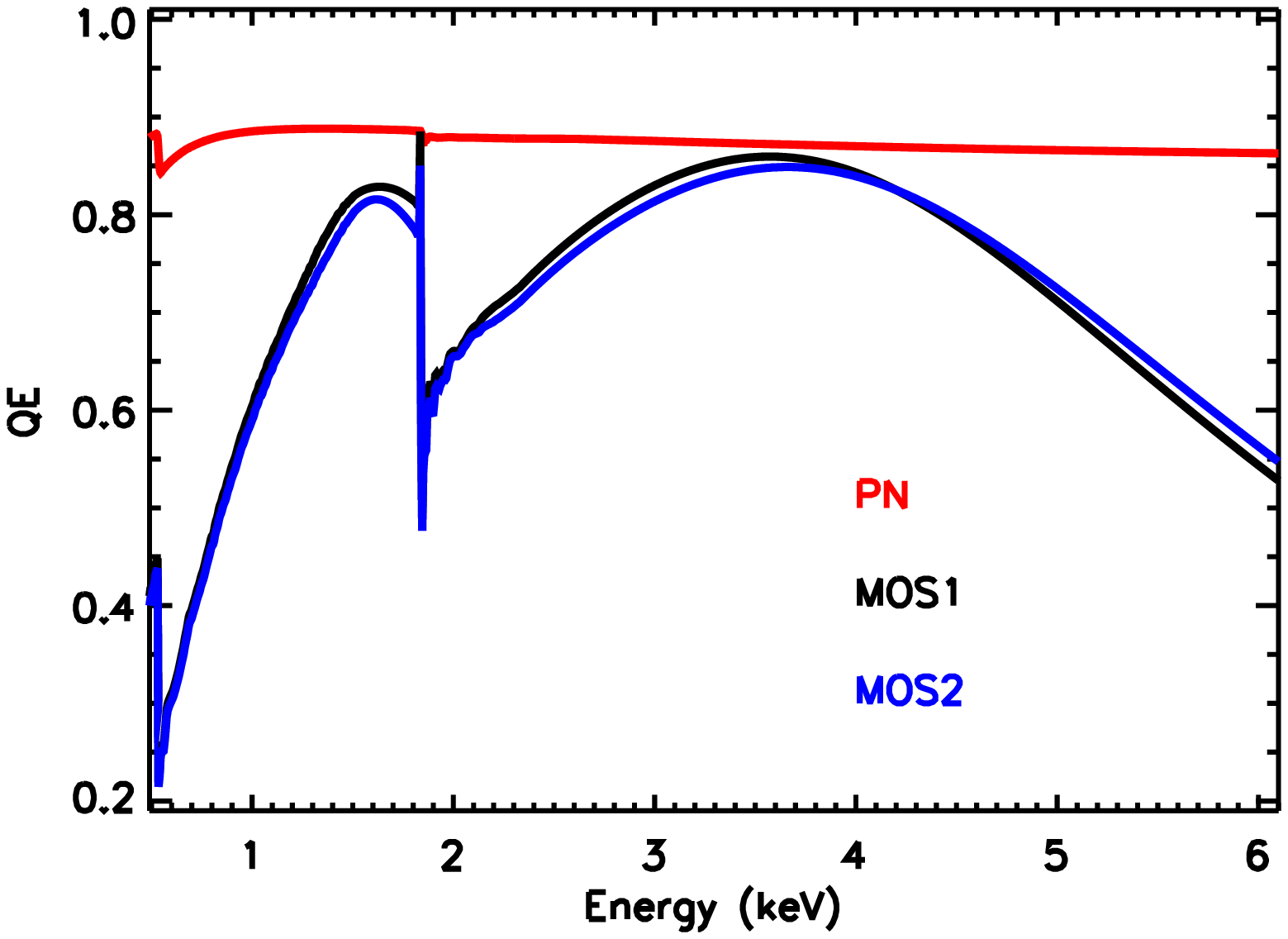}
\includegraphics[width=8cm,angle=0]{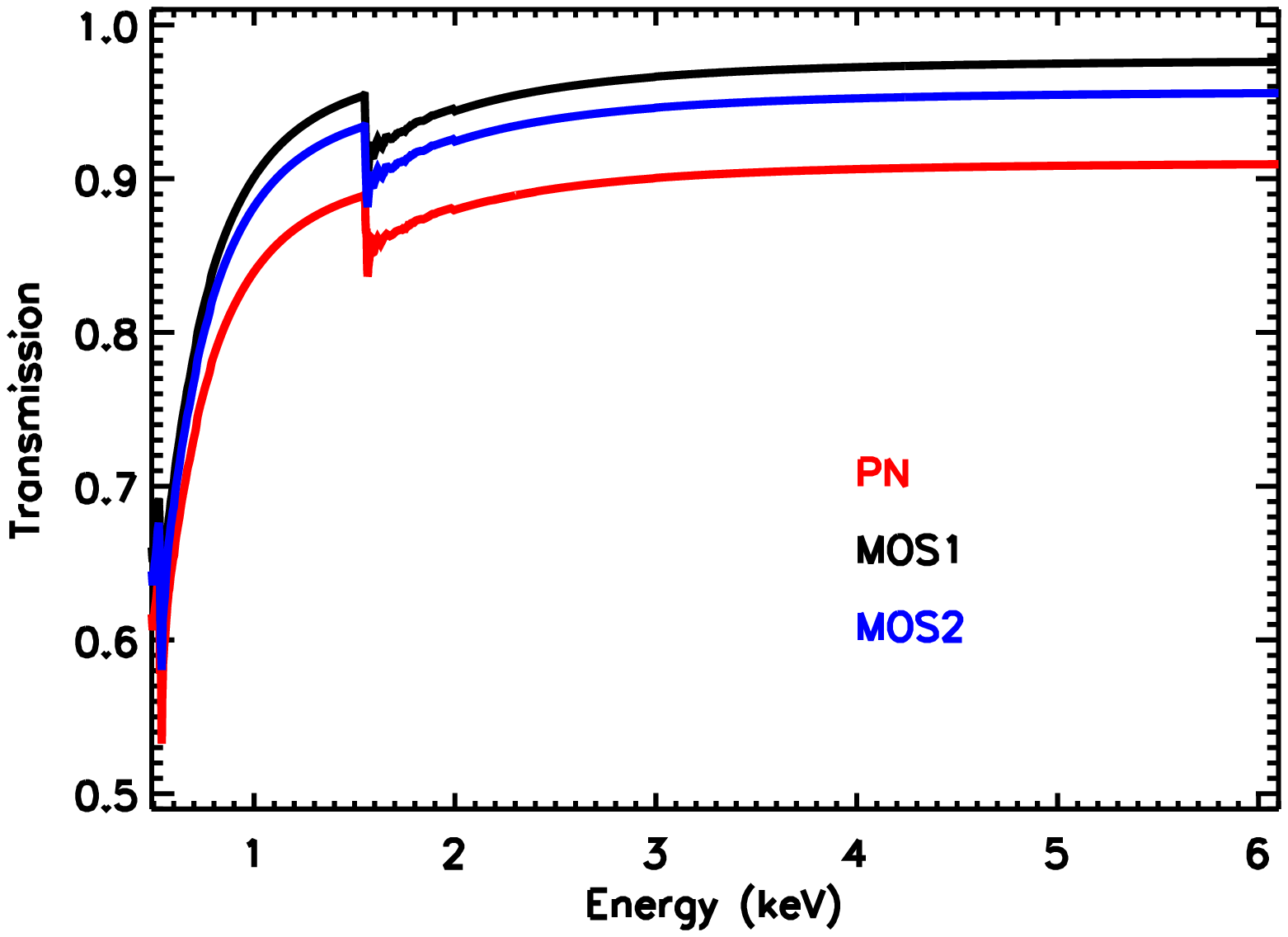}
}
\caption{Different components of instrumentation for A1795 observation 0723802101 for pn (red lines), MOS1 (black lines) and MOS2 (blue lines).
Top left panel shows the total effective area generated with $arfgen$ (solid lines, for clarity MOS1 is not shown). The dotted lines indicate the average correction to the effective area assuming that the other instrument is perfectly calibrated (see Section \ref{Application}). Other panels show the mirror-only effective area (top right), quantum efficiency of the detector (bottom left) and the transmission of the optical filter (bottom right). }
\label{aeff.fig}
\end{figure*}

To investigate the origin of the possible calibration problems, we additionally used arfgen to separate the mirror effective area, QE of the detector and optical filter transmission curves for A1795 (see Fig. \ref{aeff.fig}).

Following \cite{N05}, we estimated the total and particle background spectra utilising a sample of blank sky and closed cover observations with XMM-Newton/EPIC.
We extracted these spectra using the same detector region as applied to our cluster sample, i.e. a central circle with r = 6 arcmin and co-added them in order to obtain single 
blank sky and closed cover spectra to be applied to all clusters (pn, MOS1 and MOS2 separately).
The blank sky spectrum approximates the sum of the photon background and the quiescent particle background. 
We additionally extracted the background and cluster spectra in the full FOV in order to account for the variability of the residual particle flares enter our data despite of our flare filtering. Namely, we obtained the normalisation of the residual flare spectrum (we assumed that its shape is that of the closed cover spectrum) for a given cluster observation by matching the E > 10 keV count rate (dominated by particles due to very low mirror effective area at these energies) of the full FOV particle background spectra with that of the full FOV cluster spectrum. 
We then subtracted such normalised particle background spectrum from the cluster spectrum, together with the blank sky spectrum.

\subsection{Spectral analysis}
\label{spectral-analysis}
We used the XSPEC software \citep{xspec} in order to perform spectral analysis.
We first binned the spectra to contain a minimum of 100 counts per bin. 
We modelled the background subtracted (see Section \ref{Reduction}) 0.5-6.0 keV band data of EPIC-pn (data$_{ref}$ in Eq. \ref{R2.eq}) with a $MEKAL$ emission model absorbed by $phabs$ model in case of Method 1. When following Method 2, the thermal model is modified by the splines (see Section \ref{method2}). 
We adopted redshifts from NASA's Extragalactic Database  and the Galactic NH values from \cite{Willingale}. 
We fixed the metal abundance to 0.3 Solar.

We formed the model prediction by multiplying the trial models with the effective area file (arf, including the bad pixel correction, see Section \ref{badpix}) and convolving the product with the redistribution matrix (rmf). We fitted the data with such a model prediction and the consequent best fit model corresponds to the reference model (model$_{ref,1}$ and model$_{ref,2}$ in Eqs.
 \ref{R1.eq}, \ref{R2.eq}, \ref{J1.eq},  \ref{R3.eq} and  \ref{Jmeas.eq}.)
  
\section{Results}
\label{Results}
When discussing the sample for a given instrument pair as a whole we prefer to use the sample median over the mean since the relatively small number of the data points in our sample
does not allow robust evaluation of the shape and in particular the symmetry of the distributions. 
Possible deviations from symmetry could bias the sample mean. We thus chose to use the more general sample median as a representative value of the distribution. 
This choice also mitigates the problem arising from the inhomogeneous statistical quality of the sample: the clusters with higher number of counts would bias the weighted mean.
We adopted the error of the mean, multiplied by a factor of 1.3 \citep{sokal} as the statistical uncertainty of the sample median. 

\subsection{Comparing the methods}
\label{comparing-the-methods}
\begin{figure*}
\hbox{
\hspace{-0.5cm}
\includegraphics[width=10cm,angle=0]{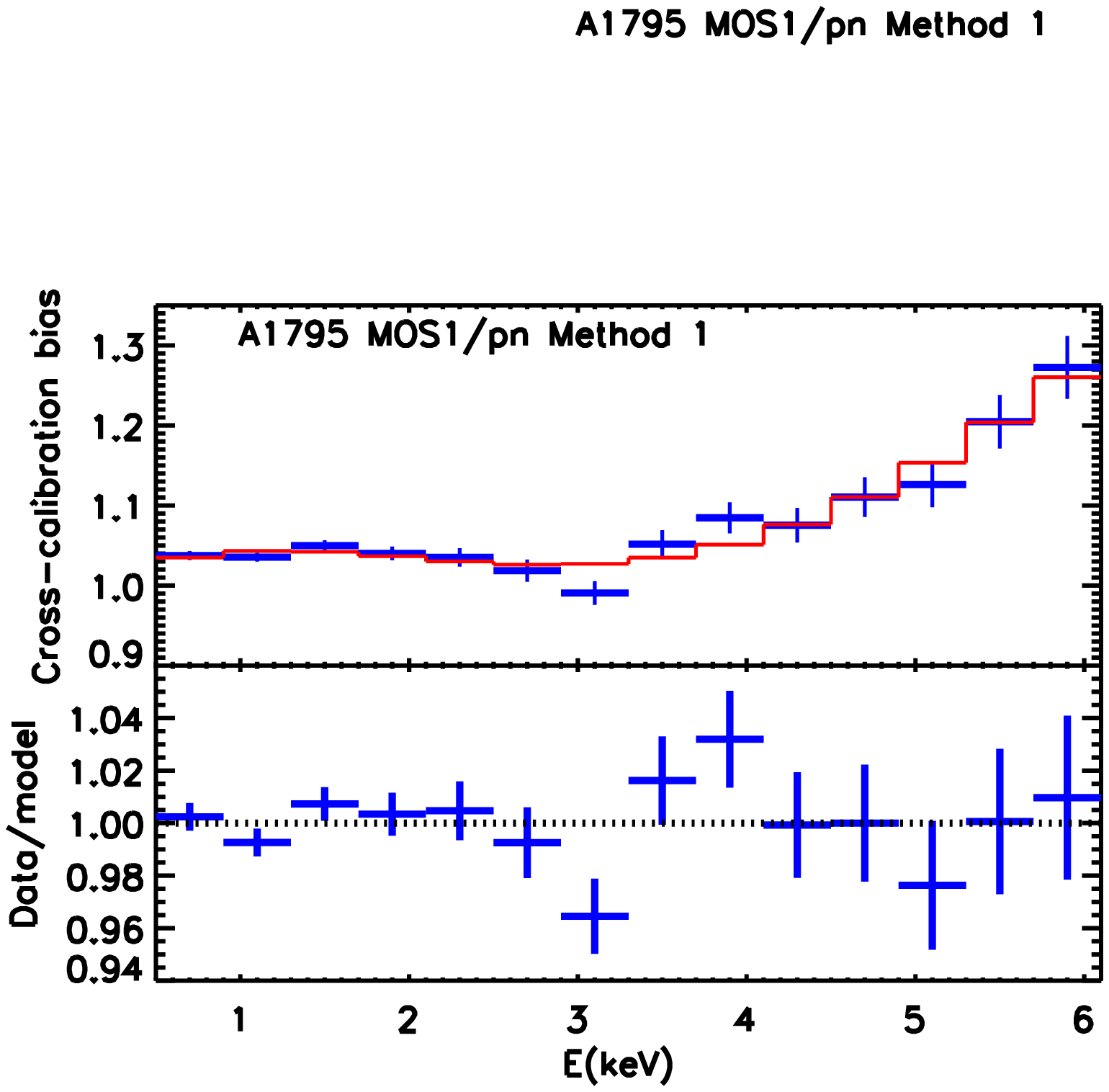}
\hspace{-1cm}
\includegraphics[width=10cm,angle=0]{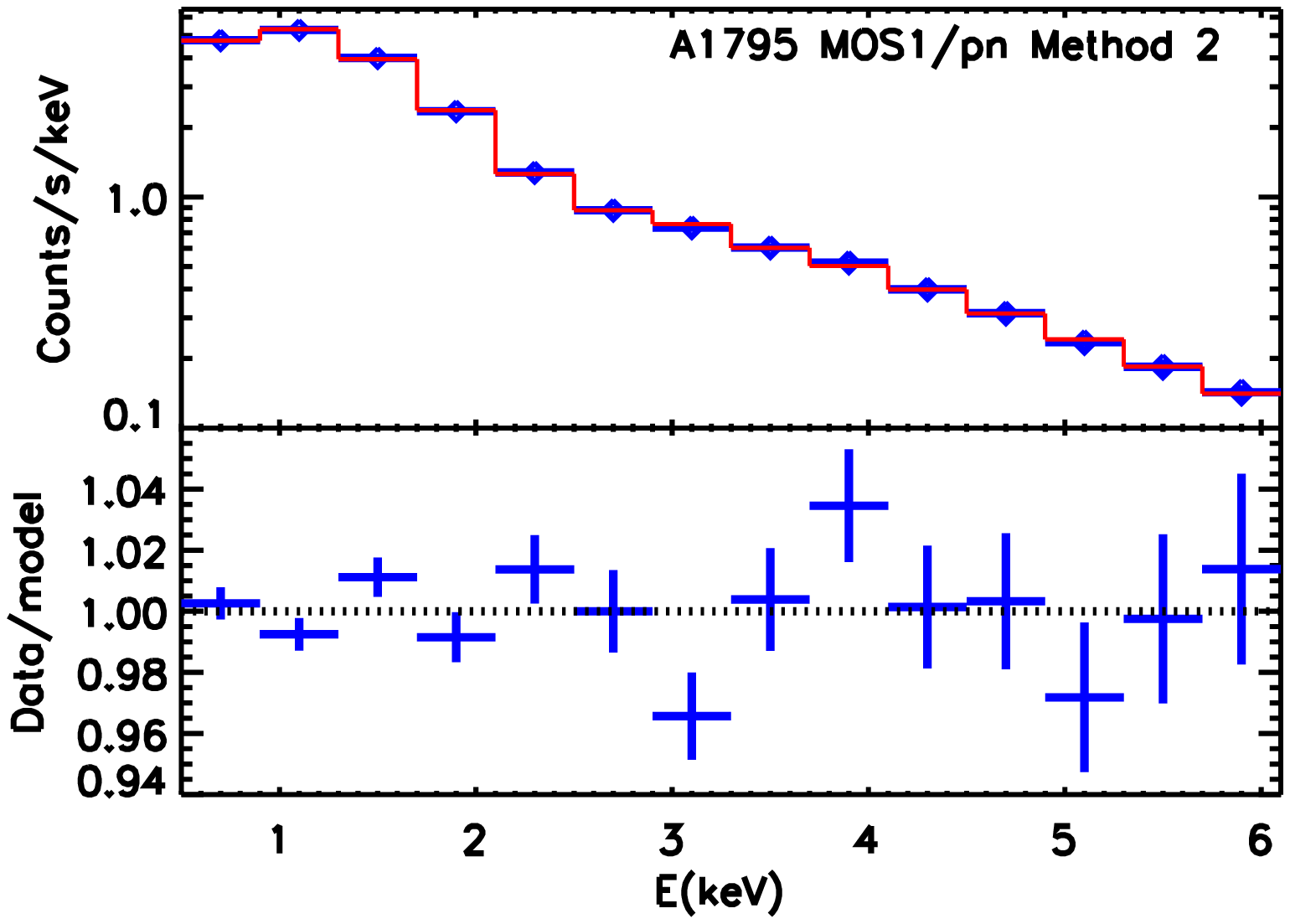}
 }
\caption{Demonstration of the quality of the modelling of the MOS1/pn cross-calibration bias for A1795 using Method 1 (left panel) and Method 2 (right panel). When applying Method 1, we evaluate Eq. \ref{J1.eq}, i.e. we approximate the cross-calibration bias after the convolution
(blue crosses in the upper left panel). The best-fit model (Eq. \ref{poly.eq}) to such data is shown as red line. When applying Method 2, we fit the MOS1 spectra (blue symbols in the upper right panel, the statistical uncertainty is smaller than the plot symbol) with a 
model consisting of a fixed component for accurate modelling of the pn data (see text for details) while the cross-calibration bias is modelled as 4th order polynomial (the total model is shown as a red line in the upper right panel, see Eq.\ref{R3.eq}).
In both cases the lower panel shows the ratio of the best-fit model to the data.
}
\label{MOS1-pn-A1795_method-comparison.fig}
\end{figure*}

Following the description in Section \ref{methods}, we evaluated the MOS1/pn cross-calibration bias using Methods 1 and 2. The 4th order polynomials yielded adequate fits to the full band (see Fig. \ref{MOS1-pn-A1795_method-comparison.fig} for the fit quality for MOS1/pn data of A1795 cluster). There are single spectral bins where the data deviate statistically significantly at a few \% level from the best-fit model. These features appear at the same energies and with similar magnitudes when using either of the two methods. They are hard to understand in terms of the effective area components (mirror effective area, filter transmission and detector QE). The apparently chaotic behaviour of the above isolated deviations with time and energy suggests that this issue could be due to the data processing and not inherent to the cross-calibration. The possible patterns of these features could be used to examine in detail the cause of such scatter which is out of the scope of this work.
 
\begin{figure*}
\hbox{
\hspace{-0.5cm}
\includegraphics[width=10cm,angle=0]{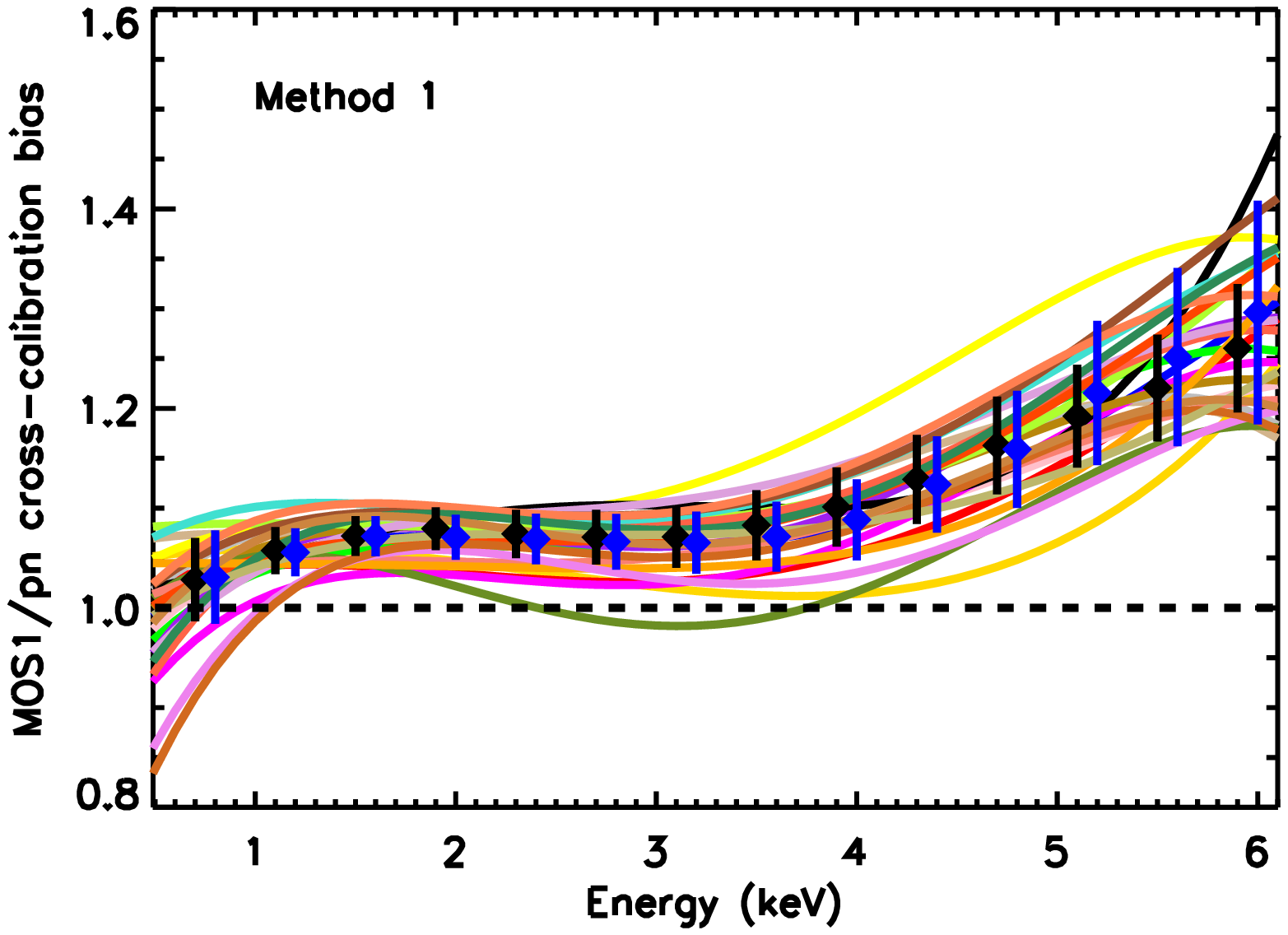}
\hspace{-1cm}
\includegraphics[width=10cm,angle=0]{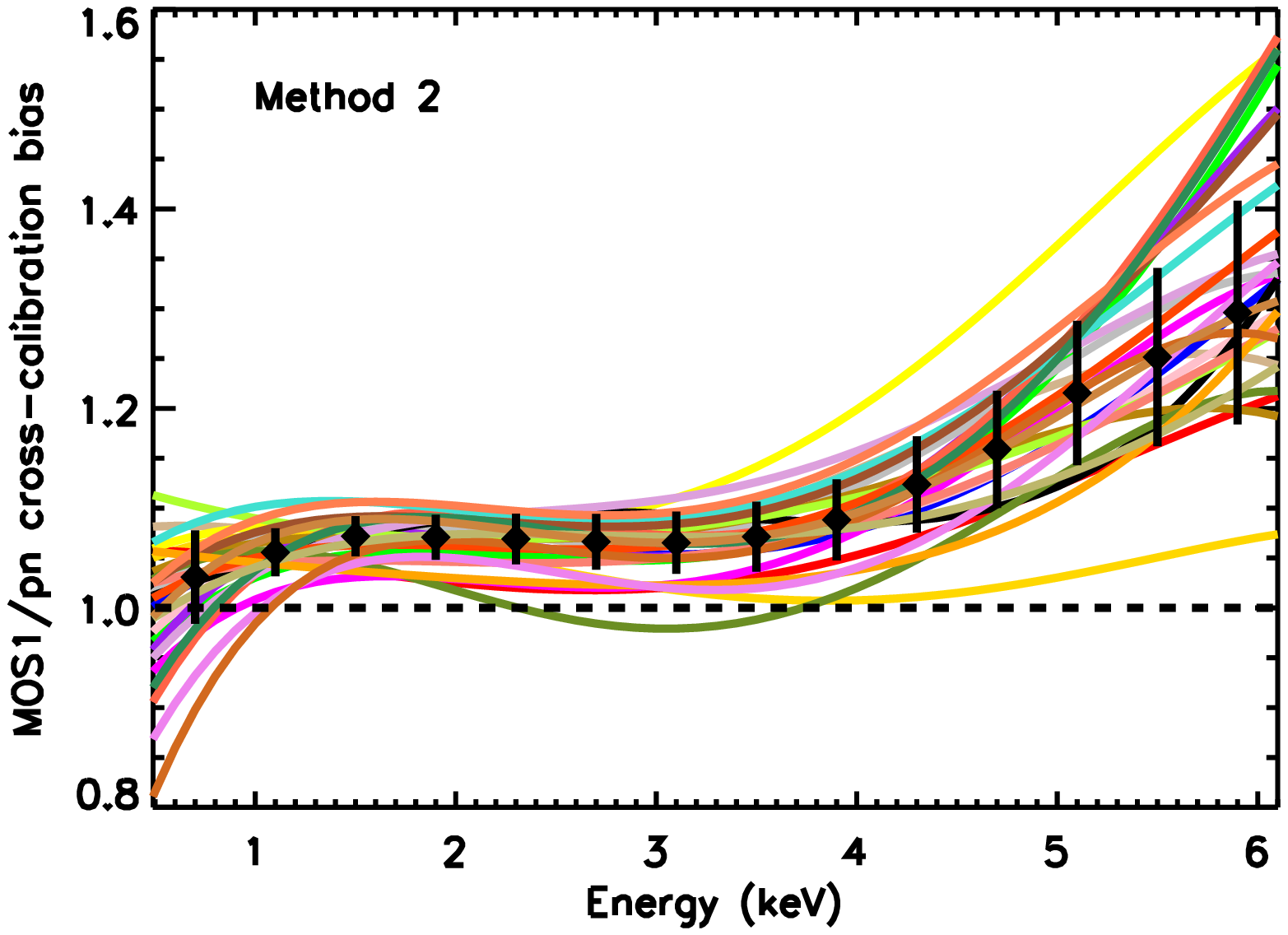}
 }
\caption{
The best-fit models for each individual cluster obtained by fitting the MOS1/pn cross-calibration data 
 using Eq. \ref{poly.eq} when applying Method 1 (left panel) or Method 2 (right panel).
 Each solid line connects the values of one cluster at the energy bin centers. The black symbols and vertical lines indicate the median and the standard deviation of the sample at each energy. The symbols from the right panel are repeated in the left panel in blue. The horizontal dashed line at unity indicates the expectation in case of no cross-calibration bias.
}
\label{MOS1-pn-method-comparison.fig}
\end{figure*}

We performed the comparison of the two methods in terms of the individual best-fit 4th order polynomials (see Fig. \ref{MOS1-pn-method-comparison.fig}).
While the median J deviates by less than 1\% at the lower energies, the deviations become larger at higher energy, reaching 3\% at 6 keV.
The models obtained via Method 2 tend to raise more sharply with energy while this is suppressed when analysing the convolved results via Method 1.
Also, the cluster-to-cluster scatter is very similar in the two approaches except at the highest energies, where Method 2 has more scatter while it is somewhat suppressed when applied Method 1. These minor effects may be due to the difference of the relative net redistributed flux in the pn and MOS, reported in Section \ref{redistribution}, which increases with energy.

While our requirement for the systematic uncertainty components is less than 1\%, and since the Method 1 may be underestimating the true scatter, we adopt Method 2 in the following when reporting most of the results. Due to practical reasons, in some instances when the small differences between Methods 1 and 2 are not important,
we use Method 1 to highlight some aspects of the cross-calibration bias.

\subsection{Measure of the cross-calibration bias}
\label{measure}
\begin{figure*}
\hbox{
\hspace{-0.5cm}
\includegraphics[width=6.5cm,angle=0]{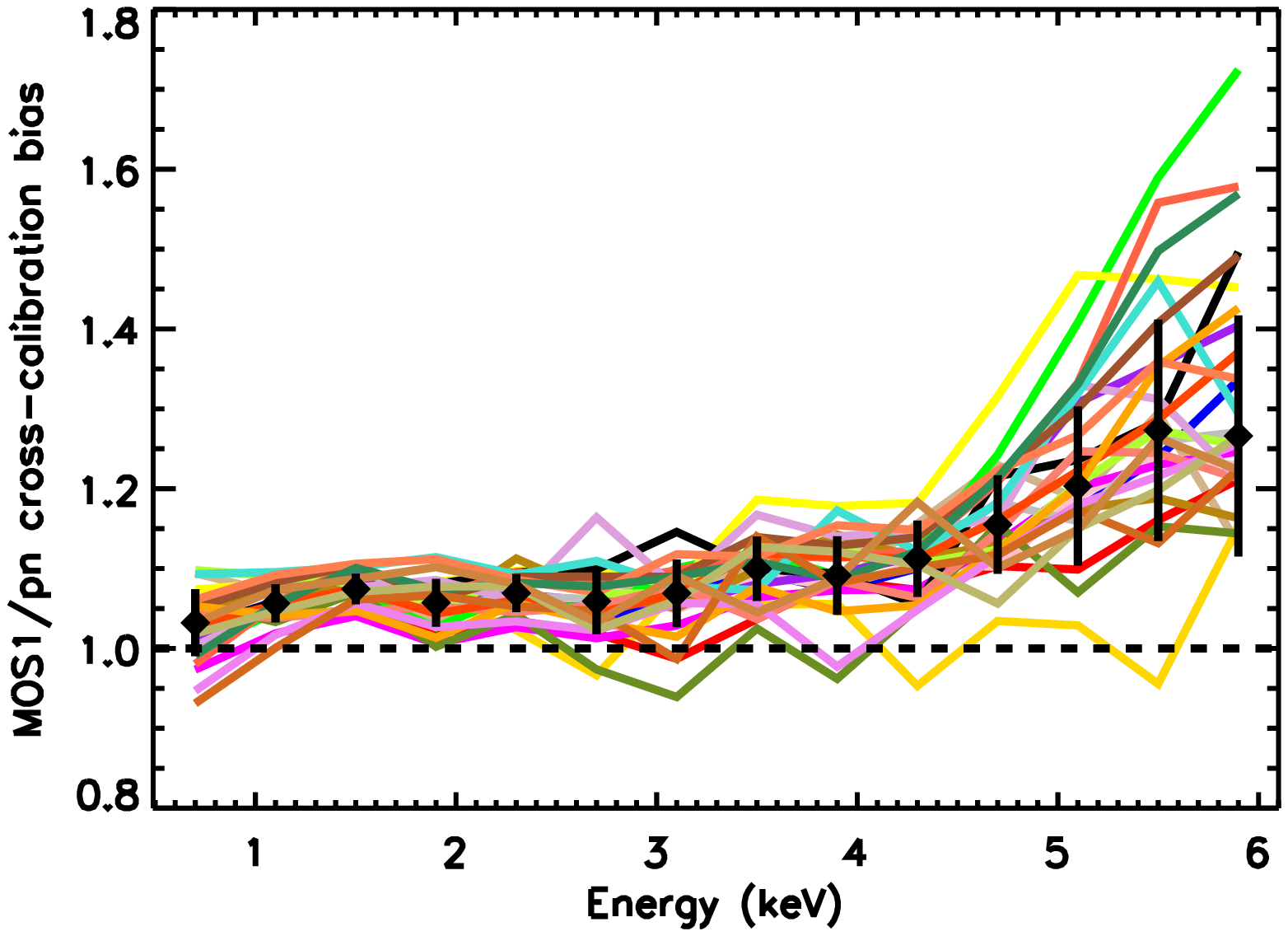}
\hspace{-0.5cm}
\includegraphics[width=6.5cm,angle=0]{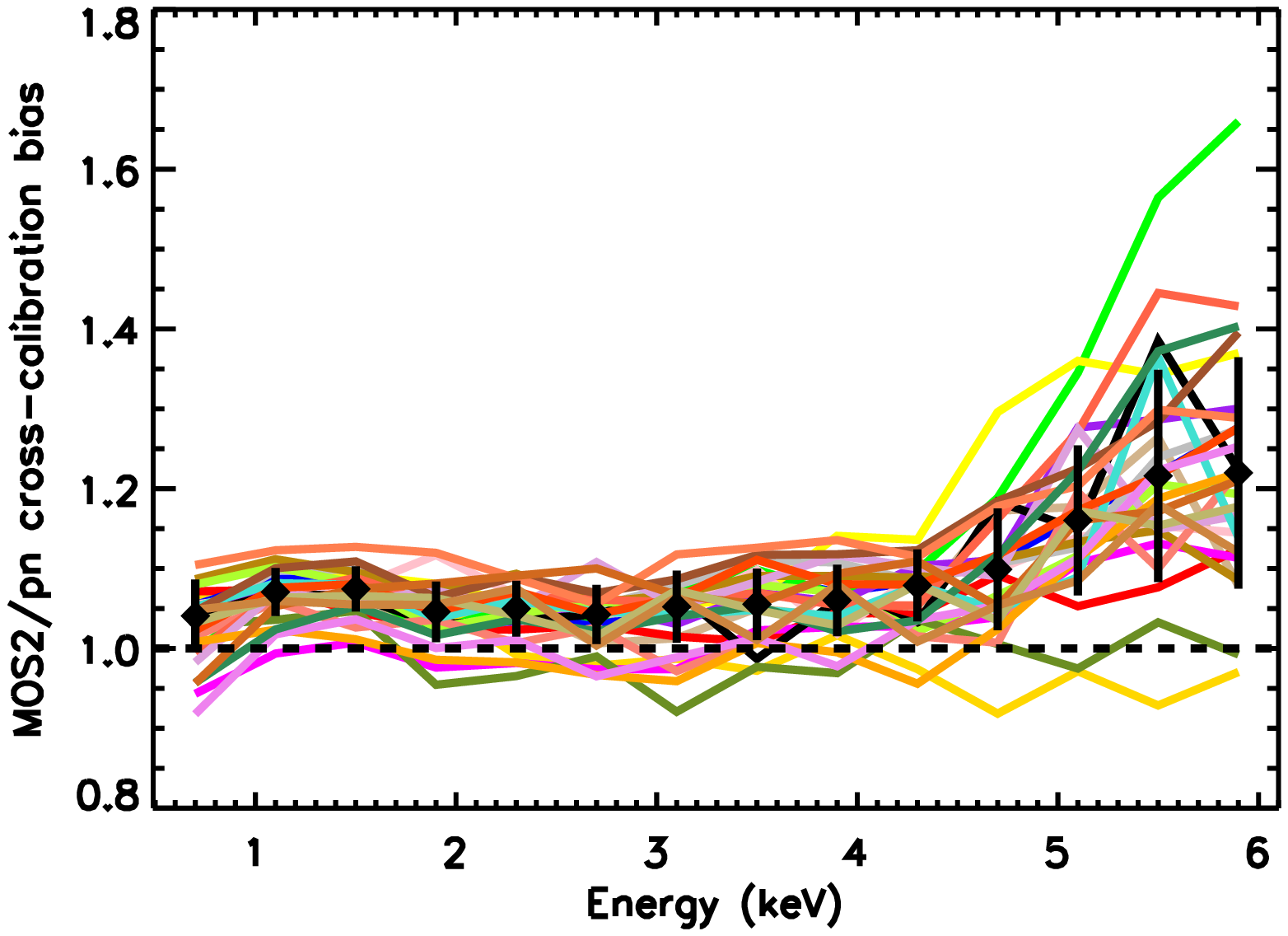}
\hspace{-0.5cm}
\includegraphics[width=6.5cm,angle=0]{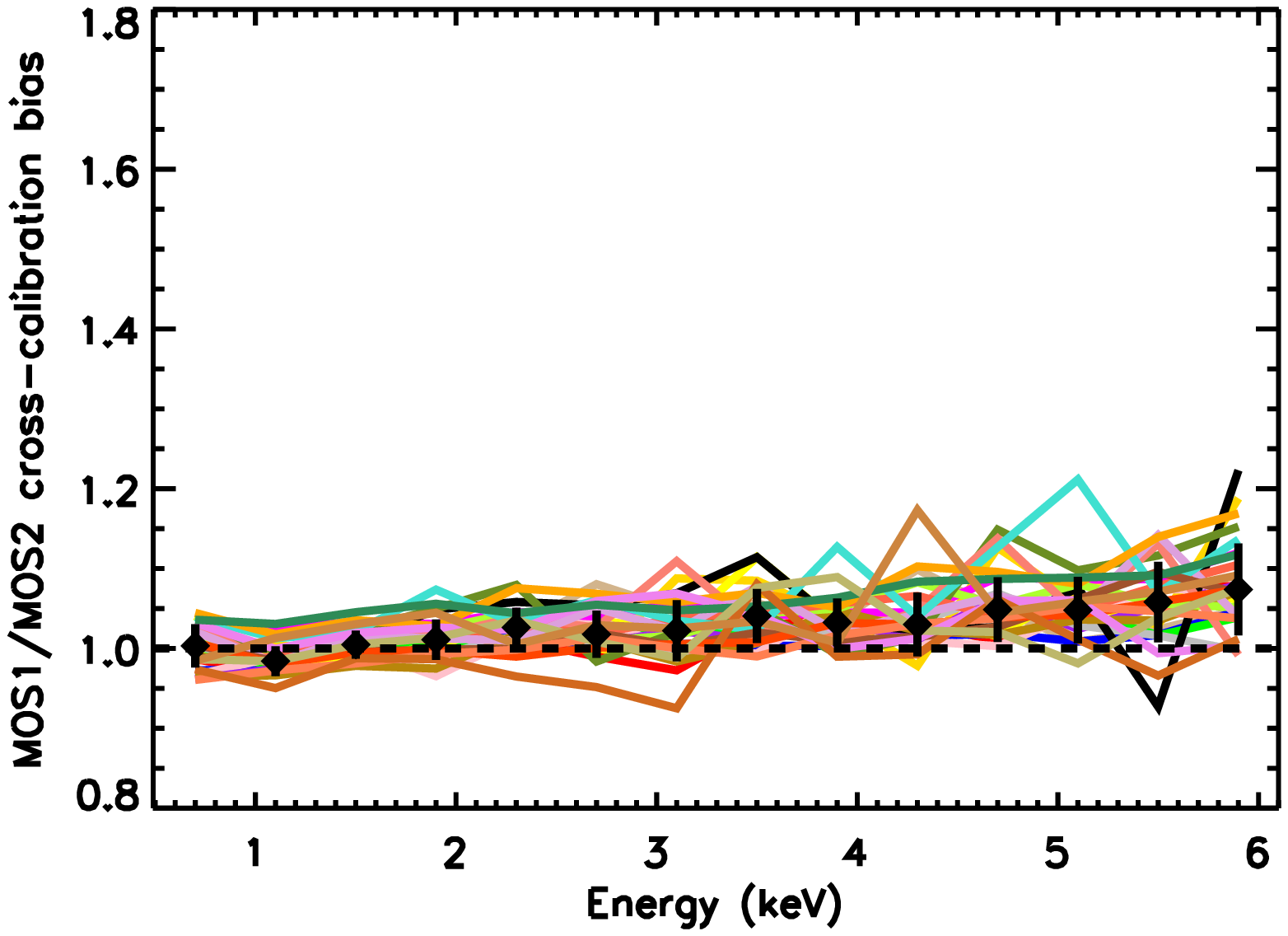}
}
\hbox{
\hspace{-0.5cm}
\includegraphics[width=6.5cm,angle=0]{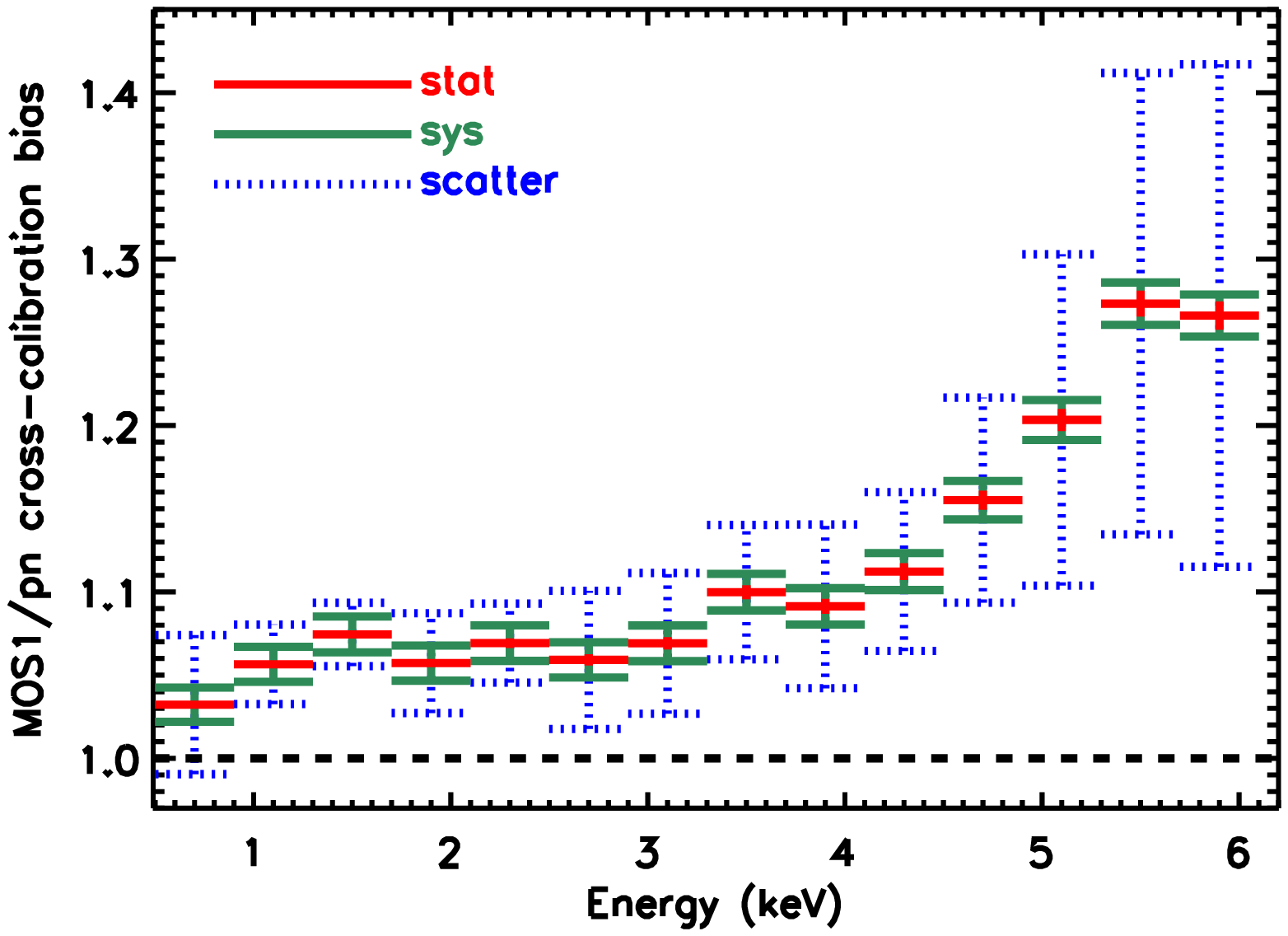}
\hspace{-0.5cm}
\includegraphics[width=6.5cm,angle=0]{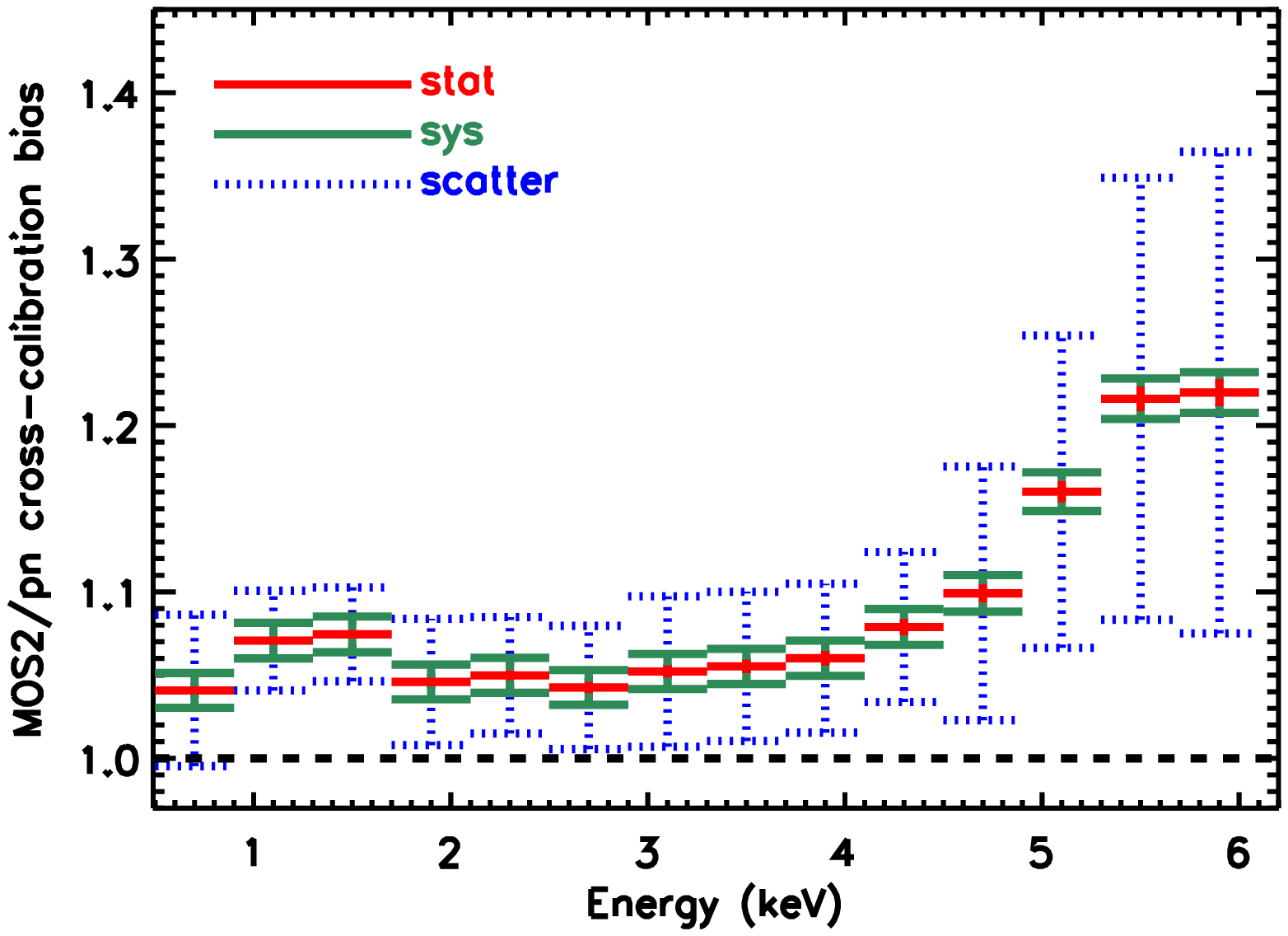}
\hspace{-0.5cm}
\includegraphics[width=6.5cm,angle=0]{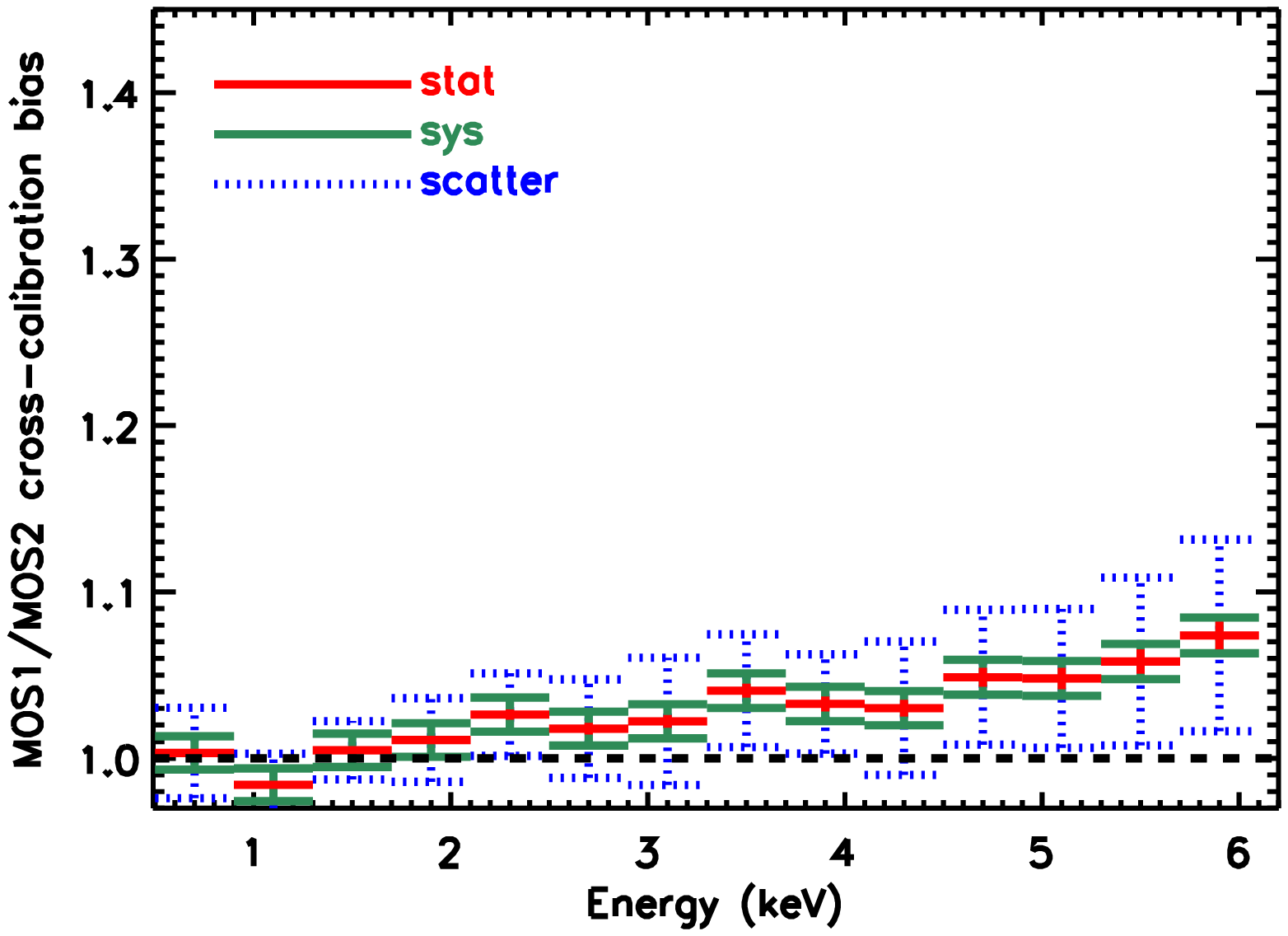}}
\caption{{\it Upper panels:} The measure of the cross-calibration bias (J$_2$, see Eq. \ref{Jmeas.eq}) for each observation for MOS1/pn (left panel) and MOS2/pn (middle panel) obtained via Method 2 (see text).
Each line connects the values of one cluster at the energy bin centers. The black symbols and vertical lines indicate the median and the standard deviation of the sample at each energy.  
{\it Lower panels:} The median of the measures of cross-calibration bias (J$_2$, see Eq. \ref{Jmeas.eq}) of the sample and their statistical (red crosses) and systematic (green crosses) uncertainties for MOS1/pn (left panel) and MOS2/pn (middle panel) pairs. The blue crosses indicate the standard deviation of the sample at a given energy (they are equal to the black horizontal lines in the upper panels). In all panels, the horizontal dashed line indicates the expectation (unity) in case of no cross-calibration bias. Please note that the scales of the vertical axes are different in upper and lower panel in order to highlight better the details.  
{\it Right panels:} The approximative results for MOS1/MOS2 pair are obtained by dividing the results of the left and middle panels.
}
\label{Jmeas.fig}
\end{figure*}

Before discussing in detail the results of modelling the cross-calibration bias with 4th order polynomials, we want to visualise the cross-calibration signal we are modelling.
The concept of the cross-calibration $data$ is complex since in order to evaluate the cross-calibration bias between two instruments following Method 2, we $fit$ two independent data sets, the X-ray spectra of the two instruments. Essentially the cross-calibration bias is a ratio of the data from the test instrument and the prediction of the model describing the reference instrument data. To obtain an observation-based measure of the cross-calibration bias we modified the function (Eq. \ref{R3.eq}) used when finding the model for the cross-calibration bias
by removing that model. The measure of the cross-calibration bias is thus given by   
\begin{equation}
J_2 = \frac{{\rm data}_{test}}{({\rm model}_{ref,2} \times  {\rm arf}_{test}) \otimes {\rm rmf}_{test}}~.
\label{Jmeas.eq}
\end{equation}
Such measure does not depend on the model as long as the model describes the data of the reference instrument accurately. Thus we will treat this measure in the following
as data in order to examine the statistical properties of the cross-calibration bias J$_2$ for 
MOS1/pn and MOS2/pn pairs (J$_{\rm 2,MOS1/pn}$ and J$_{\rm 2,MOS2/pn}$) at each energy bin for each observation (see Fig. \ref{Jmeas.fig} and Table \ref{statsys.tab}).

In this work we focus on the cross-calibration between the pn and the MOS units by treating the pn as a reference instrument and the MOS units as test instruments. In order to obtain qualitative results for the cross-calibration status between the MOS units for the discussion we divide the MOS1/pn cross-calibration bias curves with those of the MOS2/pn pair. A more robust quantitative evaluation of MOS1/MOS2 cross-calibration would require repeating the analysis reported in this work but replacing the pn as a reference instrument with one of the MOS units and using the other MOS unit as a test instrument.

The visual inspection of the results indicates that while the energy dependence of the median J$_{2,MOS/pn}$ values in the 0.5-4.5 keV is to the first order linear, there is a sharp upturn at $\sim$5 keV (see Fig. \ref{Jmeas.fig}).
The visual inspection also indicates that while the individual J$_2$ curves behave quite systematically with energy, there is a substantial scatter (see Fig. \ref{Jmeas.fig}, upper panels). 
This is not driven by some outliers but seems to be present in the full sample, increasing with energy.
 We will investigate these features in detail below. 

\begin{table*}
 \centering
  \caption{Statistical results of the measure of the cross-calibration bias obtained with Method 2 for the cluster sample (available on-line)
  \label{statsys.tab}}
    \begin{tabular}{lcccccccccccc}
  \hline\hline
        &      &  & & & & & &    & & & &                                   \\ 
        & \multicolumn{4}{c}{\bf MOS1/pn}  & \multicolumn{4}{c}{\bf MOS2/pn} & \multicolumn{4}{c}{\bf MOS1/MOS2\tablefootmark{f}}\\
E bin\tablefootmark{a}   & J$_2$\tablefootmark{b} & stat\tablefootmark{c}  & sys\tablefootmark{d} & scat\tablefootmark{e}  & J$_2$\tablefootmark{b} & stat\tablefootmark{c}  & sys\tablefootmark{d} & scat\tablefootmark{e}  & J$_2$\tablefootmark{b} & stat\tablefootmark{c}  & sys\tablefootmark{d} & scat\tablefootmark{e}  \\
\hline
0.5-0.9 &  1.032 & 0.002    & 0.010 & 0.042 & 1.041 & 0.002 &  0.010  &  0.046                & 1.003  & 0.002 & 0.010 & 0.027                              \\ 
0.9-1.3 &  1.056 & 0.002    & 0.011 & 0.024 & 1.071 & 0.002 &  0.011  &  0.030                & 0.984  & 0.002 & 0.010 & 0.019                               \\ 
1.3-1.7 &  1.074 & 0.002    & 0.011 & 0.019 & 1.074 & 0.002 &  0.011  & 0.028                 & 1.005  & 0.002 & 0.010 & 0.018                                      \\ 
1.7-2.1 &  1.057 & 0.002    & 0.011 & 0.030 & 1.049 & 0.002 &  0.010  & 0.037                 & 1.011  & 0.002 & 0.010 & 0.025                                        \\ 
2.1-2.5 &  1.069 & 0.003    & 0.011 & 0.024 & 1.050 & 0.003 &  0.010  & 0.035                 & 1.026  & 0.003 & 0.010 & 0.025                                       \\ 
2.5-2.9 &  1.059 & 0.003    & 0.011 & 0.042 & 1.043 & 0.003 &  0.010  & 0.037                 & 1.018  & 0.004 & 0.010 & 0.030                                     \\ 
2.9-3.3 &  1.069 & 0.004    & 0.011 & 0.042 & 1.052 & 0.004 &  0.011  & 0.045                 & 1.022  & 0.004 & 0.010 & 0.038                                      \\ 
3.3-3.7 &  1.100 & 0.004    & 0.011 & 0.040 & 1.055 & 0.004 &  0.011  & 0.045                 & 1.041  & 0.004 & 0.010 & 0.034                                     \\ 
3.7-4.1 &  1.091 & 0.004    & 0.011 & 0.049 & 1.060 & 0.004 &  0.011  & 0.045                 & 1.033  & 0.004 & 0.010 & 0.030                                      \\ 
4.1-4.5 &  1.112 & 0.005    & 0.011 & 0.048 & 1.079 & 0.005 &  0.011  & 0.045                 & 1.030  & 0.005 & 0.010 & 0.040                                       \\ 
4.5-4.9 &  1.116 & 0.006    & 0.012 & 0.062 & 1.099 & 0.005 &  0.011  & 0.076                 & 1.049  & 0.006 & 0.010 & 0.040                                       \\ 
4.9-5.3 &  1.203 & 0.007    & 0.012 & 0.100 & 1.160 & 0.006 &  0.012  & 0.094                 & 1.048  & 0.006 & 0.010 & 0.042                                      \\ 
5.3-5.7 &  1.273 & 0.008    & 0.013 & 0.139 & 1.216 & 0.007 &  0.012  & 0.133                 & 1.058  & 0.007 & 0.011 & 0.050                                      \\ 
5.7-6.1 &  1.266 & 0.008    & 0.013 & 0.151 & 1.220 & 0.008 &  0.012  & 0.14                  & 1.074  & 0.008 & 0.011 & 0.056                                  \\ 
\hline 
\end{tabular}
\tablefoot{\\
\tablefoottext{a}{The boundaries of the energy bins in keV.} 
\tablefoottext{b}{The sample median measure of the cross-calibration bias (Eq. \ref{Jmeas.eq}).}
\tablefoottext{c}{The photon counting statistics uncertainty of J$_2$.} 
\tablefoottext{d}{The estimate for the level of the systematic uncertainties of J$_2$.} 
\tablefoottext{e}{The standard deviation of the J$_2$ data points at a given energy bin.} 
\tablefoottext{f}{The approximative results for MOS1/MOS2 pair are obtained by dividing the J$_2$ curves of MOS1/pn and MOS2/pn pairs for each cluster.}
}
\end{table*}

 \subsection{Scatter}
  \label{Scatter}
\begin{figure}
\includegraphics[width=9cm,angle=0]{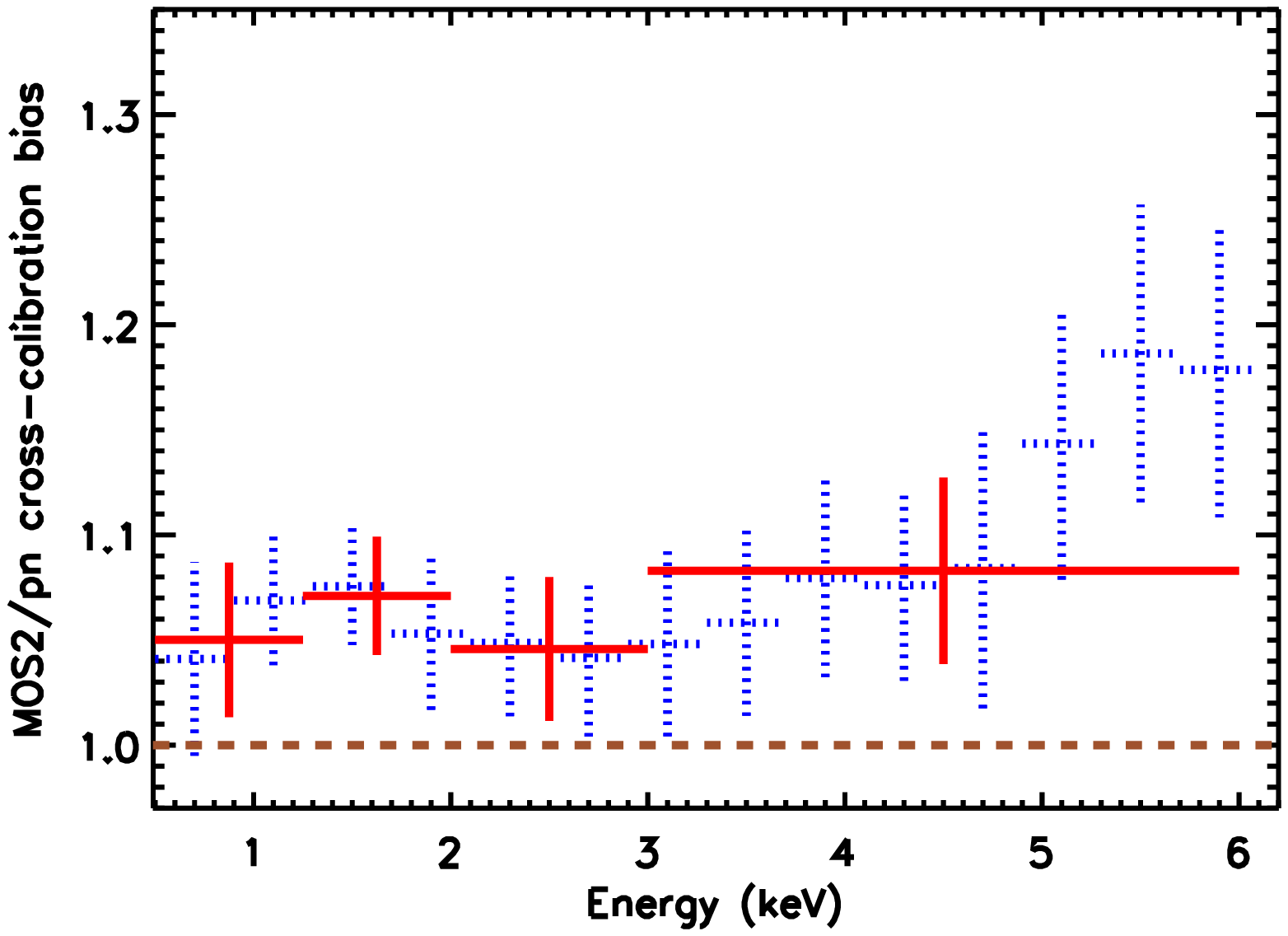}
\caption{The median cross-calibration bias parameter J$_1$ (Eq. \ref{J1.eq}) obtained with Method 1 of different observations and their standard deviation for MOS2/pn pair for 
the standard energy binning used in the paper (blue crosses) and for a coarser binning (red crosses). In the latter case the statistical uncertainties of each individual observation are below 1\% at each bin.}
\label{ebintest.fig}
\end{figure}

We discuss next the substantial cluster-to-cluster scatter of the measure of the cross-calibration bias (see Fig.  \ref{Jmeas.fig}). The standard deviation of the individual J$_2$ values of the full cluster sample amounts to $\sim$2-13\% of the median value in the studied energy range with an tendency to increase with the photon energy (see Fig. \ref{Jmeas.fig} and Table \ref{statsys.tab}). The scatter exceeds the level of the known systematic uncertainties (see Section \ref{systematics}) by a factor of $\sim$2-13. The statistical uncertainties of the individual clusters are much smaller than the scatter and may thus explain only a small fraction of the scatter.
We demonstrated this by repeating the cross-calibration bias analysis using Method 1 for MOS2/pn, but this time using much bigger energy bins so that the statistical uncertainty of each individual cluster observation is below 1\% at each bin (see Fig. \ref{ebintest.fig}). Yet, at a given photon energy, the scatter remains almost intact.

\begin{figure*}
\hbox{
\includegraphics[width=6cm,angle=0]{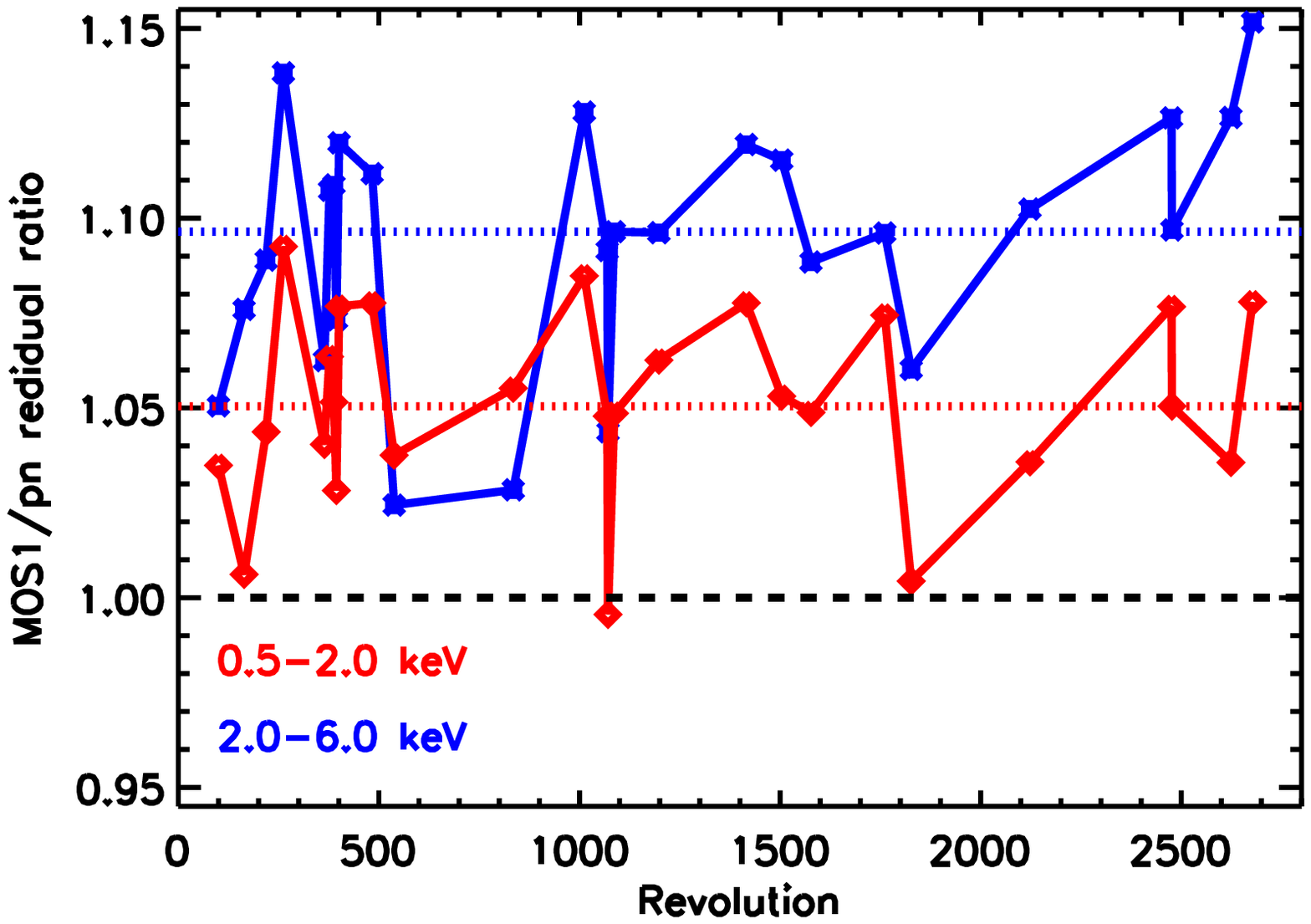}
\includegraphics[width=6cm,angle=0]{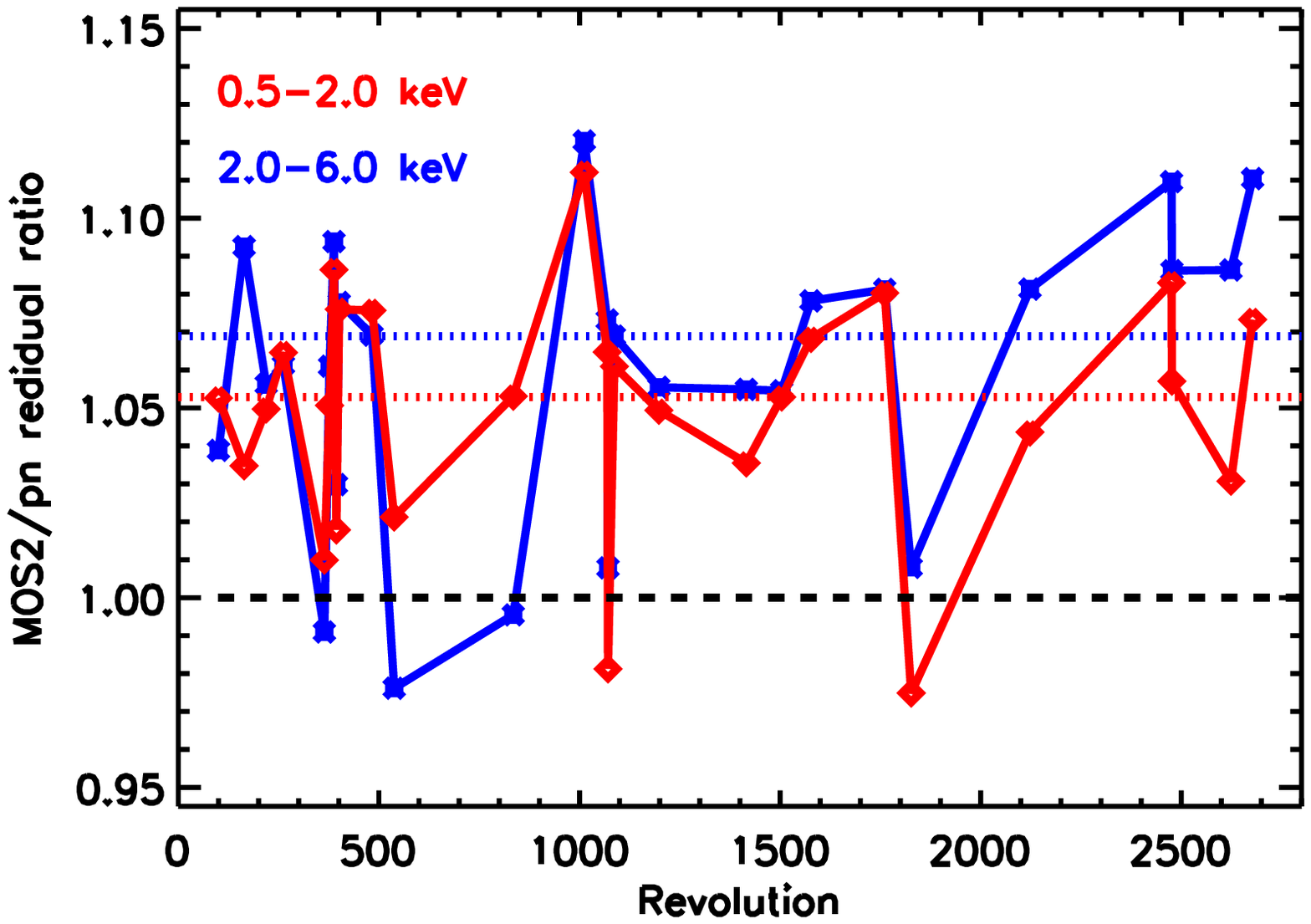}
\includegraphics[width=6cm,angle=0]{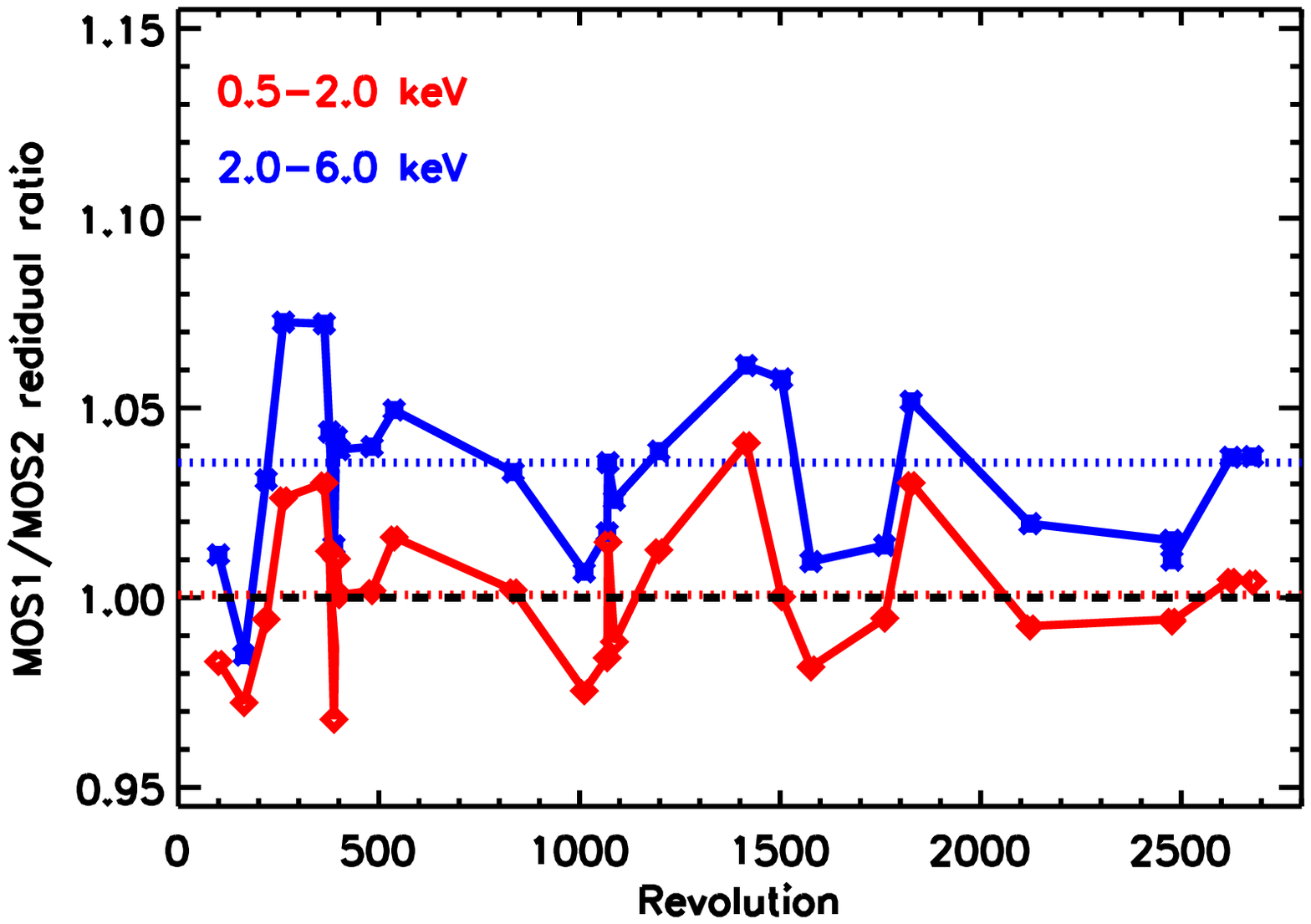}
}
\caption{The cross-calibration bias factor J$_1$ using Method 1 (Eq. \ref{J1.eq}) in the 0.5-2.0 keV band (red lines and symbols) and 2.0-6.0 keV band (blue lines and symbols) band as a function of the time (revolution number during which the given observation has been performed) for MOS1/pn (left panel), MOS2/pn (middle panel) or MOS1/MOS2 (right panel) pairs. The dotted lines indicate the sample median.}
\label{rev.fig}
\end{figure*}
 
\begin{figure*}
\hbox{
\includegraphics[width=6cm,angle=0]{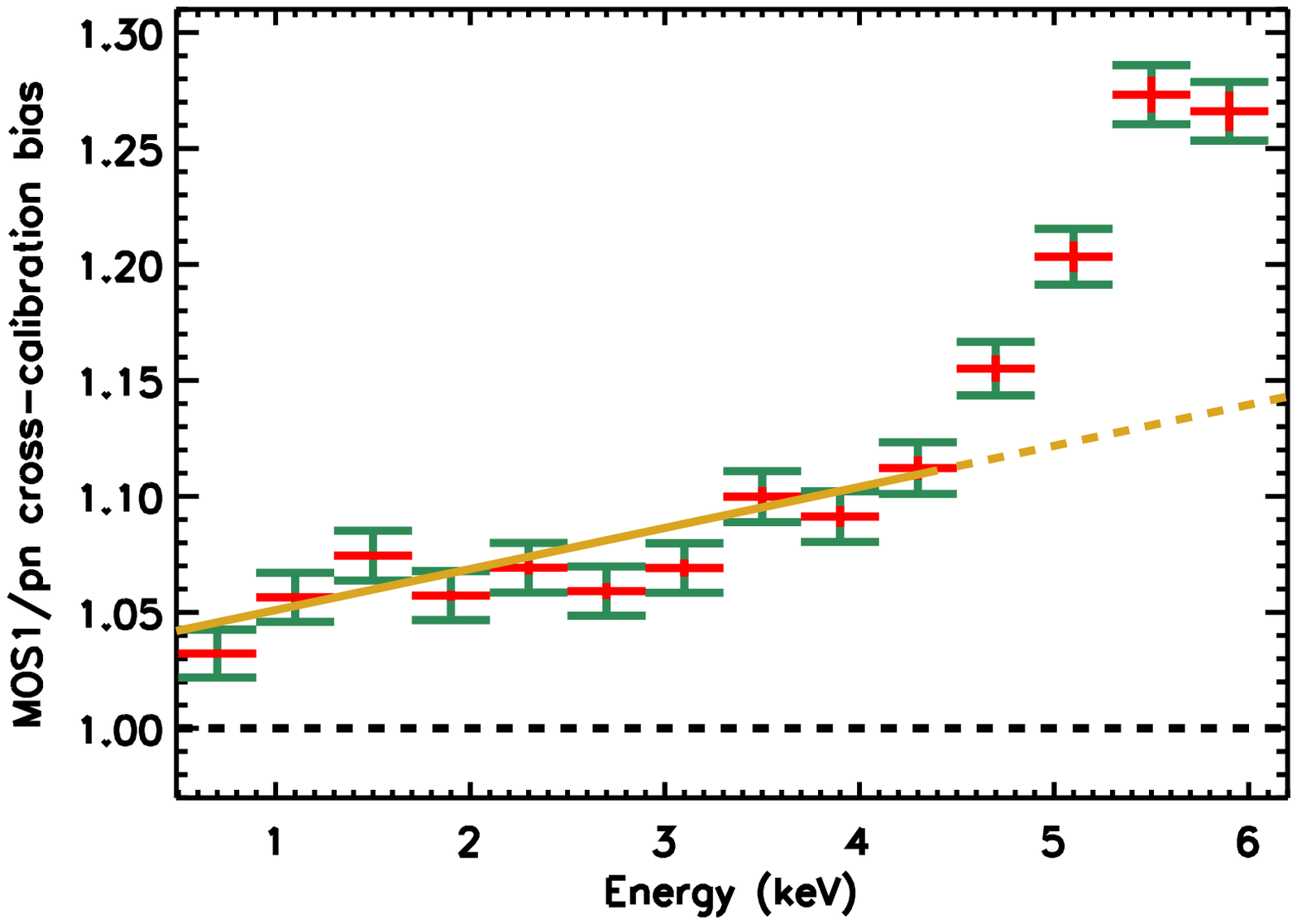}
\includegraphics[width=6cm,angle=0]{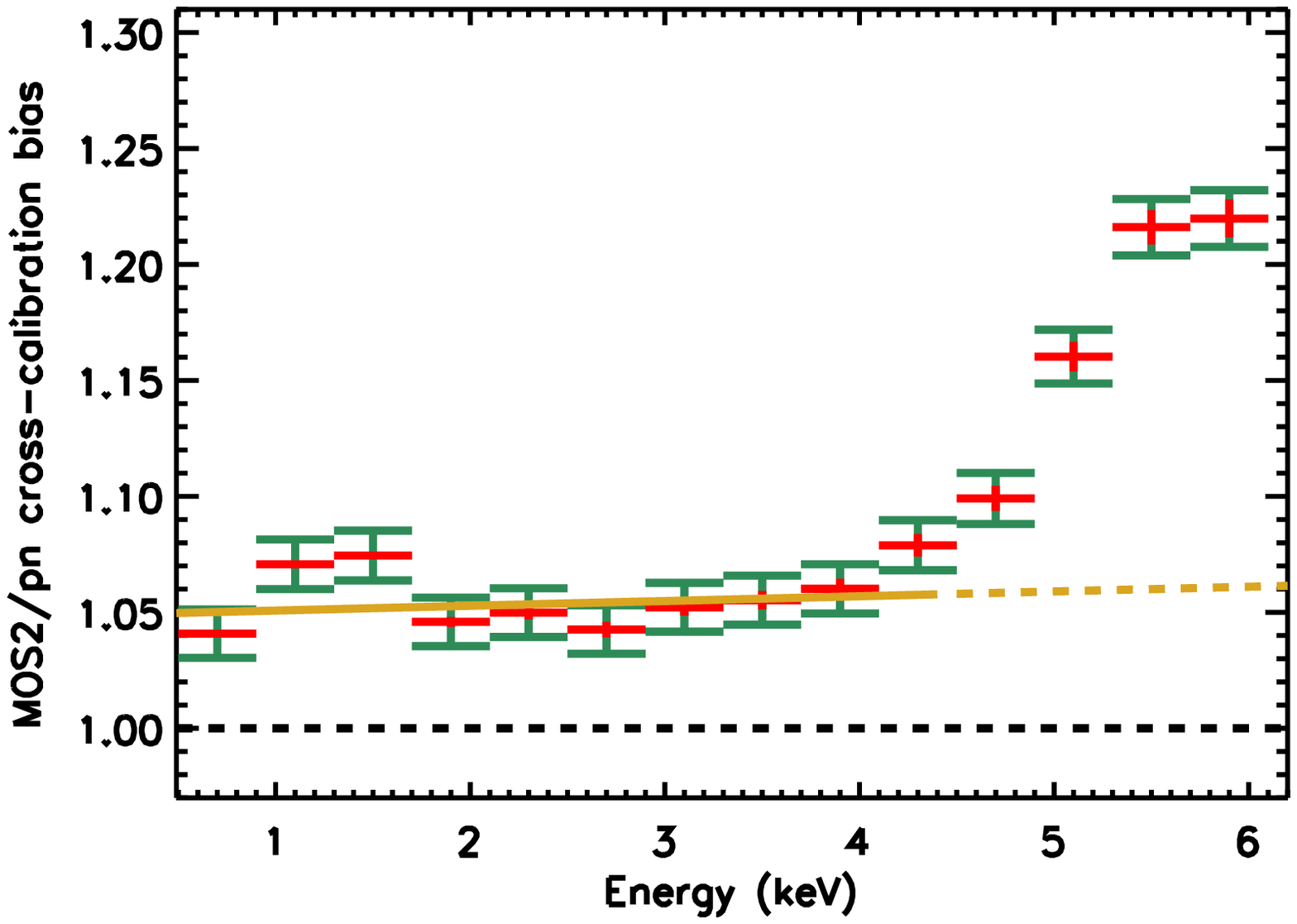}
\includegraphics[width=6cm,angle=0]{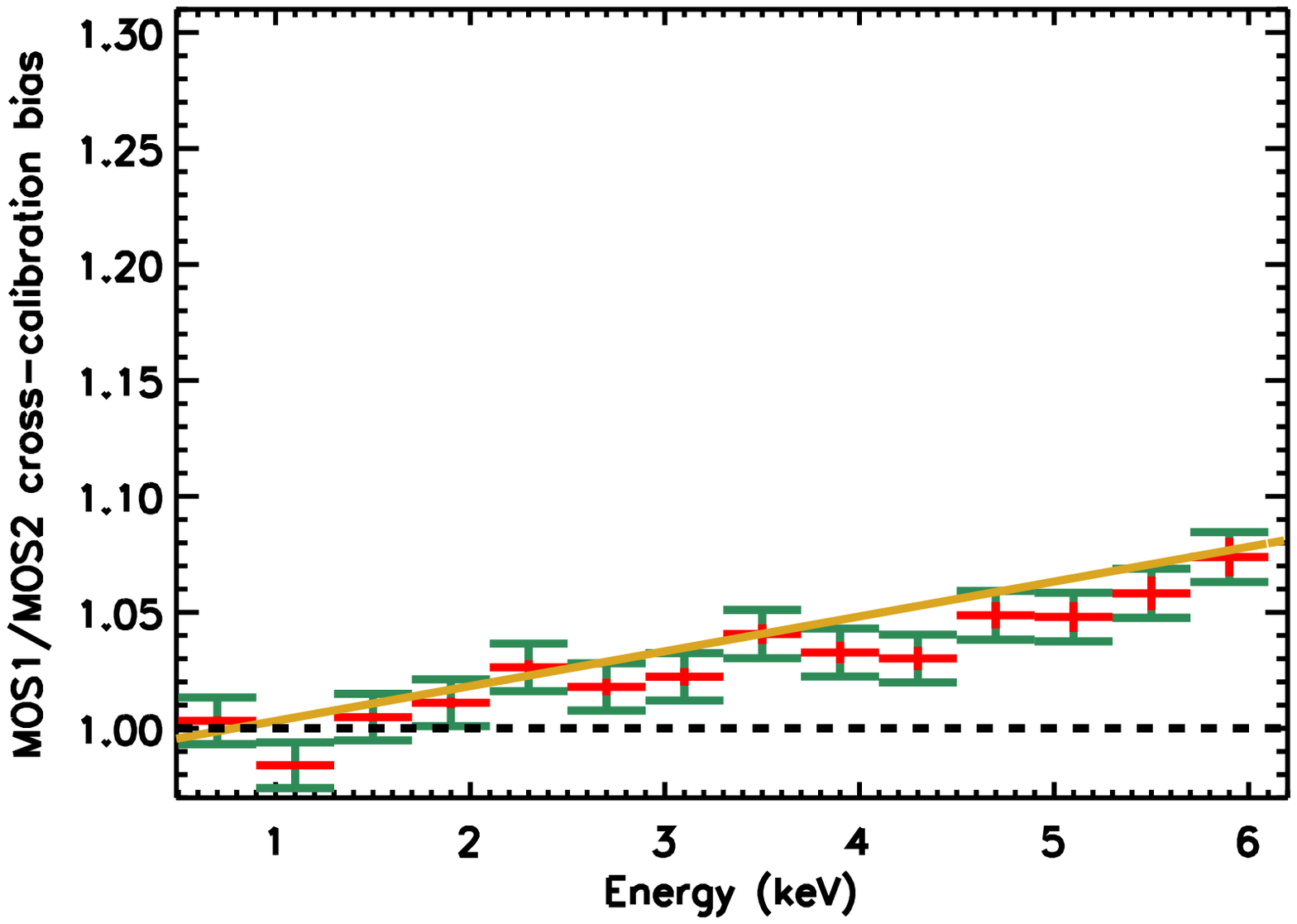}
}
\caption{
The median of the measures of cross-calibration bias, with statistical and systematic uncertainties 
(red and green symbols, repeated from the bottom panel of Fig. \ref{Jmeas.fig}) for MOS1/pn pair (left panel) and MOS2/pn pair (middle panel) and MOS1/MOS2 pair.
The solid golden lines in the left and middle panels indicate such linear models whose parameters are the medians of those obtained by fitting the individual
cluster data for MOS/pn pairs in the 0.5-4.5 keV band with a linear model. The dashed golden lines indicate the extrapolation of the above models to 4.5-6.1 keV band. The MOS1/MOS2 data were fitted in the 0.5-6.1 keV band (right panel).
In all panels, the horizontal dashed line indicates the expectation (unity) in case of no cross-calibration bias. 
}
\label{linfit.fig}
\end{figure*}
 
We investigated the possibility that the scatter is due to the random sampling of the parent population with a limited sample size.  
For a given statistical error of the sample mean ($\mu$)\footnote{We replace the error of the mean with the error of the median as discussed in Section \ref{Results}.} and the number of data points in the sample (N), the expected standard deviation ($\sigma$) due to the statistical fluctuations in the sample
is given by $\sigma = \mu \times \sqrt{N}$.  In our case, 
the standard deviation exceeds the statistical uncertainty of the sample median by a factor of 8-26 (MOS1/pn) and 9-28 (MOS2/pn) 
in the 0.5-6.0 keV band. Since we have N = 27, the standard deviation in our sample exceeds the expectation ($\mu \times \sqrt{N}$) by a factor of 
1.6-5.0 (MOS1/pn)
and
1.8-5.4 (MOS2/pn), depending on the photon energy.
Thus, the scatter is significantly larger than expected by the sampling uncertainties.
  
Our analysis of the bias factor using Method 1 in the 0.5-2.0 and 2.0-6.0 keV bands indicates no regular patterns in the behaviour as a function of the observation date (see Fig. \ref{rev.fig}). 
A more detailed analysis of the problem is out of the scope of this work. Expert knowledge on the XMM-Newton/EPIC instruments and SAS software
would be needed to probe the cause of the substantial and significant scatter of the XMM-Newton/EPIC cross-calibration 
(which is not inherent to clusters, see Section \ref{Scatter2}). 
 
\subsection{Normalisation}
We started the analysis of the measures of the cross-calibration bias J$_2$ (Eq. \ref{Jmeas.eq}) by 
investigating the normalisation of the cross-calibration using the sample medians. In particular we examined deviations from the unity 
since J = 1 in the case of no bias in the normalisation of the effective area cross-calibration.
  
The approximative results for the MOS1/MOS2 pair (see Section \ref{measure}) indicates in general J$_2$ values closest to unity 
compared to pairs involving the pn (see Fig. \ref{Jmeas.fig}). This is expected since the full light path entering MOS1 and MOS2 detectors are very similar and, apart from the mirror, very different from that of pn. 
 
At all energies the bias factor between MOS and pn (J$_{2,\rm MOS1/pn}$ and J$_{2,\rm MOS2/pn}$) 
 is above unity. On average J$_{2,\rm MOS1/pn}$ $\sim$1.12 and J$_{2,\rm MOS2/pn}$ $\sim$1.09, indicating a systematic normalisation bias in the effective area calibration in pn or MOS, or both (see Fig. \ref{Jmeas.fig}).
 The MOS1/pn (MOS2/pn) deviations amount to $\sim$27 \% ( $\sim$22 \%) in maximum in the 0.5-6.0 keV band, indicating a similar maximal level of MOS/pn effective area normalisation problems.

Let's assume first for the sake of the argument that the normalisation of the effective area of pn is accurately calibrated, i.e. 
the normalisation of $model_{ref,2}$ in Eq. \ref{Jmeas.eq} is correct. Since the MOS/pn measure of the cross-calibration (J$_{2,\rm MOS/pn}$) is above unity, it follows (see Eq. \ref{Jmeas.eq}) that the MOS prediction is too low, i.e. MOS effective area normalisation is calibrated too low in this scenario (see Fig. \ref{aeff.fig}). Assuming instead that the normalisation of the MOS effective area is accurately calibrated, that of pn is calibrated too high. In this scenario the standard spectral analysis of pn spectra would yields fluxes which are biased low. 

The comparison with different works on the XMM-Newton/EPIC effective area cross-calibration is complicated because the implemented calibration information changes with time. Also, different methods and X-ray sources have different levels of systematics which are not always explicitly evaluated and reported. However, it appears that the above qualitative feature, i.e. the MOS yielding higher fluxes than the pn, has been consistently reported for the past 14 years (e.g. \cite{Mateos}, \cite{N10}, \cite{Tsujimoto}, \cite{Andy}, \cite{Gerrit},  \cite{Madsen}, \cite{Plucinsky}, \cite{Marshall}), Fuerst, 2022\footnote{https://xmmweb.esac.esa.int/docs/documents/CAL-TN-0230-1-3.pdf}.
  
\subsection{Energy dependence}
We then studied the issue of the energy dependence as indicated by the measure of the cross-calibration. 

\subsubsection{MOS2/pn} 
Due to the issues of the MOS/pn pairs at the highest energies (see Section \ref{measure}) we study first MOS/pn results in the 0.5-4.5 keV band. Visual investigation of the data indicated that the cross-calibration bias between MOS2 and pn is almost constant
(see Fig. \ref{linfit.fig}, middle panel).
We examined this in more detail by fitting the individual curves of the measure of the cross-calibration (J$_{2,MOS2/pn}$) for each cluster in the 0.5-4.5 keV band with a linear model. We used these models to construct a model describing the full sample by 
adopting such a linear model, whose parameters are the medians of those in the individual best-fit models.  We approximated the uncertainty of the above parameters with the standard deviation of the best-fit parameters in the individual fits. The resulting median value of the best-fit linear coefficients is 0.002 $\pm$ 0.012, i.e. it does not deviate significantly from zero. 

The median J$_{\rm 2,MOS2/pn}$ varies in the range $\sim$1.05-1.08 in this band (see Fig. \ref{linfit.fig}, middle panel).
In most of the 0.5-4.5 keV band the linear model described above agrees with the data within the estimates of the systematic and statistical uncertainties. At 0.9-1.7 keV the data exceed the linear model by more than the estimated uncertainties (see Fig. \ref{linfit.fig}, middle panel). However, the deviation amounts only to $\sim$2-3\%.

 In such a wide band (0.5-4.5 keV) all the components of the effective area (filter transmission, detector QE and mirror effective area) vary substantially with the energy, and differently from each other (see Fig. \ref{aeff.fig}). Thus it is very unlikely that these components have such calibration biases which cancel out the energy dependence of the cross-calibration bias.   
It is more likely that the energy dependencies of the effective area components of MOS2 and pn in most of the 0.5-4.5 keV band as implemented in the public XMM-Newton calibration on Nov 2021 are very accurately modelled.

\subsubsection{MOS1/pn} 
The situation is different for the MOS1/pn pair.
J$_{\rm 2,MOS1/pn}$ increases approximately linearly from $\sim$1.04 at 0.5 keV to $\sim$1.11 at 4.5 keV (Fig. \ref{linfit.fig}, left panel). 
The best-fit linear co-efficient is 0.018$\pm$0.013, i.e. $\sim$10 times larger than that in the case of MOS2/pn. 
Given the above MOS2/pn consistence this suggests that the MOS1 effective area energy dependency in the 0.5-4.5 keV band, as implemented in the XMM-Newton calibration on Nov 2021, has an approximately linear bias increasing with the photon energy (see Fig. \ref{linfit.fig}, left panel). However, due to the large scatter described above, the linear coefficient of the best-fit linear model describing the MOS1/pn cross-calibration bias does not deviate very significantly from zero. Thus, the confirmation of this feature requires more work in the future.

\subsubsection{MOS1/MOS2} 
Given that the MOS units behave differently in comparison to pn, there is some level of cross-calibration bias between MOS1 and MOS2. 
Using the approximative approach for the MOS1/MOS2 pair (see Section \ref{measure}),
the linear coefficient of the best-fit linear model to J$_{2,\rm MOS1/MOS2}$ data in the 0.5-6.0 keV band (0.015$\pm$0.005) deviates from zero by $\sim$3 $\sigma$ (Fig. \ref{linfit.fig}, right panel). The deviation from zero is larger than that between MOS2 and pn derived earlier. This suggests a bit surprisingly that the cross-calibration of the energy dependence between MOS2 and pn is in better agreement than that between MOS1 and MOS2.

\subsubsection{4.5-6.0 keV band}
 In the 4.5-6.1 keV band the MOS1/pn and MOS2/pn bias factors stand out from the linear trend extrapolated from the lower energies by $\sim$30\% at the maximum (see Fig. \ref{linfit.fig}, left and middle panels). This feature is not present in the MOS1/MOS2 pair (see Fig. \ref{linfit.fig}, right panel)
Inspection of the individual J curves indicates that the above jump is a systematic feature and not driven by some extreme outliers. 
 
While the mirror effective areas and filter transmissions of MOS and pn have similar energy dependence at these wavelengths, the quantum efficiencies (QE) differ substantially (see Fig. \ref{aeff.fig});
the QE of pn maintains at a rather constant level of 0.8 while that of MOS drops from 0.8 at 4 keV to 0.6 at 6 keV. If the drop has been overestimated so that the actual QE of MOS1 and MOS2 would be harder, the MOS data/model ratio would actually be softer, and thus J$_{\rm 2,MOS1/pn}$ and J$_{\rm 2,MOS2/pn}$ would approach unity. This suggests there may be an issue in the QE calibration of MOS2 and even more so of MOS1.

 \begin{figure*}
\hbox{
\hspace{-0.5cm}
\includegraphics[width=10cm,angle=0]{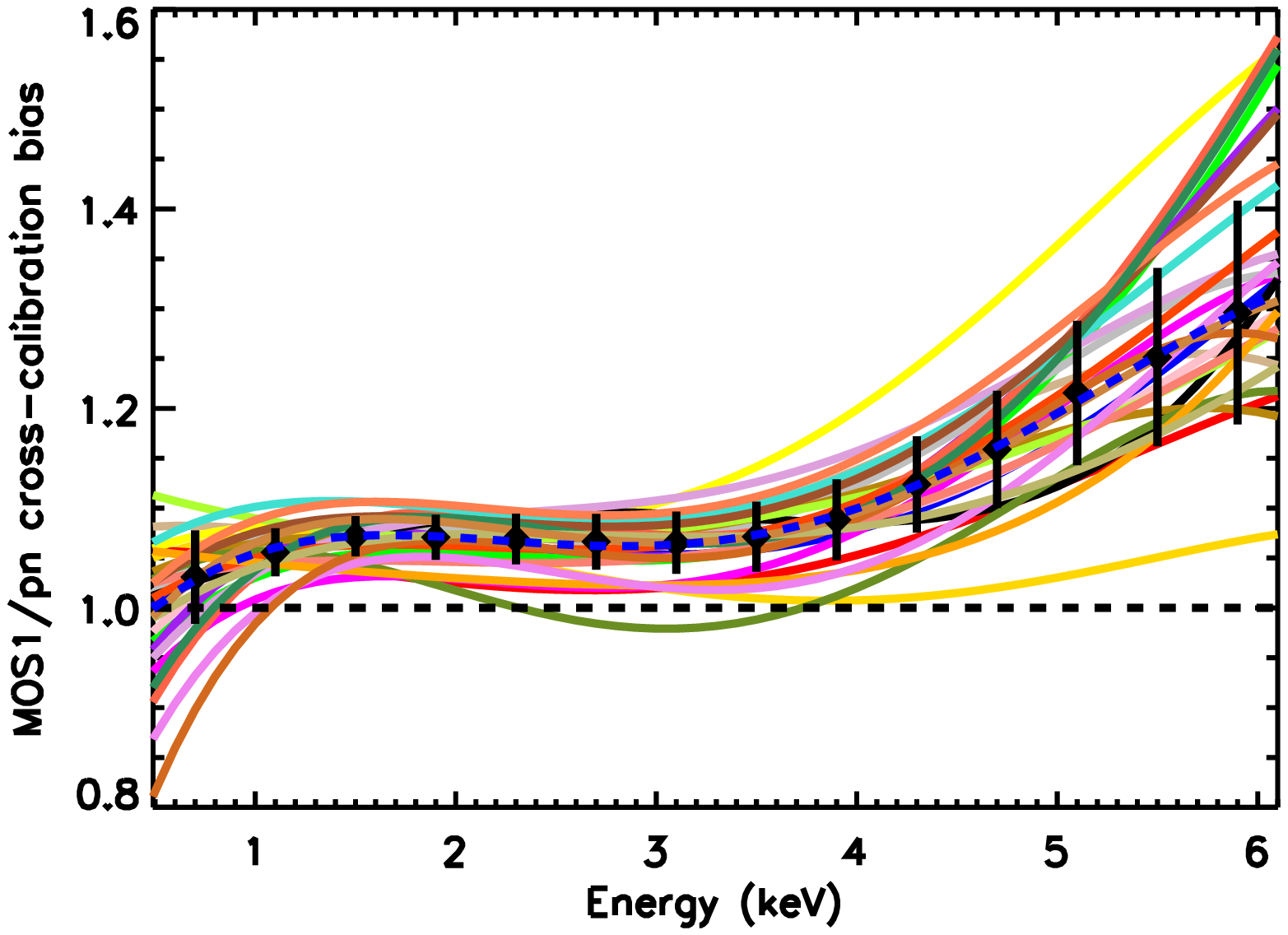}
\hspace{-1cm}
\includegraphics[width=10cm,angle=0]{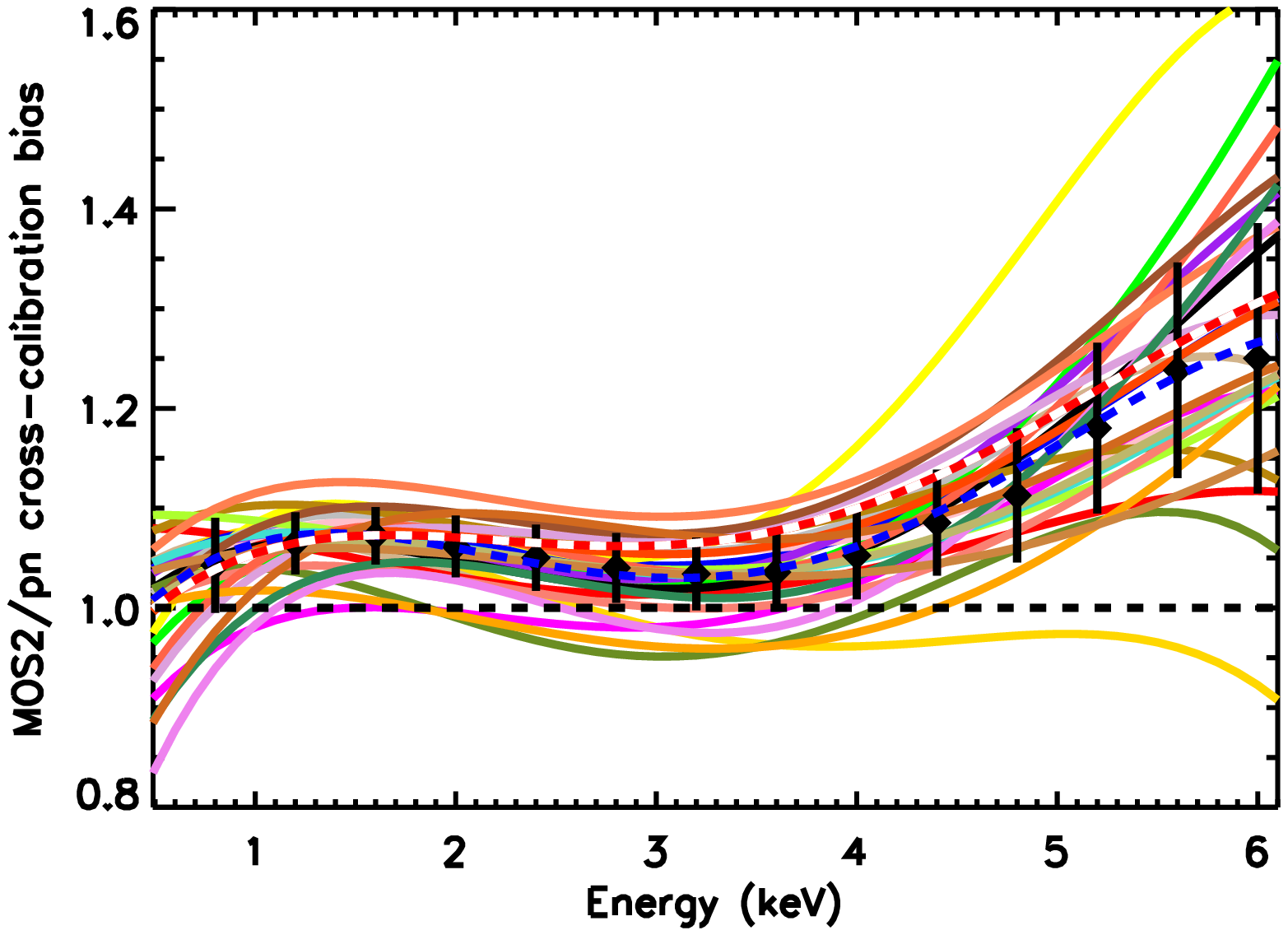}
 }
\caption{The best-fit models for each individual cluster obtained by fitting the cross-calibration data of MOS1/pn pair (left panel) or MOS2/pn pair (right panel, repeated from Fig. 3, right panel), using Method 2. Each solid line connects the values of one cluster at the energy bin centers. The black symbols and vertical
lines indicate the median and the standard deviation of the sample at each energy.
The blue dashed lines indicate the 4th order approximation of the median. The MOS1/pn curve is repeated in the right panel as a white-red dashed line. 
}
\label{polyfits.fig}
\end{figure*}

 \subsection{Modelling the bias using Method 2}
\label{results-method2}
We then proceed into detailed analysis of the energy dependence of the cross-calibration bias obtained via Method 2, i.e. 
by modelling the MOS1/pn and MOS2/pn bias by 4th order polynomial. 
Given the larger than expected scatter (see Section \ref{Scatter}), we fitted all the individual clusters separately
(see Fig. \ref{polyfits.fig}).
In this approach the statistical uncertainties dominate over the known systematic uncertainty level of 1\% (discussed in Section \ref{systematics}). The best-fit MOS/pn models indicate a general behaviour whereby there is a small bump at 1-2 keV, a dip at 3 keV, and a sharp upturn above $\sim$ 5 keV (see Fig. \ref{polyfits.fig}).

\begin{figure*}
\hbox{
\hspace{-0.5cm}
\includegraphics[width=9.5cm,angle=0]{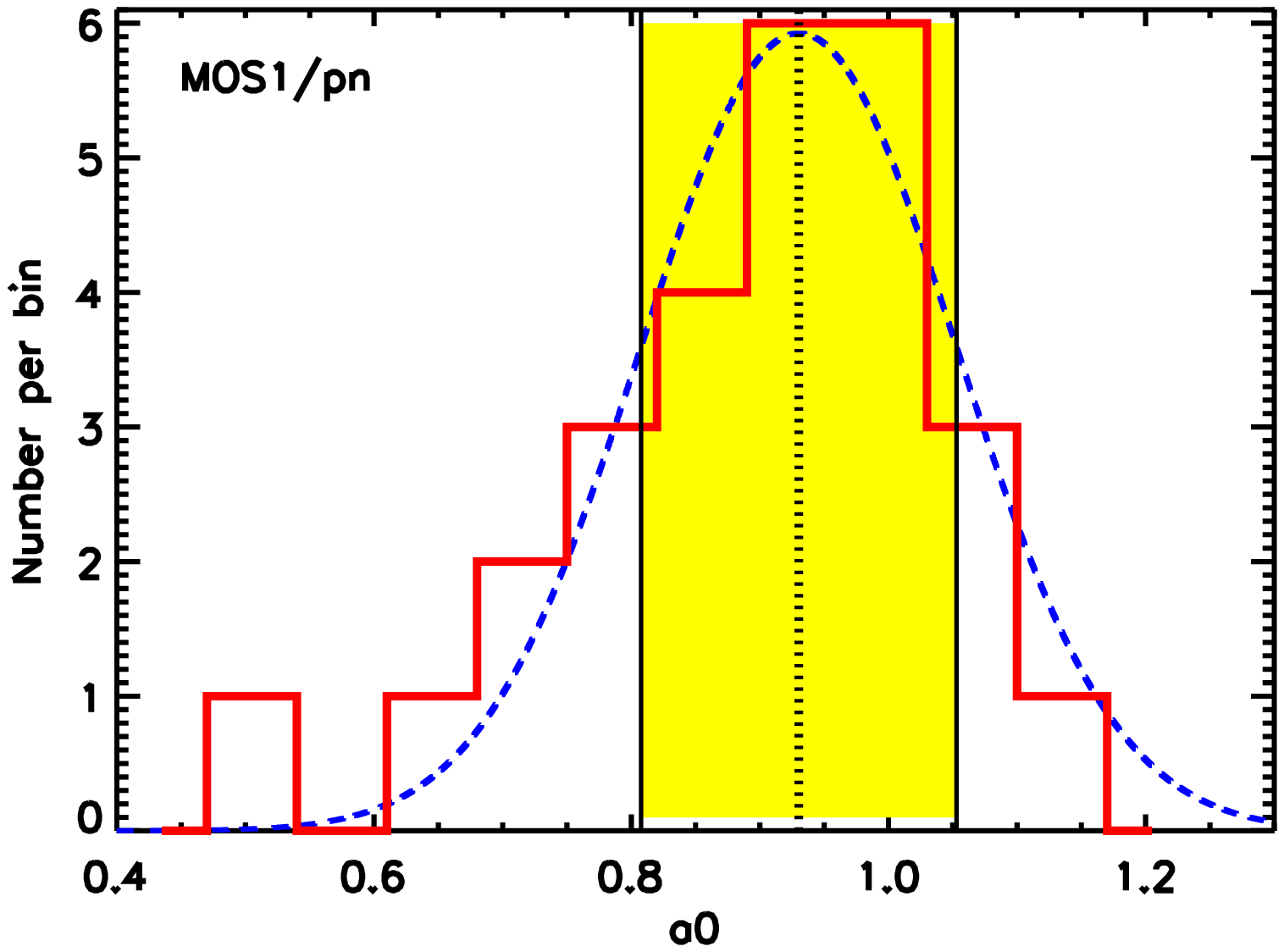}
\hspace{-0.5cm}
\includegraphics[width=9.5cm,angle=0]{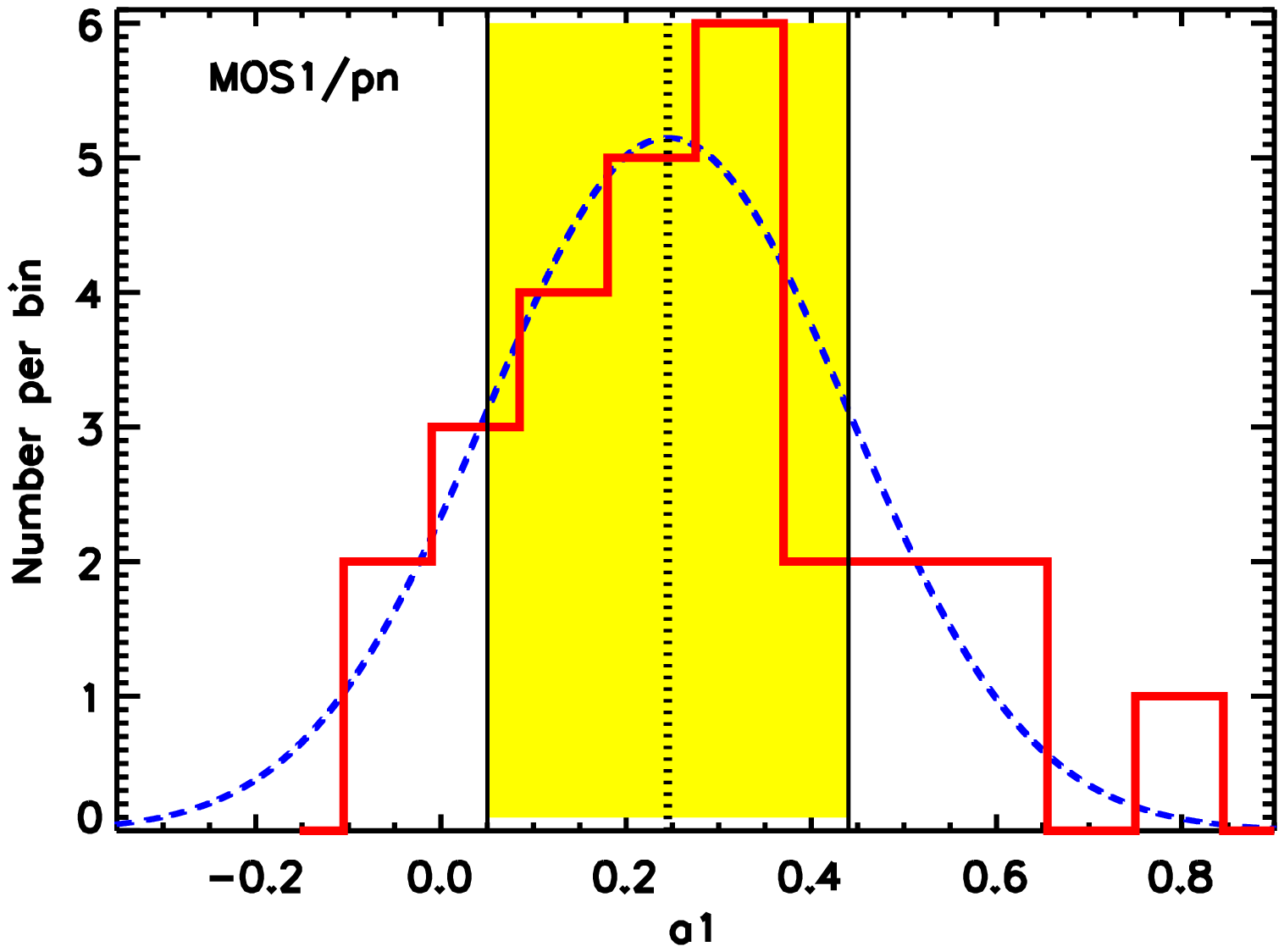}
}
\hbox{
\hspace{-0.5cm}
\includegraphics[width=9.5cm,angle=0]{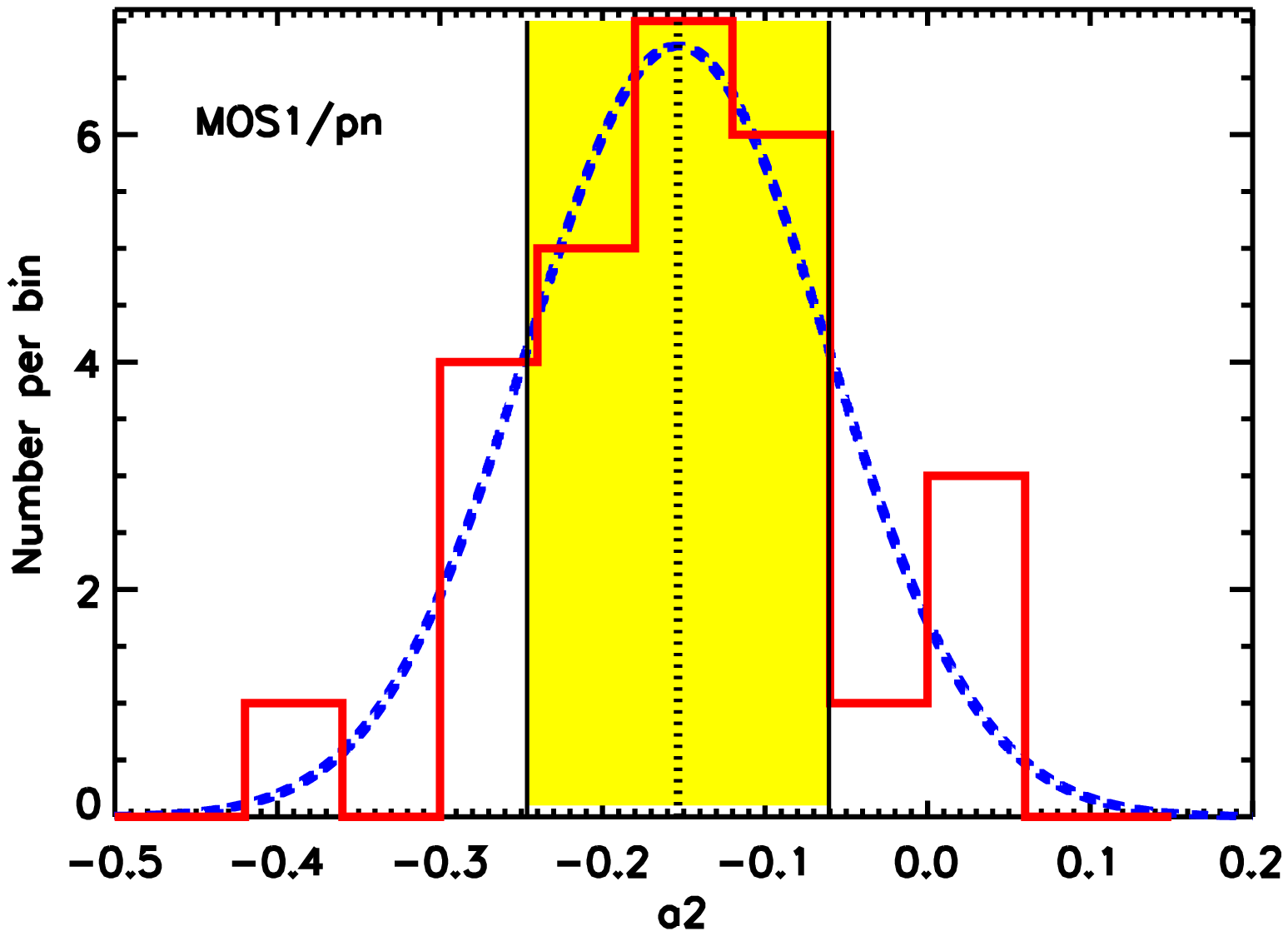}
\hspace{-0.5cm}
\includegraphics[width=9.5cm,angle=0]{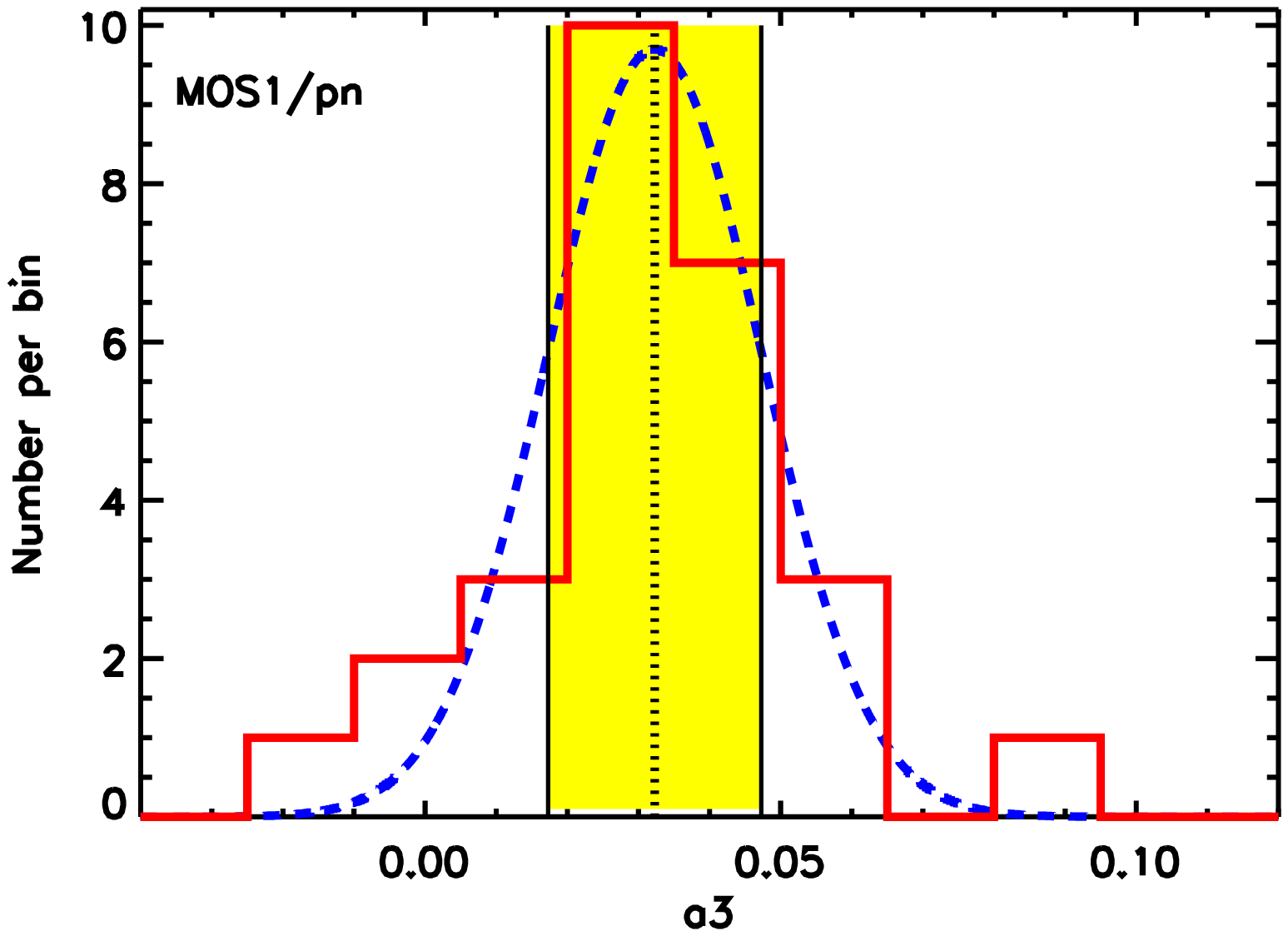}
}
\hbox{
\hspace{-0.5cm}
\includegraphics[width=9.5cm,angle=0]{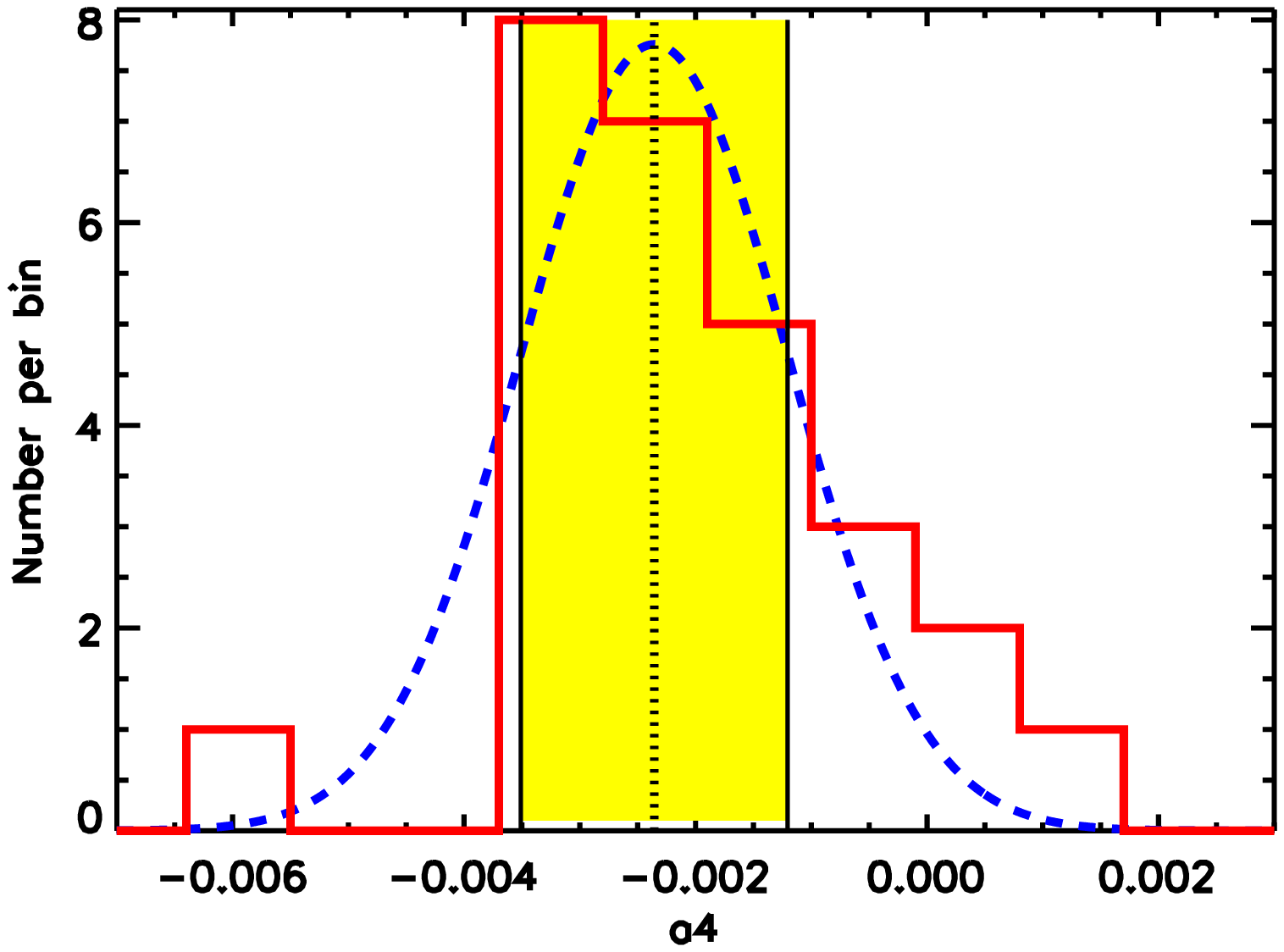}
 }
 \caption{The distributions of the best-fit parameters of the 4th order polynomials used for modelling the cross-calibration bias of single clusters for MOS1/pn pair are shown with the red histogram. 
 The best-fit Gaussians to the above distributions are shown as a blue dashed line. The best-fit centroid of the Gaussian and the central interval containing 
68\% of the probability interval are indicated with a black dashed line and the yellow band, respectively.  
}
\label{polydistr-m1pn.fig}
\end{figure*}

\begin{figure*}
\hbox{
\hspace{-0.5cm}
\includegraphics[width=9.5cm,angle=0]{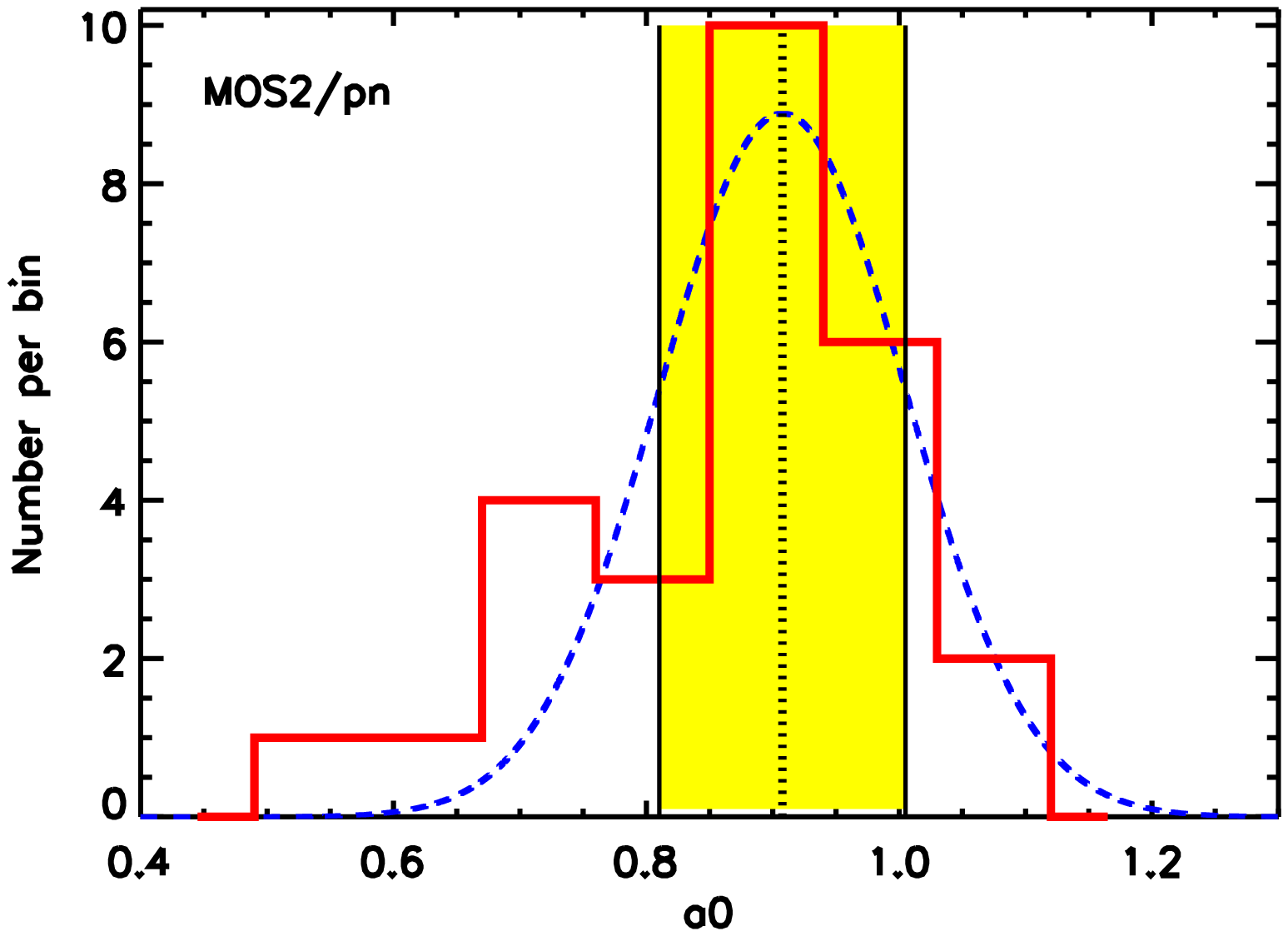}
\hspace{-0.5cm}
\includegraphics[width=9.5cm,angle=0]{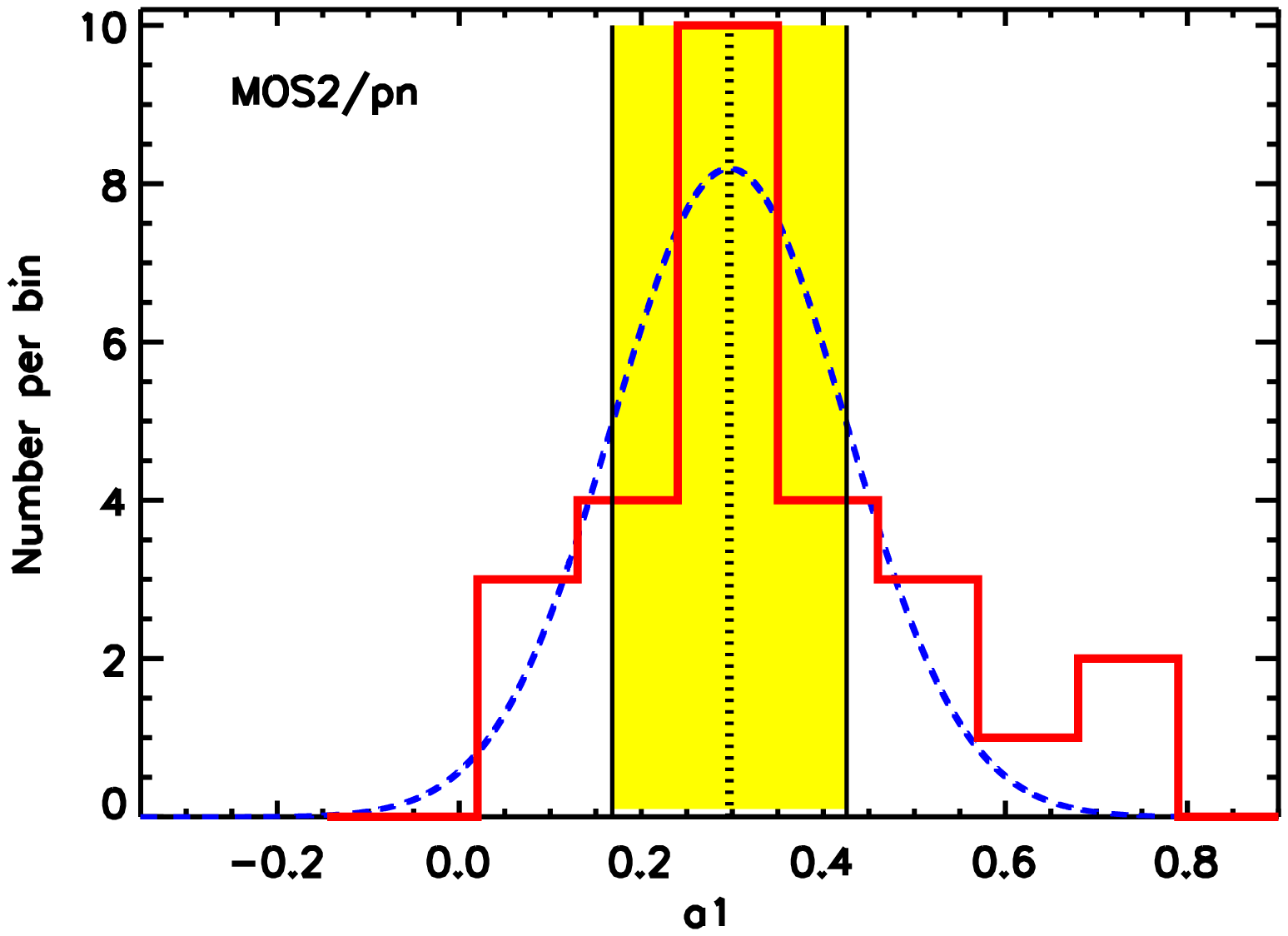}
}
\hbox{
\hspace{-0.5cm}
\includegraphics[width=9.5cm,angle=0]{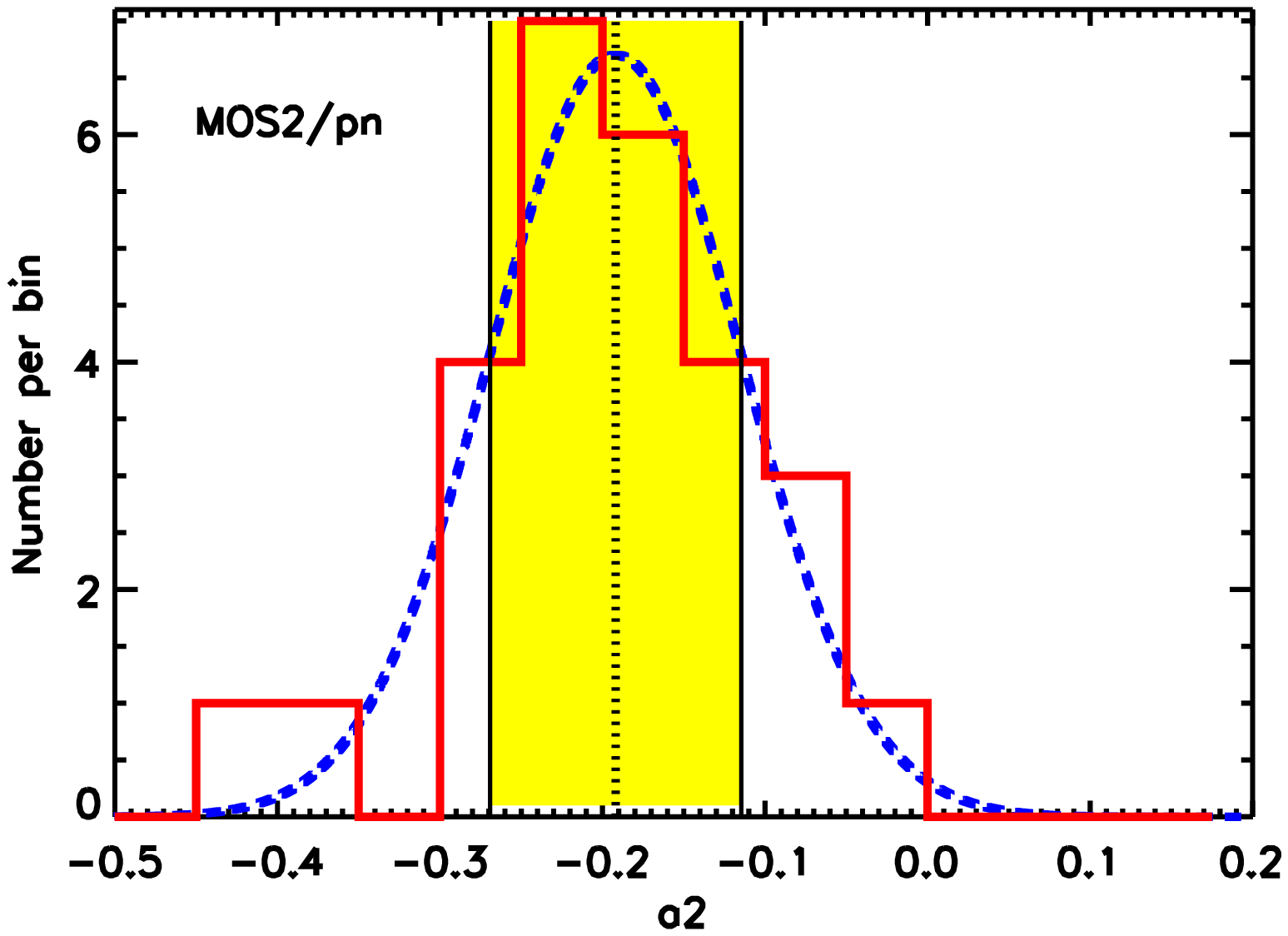}
\hspace{-0.5cm}
\includegraphics[width=9.5cm,angle=0]{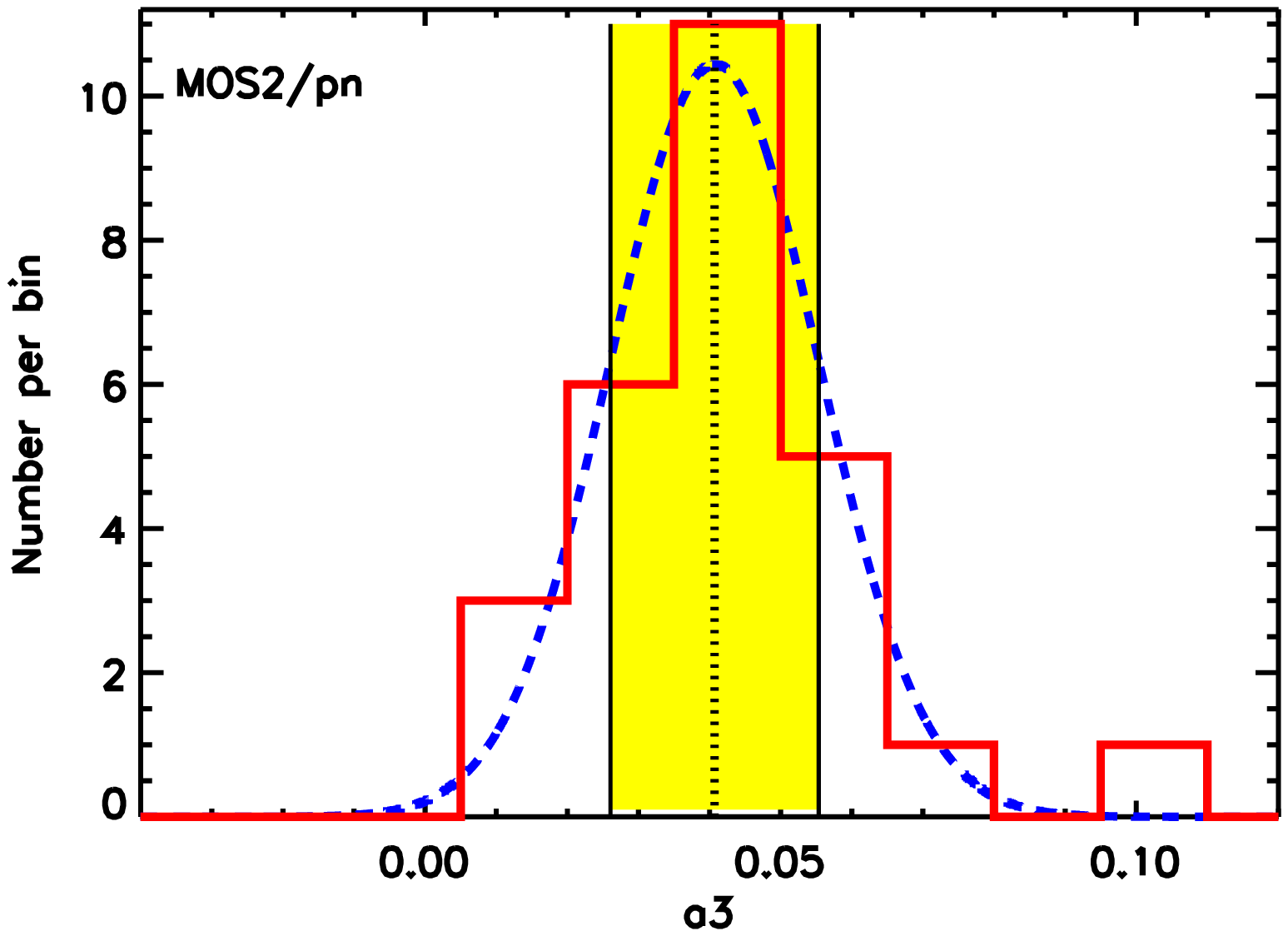}
}
\hbox{
\hspace{-0.5cm}
\includegraphics[width=9.5cm,angle=0]{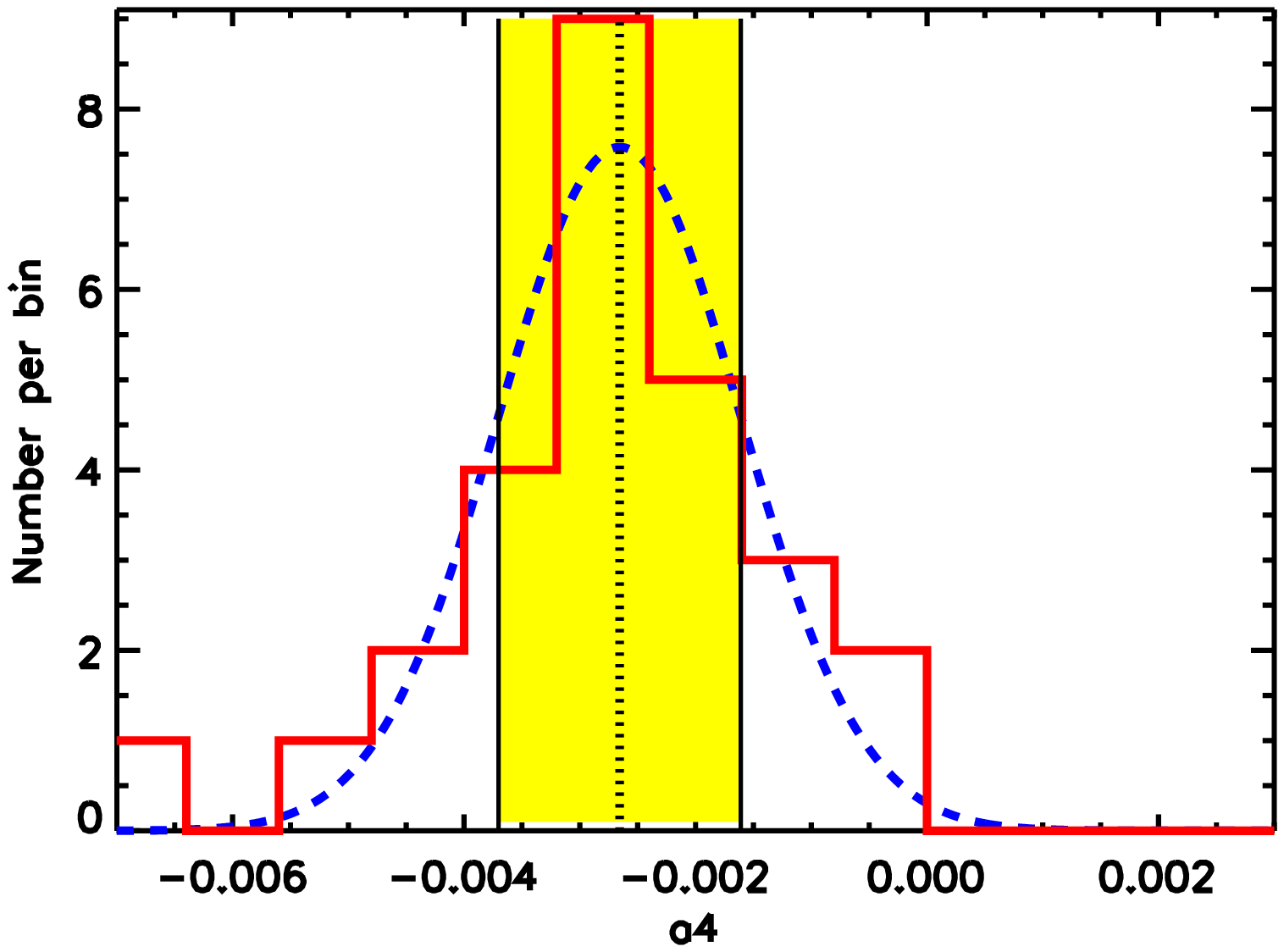}
 }
 \caption{The same as for Fig. \ref{polydistr-m1pn.fig} but for MOS2/pn pair.
}
\label{polydistr-m2pn.fig}
\end{figure*}

We then examined the resulting distributions of the best-fit coefficients of the 4th order polynomial.
This approach serves as an empirical evaluation of the effect of the unknown scatter to the cross-calibration bias.
The distributions can be reasonably well approximated with Gaussian models (see Figs. \ref{polydistr-m1pn.fig} and \ref{polydistr-m2pn.fig}
and Table \ref{results.tab}).
 
There is significant correlation between the parameters (see Appendix \ref{corr}) and thus the best-fit parameters and the Gaussians widths of the parameter distributions reported in 
Figs. \ref{polydistr-m1pn.fig} and \ref{polydistr-m2pn.fig} and Table \ref{results.tab}
are not alone sufficient to describe the statistical properties of the sample. We explore this further in Appendix \ref{corr}. 

In order to produce useful information for the general user, we approximated the median curves of the cross-calibration bias for MOS1/pn and MOS2/pn pairs with 4th order polynomials (see Fig. \ref{polyfits.fig} and Table \ref{best-fit-param.tab}). These models describe the average MOS1/pn and MOS2/pn cross-calibration bias. They provide a quick quantitative way of understanding the effects of the cross-correlation bias to the first order. Please note that the proper treatment of the uncertainties requires the usage of parameter correlations which is explored further in Appendix \ref{corr}.  \\

\begin{table*}
\small
 \centering
  \caption{Statistical properties of the cross-calibration bias parameters
  \label{results.tab}}
    \begin{tabular}{lccccc}
  \hline\hline
        &     &    &    &     &                                                   \\ 
Pair    & a$_0$  & a$_1$ & a$_2$ &  a$_3$ & a$_4$ \\
\hline
MOS1/pn            & 0.93$\pm$0.12  &  0.25$\pm$0.19  &  -0.15$\pm$0.09  &  0.03$\pm$0.01  &  -0.002$\pm$0.001 \\   
MOS2/pn            & 0.91$\pm$0.10  &  0.30$\pm$0.13  &  -0.19$\pm$0.08  &  0.04$\pm$0.01  &  -0.003$\pm$0.001 \\   
\hline 
\end{tabular}
\tablefoot{\\
The centroids and widths of the Gaussian fits to the distributions of the 4th order polynomials we used to model the cross-calibration bias
in the 0.5-6.1 keV band for each individual cluster. The uncertainties correspond to the 68\% confidence level and 
contain both the statistical uncertainties and the effect of the scatter of unknown origin discussed in Section \ref{Scatter}
}
\end{table*}

\begin{table*}
\small
 \centering
  \caption{Best-fit cross-calibration bias functions
  \label{best-fit-param.tab}}
    \begin{tabular}{lccccc}
  \hline\hline
        &     &    &    &     &                                                   \\ 
Pair    & a$_0$  & a$_1$ & a$_2$ &  a$_3$ & a$_4$ \\
\hline
MOS1/pn            & 0.8895 &  0.2955  &  -0.1629  &  0.03488  &  -0.002335 \\   
MOS2/pn            & 0.8807 &  0.3522  &  -0.114   &  0.04655  &  -0.003217 \\   
 \hline 
\end{tabular}
\tablefoot{\\
The parameters of the coefficients of the 4th order polynomials which best describe the median cross-calibration bias for MOS1/pn and MOS2/pn samples in the 0.5-6.1 keV band.  
}
\end{table*}

\section{Application to scientific analysis}
\label{Application}
In this work we have evaluated and reported the significant systematic uncertainties in the XMM-Newton/EPIC instrument calibration. 
They will bias the derived physical parameters of the source under study at some level. This sets a limit to the accuracy of the parameters derived with XMM-Newton spectra which cannot be improved with higher photon counts obtained via longer exposures or by combining the pn and MOS data. 
In the case of Chandra/ACIS-S3 the limit is reached with $\sim 10^4$ counts \citep{Drake}. 

We approach the problem by first playing the game of modifying the effective area of one instrument assuming that the other instrument is perfectly calibrated. This can be done by multiplying or dividing the effective area produced by SAS with the appropriate average bias function (i.e. 4th order polynomial with parameters reported in Table \ref{best-fit-param.tab}).
For this illustration we use the data from the  observation of A1795 cluster
(see Fig. \ref{aeff.fig} top left panel for MOS2/pn pair). If pn (MOS) was perfectly calibrated, the effective area of MOS (pn) should be increased (reduced), with a factor that increases with the photon energy.  
   
In principle a general XMM-Newton user could apply the cross-calibration bias information presented in this paper by modifying the effective area column in an XMM-Newton/EPIC arf-file produced by the SAS software as demonstrated above and re-fitting the data.  
The change in the best-fit parameters would indicate the median systematic effective cross-calibration effect.
However, in this this approach the scatter (discussed in Section \ref{Scatter}) and the parameter correlations (see Appendix \ref{corr}) are omitted. 
The rms scatter is a substantial fraction of the bias itself, in some cases even exceeding it. 
Thus a proper propagation of the substantial systematic uncertainties related to the effective area cross-calibration to the general XMM-Newton spectral analysis is not straightforward.
It would include modifying the standard effective area with a large number of J curves derived via randomisation of the data according to the relevant cross-calibration model and the uncertainties of the
model parameters, presented in Table \ref{results.tab}.
The resulting sample of modified effective areas could then be used to fit the user data. The scatter of the best-fit parameters would then yield the estimate for the systematic effect of the cross-calibration uncertainties
(see \cite{Drake} for such application for Chandra/ACIS). We describe such procedure in more detail in Appendix \ref{corr}.

\begin{figure*}
\hbox{
\hspace{-0.5cm}
\includegraphics[width=10cm,angle=0]{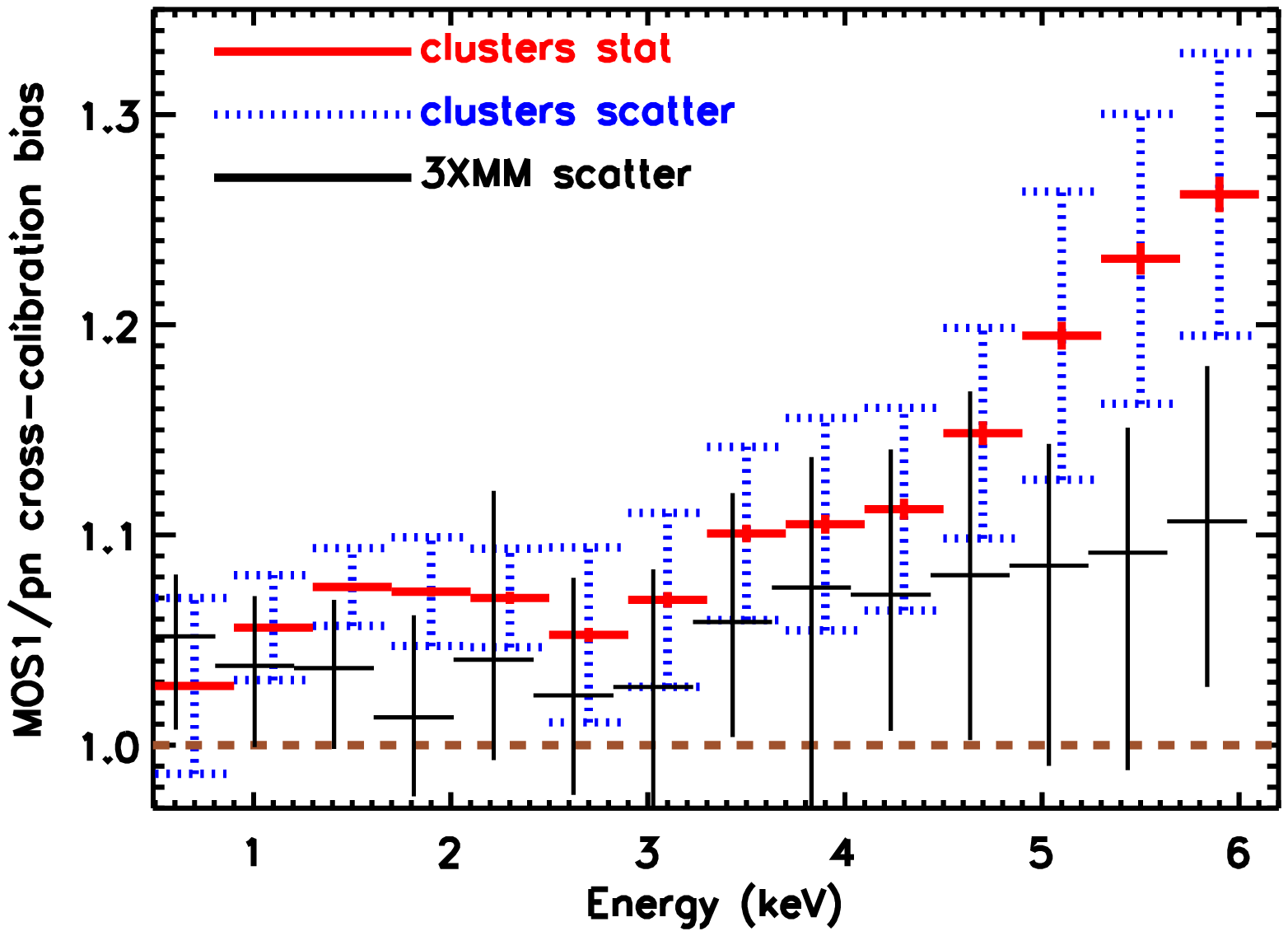}
\hspace{-1cm}
\includegraphics[width=10cm,angle=0]{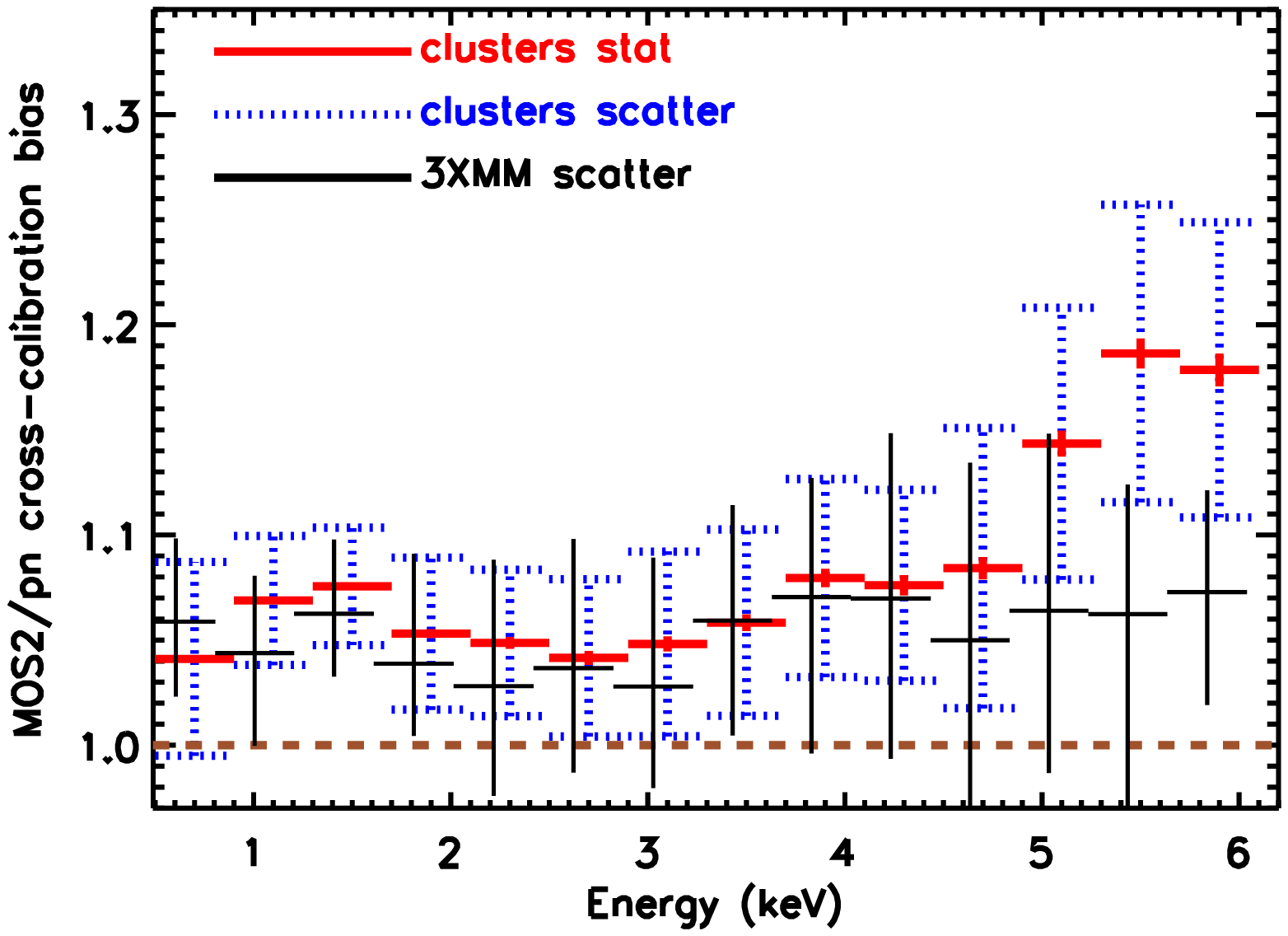}
}
 \caption{The median cross-calibration bias parameter J$_1$ (Eq. \ref{J1.eq})  of different observations and their statistical uncertainties obtained using Method 1 are shown as red crosses for MOS1/pn (left panel) and MOS2/pn (right panel) pairs (at the lowest energies the values are smaller than the plot symbol width). The blue crosses indicate the standard deviation of the cluster sample at a given energy. The black crosses indicate the medians and the range containing 68\% of the values of the 3XMM sample. The horizontal dashed line indicates the expectation (unity) in case of no cross-calibration bias.}
\label{clust_3xmm.fig}
\end{figure*}

 \section{Comparison with 3XMM}
 \label{3XMM}
An essential aspect of the current work is a strong control of the systematic uncertainties and their minimisation.
In order to make sure that we have not underestimated the effect of the studied systematics and that we have not overlooked some remaining important cluster-related systematics, we repeated the analysis presented in this paper on a subsample (3XMM in short in the following) of 120 observations of point sources from the 3XMM-DR7 catalogue \citep{Rosen}. 
Due to practical reasons, we performed the comparison using Method 1. The slight difference at the highest energies compared to our preferred Method 2 does not affect significantly the comparison of the cluster and 3XMM samples.
XMM-Newton calibration team has used 3XMM to investigate the cross-calibration between EPIC instruments, see Smith et al., 2021\footnote{https://xmmweb.esac.esa.int/docs/documents/CAL-SRN-0382-1-1.pdf} for the details of the sample.  
 
To ensure a meaningful comparison of the two data sets, we kept the data processing of 3XMM equal to that applied for the clusters to the maximal possible extent. Namely, we used the same SAS version, stage of public calibration, pattern choice, spectral bin size etc. (see Section \ref{xraydata}). The essential difference between 3XMM and the cluster sample is that the objects in the former are point sources while the latter consists of extended sources. 
 
The qualitative comparison of the cross-calibration bias parameter J$_1$ (Eq. \ref{J1.eq}) indicates similar features in 3XMM and cluster samples in the 0.5-4.5 keV band (see Fig. \ref{clust_3xmm.fig}): 1) the MOS/pn data-to-model ratios are systematically above unity, 2) MOS1/pn bias increases with energy and 3) MOS2/pn bias is energy-independent.

\subsection{MOS2/pn 0.5-4.5 keV band}
Quantitatively, very good news is brought by the excellent agreement on the MOS2/pn cross-calibration bias with clusters in the 0.5-4.5 keV band (see Fig. \ref{clust_3xmm.fig}, right panel). The median J$_1$ values of the cluster sample and 3XMM in each band differ on average by only $\sim$1\% and the differences are mostly consistent within the statistical and systematic uncertainties of the cluster data only. Rather than being a co-incidence we think this agreement serves as a proof that the systematics of both the clusters and 3XMM related to pn and MOS2 instruments affecting this band are well in control and minimised below 1\% of the cross-calibration bias signal. 
This agreement gives strength to our cluster-based argumentation that the shape of the effective areas of pn and MOS2 are currently absolutely calibrated with accuracy of $\sim$1\%.

\subsection{MOS1/pn 0.5-4.5 keV band}
The situation is less clear for MOS1/pn pair. The median J$_1$ values of 3XMM are systematically lower than those of the cluster sample in the 0.5-4.5 keV band, by $\sim$3\% on average, $\sim$7\% at maximum (see Fig. \ref{clust_3xmm.fig}, left panel). The deviations are significant when considering the uncertainties of the cluster data. 
 Currently we do not understand why the 3XMM and cluster sample results provide consistent results for the MOS2/pn analysis and significantly different results for the MOS1/pn analysis.
 
\subsection{5-6 keV band}
Towards the high end of our studied energy band, $\sim$5-6 keV, there is another problem. The 3XMM data do not exhibit the steep rise of J values for MOS/pn pairs found in the cluster data (see Fig. \ref{clust_3xmm.fig}). The estimates of the systematics related to the cluster sample, as presented in this work, are much smaller than the above difference. Similarly, the current understanding of the systematics related to 3XMM does not explain the difference. 
The solution to this problem requires more work which is beyond the scope of this paper. 
 
\subsection{The scatter}
\label{Scatter2}
The source-to-source scatter of the 3XMM data for MOS/pn pairs is substantial\footnote{The scatter would be even larger when applying Method 2} and comparable to that in the cluster sample (see Fig. \ref{clust_3xmm.fig}). 
This confirms that the relatively large scatter we reported based on the cluster sample is not an inherent problem of clusters. It
rather reflects some remaining problems in the common part of the analysis of the cluster sample and 3XMM when processing the data and/or estimating the effective area for a given observation.
    
\section{Discussion}
\label{discussion}
When producing the effective areas with the \texttt{arfgen} tool in the SAS software, the user has an option (applyxcaladjustment=yes) of adjusting the MOS effective area so that the spectral fit of the MOS data would yield results consistent with the pn on average. This correction has been derived by the XMM-Newton calibration team using the 3XMM sample (see the XMM-Newton calibration note CAL-SRN-0382). In order to avoid confusion between the above approach and the one we present in this paper, we discuss and compare the two below.

In Section \ref{3XMM} we showed that the cluster sample and the 3XMM sample both indicated remaining inconsistencies of the calibration of the effective areas of MOS and pn (Fig. \ref{clust_3xmm.fig}). We consider that the origin of the cross-calibration bias between the EPIC instruments needs to be understood before there can be scientifically justified progress on the issue of the inconsistence of the spectral fit parameters derived with the data from different EPIC instruments. Since the problem of the effective area cross-calibration bias is currently unresolved, we apply a scientific approach of considering it as a source of systematic uncertainty affecting the spectral fits. Our tools provide a practical way of evaluating and propagating these uncertainties to the results of the spectral analysis of the XMM-Newton/EPIC data.

The effective area cross-calibration bias measurements using the 3XMM and the cluster samples agree best at the lowest energies. However, the SAS implementation does not modify the effective area below 2 keV while our corrections are applicable down to 0.5 keV. On the other hand, there is a substantial difference between the medians of the two samples at the higher energies (above 4.5 keV). Thus, the average effective area correction for MOS will be different in the two approaches at some level.

The SAS implementation effectively assumes that the effective area calibration of the pn is very accurate and thus the effective area of MOS should be modified. Since there is no solid evidence for the above assumption, our tools allow the user to modify the effective area of any of the EPIC instruments.

Both the 3XMM sample and the cluster sample show that there is a substantial scatter in the evaluated cross-calibration bias between the individual objects (Section \ref{Scatter2}). The scatter is much larger than that allowed by the statistical uncertainties and the known systematic uncertainties. Thus, a single average correction for a random observation is not robust due to the ignored scatter.
The reason for this scatter remains unknown and thus we consider it as an additional source of systematic uncertainties. 
Since there is currently no method for estimating accurately the cross-calibration bias for a given individual observation, we devised practical tools which allow the user to propagate the measured scatter to the spectral fit parameter uncertainties.

\section{Summary and conclusions}
 We tested two methods for evaluating the effective area cross-calibration bias for CCD-type X-ray instruments. We applied the methods to a sample of clusters of galaxies observed with XMM-Newton/EPIC instruments and examined the  cross-calibration of the total effective area as implemented in SAS19.1.0 on Nov 2021. We found and quantified the bias in the current effective area cross-calibration between the EPIC instruments. Due to the very high signal to noise of the data of the cluster sample the discrepancies are extremely significant. We repeated the analysis to objects in the 3XMM-DR7 catalogue for verification of the robustness of the results. We derived suggestions on the most likely causes of the biases. We list and discuss below the most important results.

\begin{itemize}

\item 
We examined and compared the effect of performing the evaluation and analysis of the cross-calibration bias using the data before or after the convolution of the spectral models with the redistribution matrix. While the median cross-calibration bias obtained using the two methods deviates by less than 1\% at the lower energies, the deviations become larger at higher energy, reaching 3\% at 6 keV. Also, the cluster-to-cluster scatter is very similar in the two approaches except at the highest energies, where the post-convolution method suppresses the scatter somewhat.

\item 
The MOS/pn effective area cross-calibration bias factor in the 0.5-6.1 keV band deviates substantially from unity.  
The bias is systematic suggesting that MOS (pn) effective area may be calibrated too low (high), by $\sim$3-27\% depending on the instrument and energy band. 
 
\item 
The analysis suggests that the absolute energy dependencies (i.e. shapes) of the effective area components of MOS2 and pn are correctly calibrated within $\sim$1\% in the 0.5-4.5 keV band. This is confirmed by the consistence with the 3XMM results.
     
\item 
The cluster analysis indicates that MOS1 stands out in the energy-dependence comparison, suggesting that the absolute MOS1 effective area calibration has a significant bias which increases linearly with energy, amounting to $\sim$5-10\% in the 0.5-4.5 keV. 
  
\item 
The above results suggest a bit surprisingly that the effective area cross-calibration of the energy dependence between MOS2 and pn is in better agreement than that between MOS1 and MOS2. 
 
\item 
In the 5-6 keV band the abrupt behaviour of the bias in the cluster data between MOS and pn suggests that the drop of the MOS QE in this band has been overestimated. However, the 3XMM data does not confirm this.

\item 
The cluster-to-cluster rms scatter of the bias is substantial compared to the median bias itself. This is confirmed with 3XMM and thus the scatter is not cluster-specific. The reasons for this effect and its random nature are beyond the scope of this work. Thus, a statistically robust implementation of the cross-calibration uncertainties to a scientific analysis of XMM-Newton/EPIC data would include the propagation of the scatter to the fit results (see Appendix \ref{Application}) for the implementation) instead of simple average bias correction.
  
\item 
The studies methods are powerful diagnostics tools for the effective area cross-calibration. We demonstrated their applicability to galaxy cluster X-ray data which reveals their potential for cross-mission effective area investigations: since clusters are stable over human time scales, non-simultaneous observations of same clusters with different X-ray missions can be utilised.

\end{itemize}

\begin{acknowledgements} The research leading to these results has received funding from the European Union’s Horizon 2020 Programme under the AHEAD2020 project (grant agreement n. 871158). Thanks to members of the XMM SOC for useful discussions and for making the 3XMM results available. We acknowledge the support by the Estonian Research Council grants PRG1006, and by the European Regional Development Fund (TK133).   
\end{acknowledgements}

%
%
 
\bibliographystyle{aa}
\bibliography{refs}
 
\appendix

\section{Parameter correlations}
\label{corr}
In this paper we have characterised the MOS/pn effective area cross-calibration bias and its statistical nature.
In this section we provide practical information and procedures for the general user for treating the MOS/pn cross-calibration bias as a systematic uncertainty in the spectral analysis of the XMM-Newton/EPIC data.
  
The procedure is complicated in case of significant correlations between the coefficients of the best-fit 4th order polynomials we used to model the MOS/pn cross-calibration bias of the individual clusters (see Section \ref{results-method2} and Fig. \ref{polyfits.fig}).  In order to understand better the correlations we computed the Pearson correlation coefficients between different pairs of the coefficients (a$_i$, a$_j$, where i,j=0..4) of the above described 4th order polynomials. The correlation matrices of the form
\[
   {\rm corr}(a_i,a_j)=
  \left[ {\begin{array}{ccc}
    a_{0} & \cdots & a_{4} \\
   \vdots & \ddots & \vdots  \\
    a_{4} & \cdots & a_{4} \\
  \end{array} } \right]
\]
 for MOS1/pn and MOS2/pn pairs yield
\[
   {\rm corr_{MOS1/pn}}(a_i,a_j)=
 \left[ {\begin{array}{ccccc}
      1.00  &   -0.95 & 0.87  & -0.75 &  0.61 \\
      -0.95 &    1.00 & -0.97 &  0.89 & -0.78 \\
       0.87 &   -0.97 &  1.00 & -0.97 &  0.91 \\
      -0.75 &    0.89 & -0.97 &  1.00 & -0.98 \\
       0.61 &   -0.78 &  0.91 & -0.98 &  1.00  \\
  \end{array} } \right]
\]
and
 \[
   {\rm corr_{MOS2/pn}}(a_i,a_j)=
 \left[ {\begin{array}{ccccc}
      1.00  &   -0.98 &  0.93 &  -0.85 &  0.74  \\
     -0.98  &    1.00 & -0.98 &   0.92 & -0.84  \\
      0.93  &   -0.98 &  1.00 &  -0.98 &  0.93  \\
     -0.85  &    0.92 & -0.98 &   1.00 & -0.98  \\
      0.74  &   -0.84 &  0.93 &  -0.98 &  1.00  \\
  \end{array} } \right]
\]

(respectively) indicating significant correlations between most of the parameters.

\begin{figure*}
\hbox{
\includegraphics[width=4.5cm,angle=0]{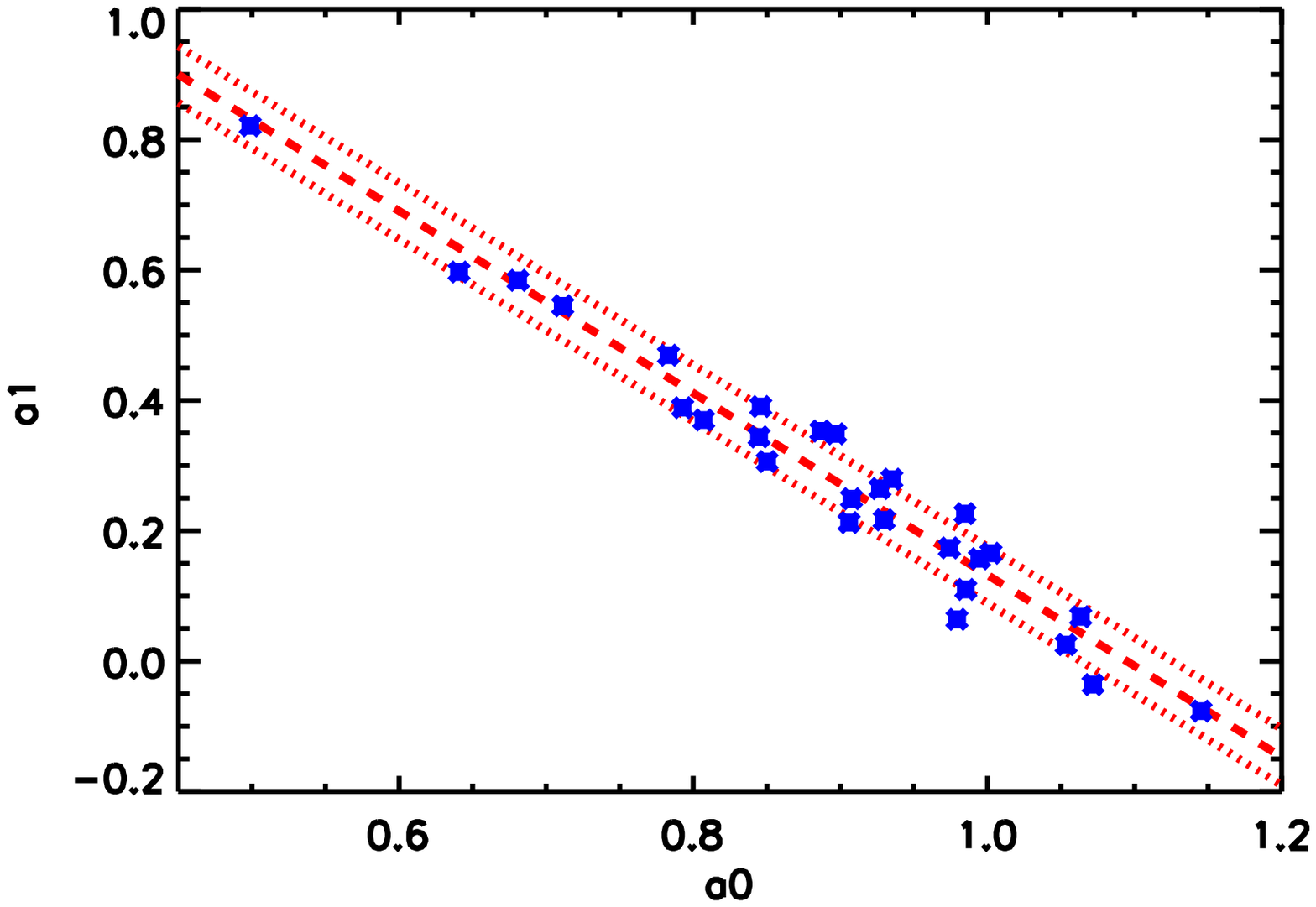}
\includegraphics[width=4.5cm,angle=0]{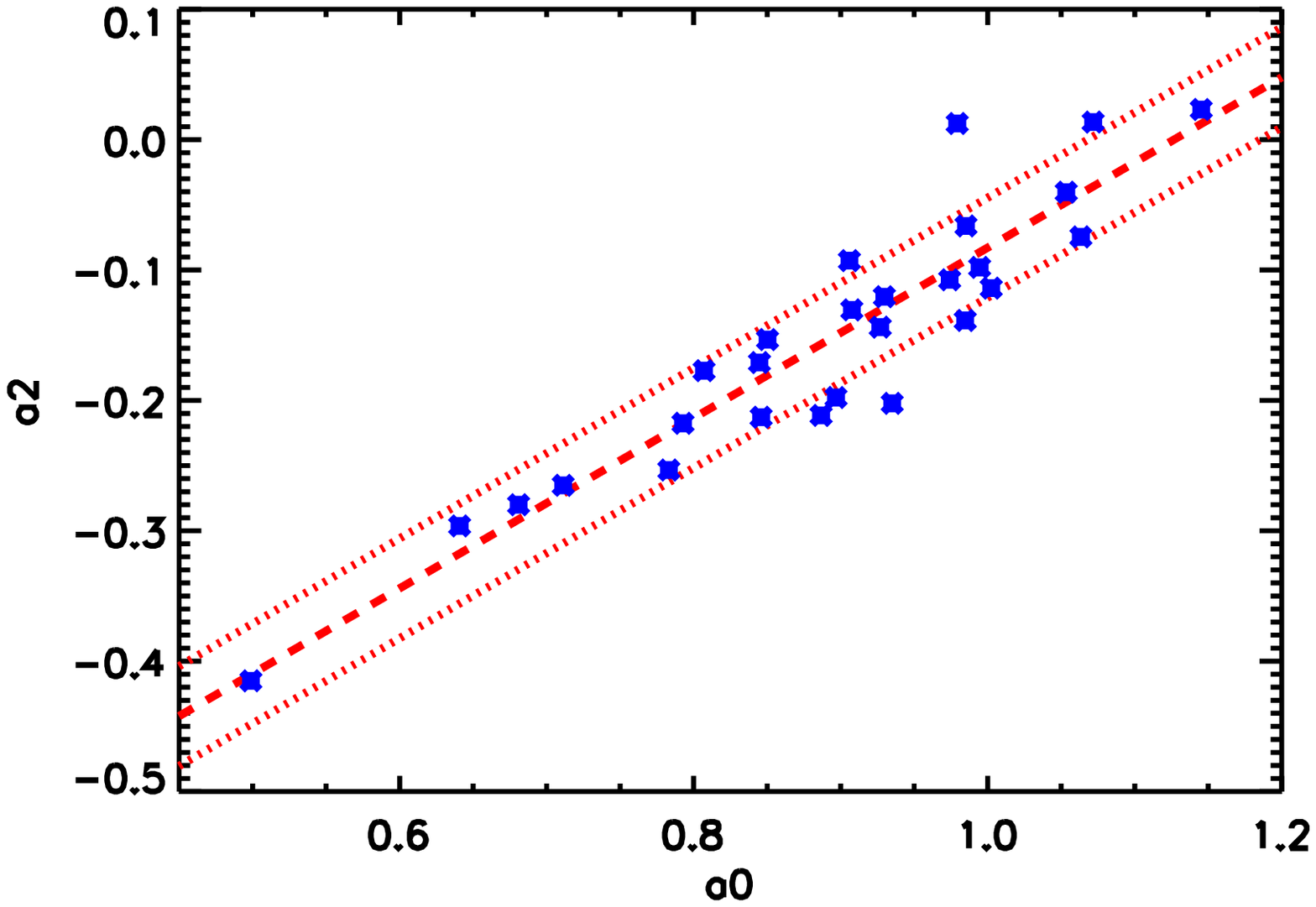}
\includegraphics[width=4.5cm,angle=0]{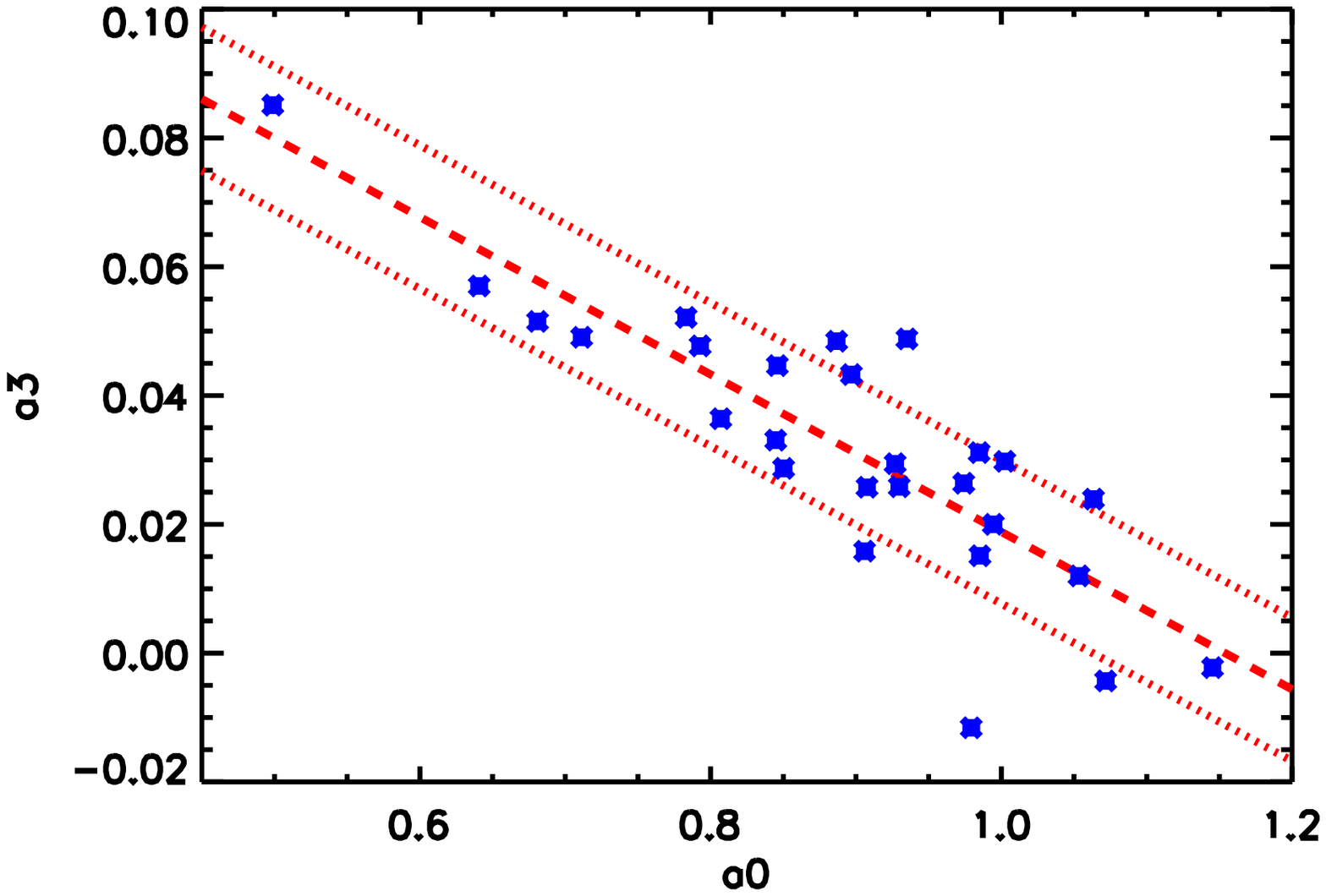}
\includegraphics[width=4.5cm,angle=0]{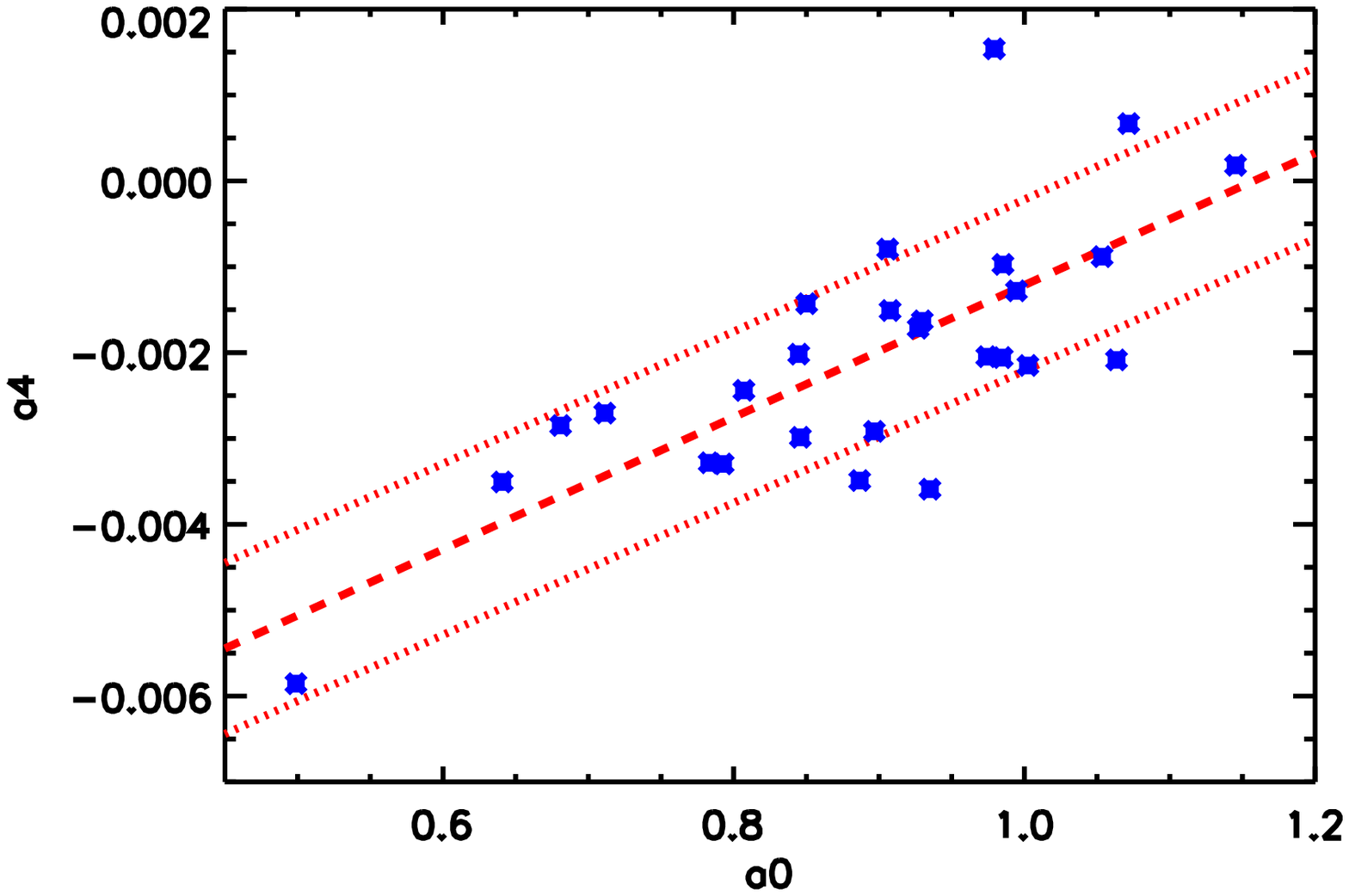}
}
\hbox{
\includegraphics[width=4.5cm,angle=0]{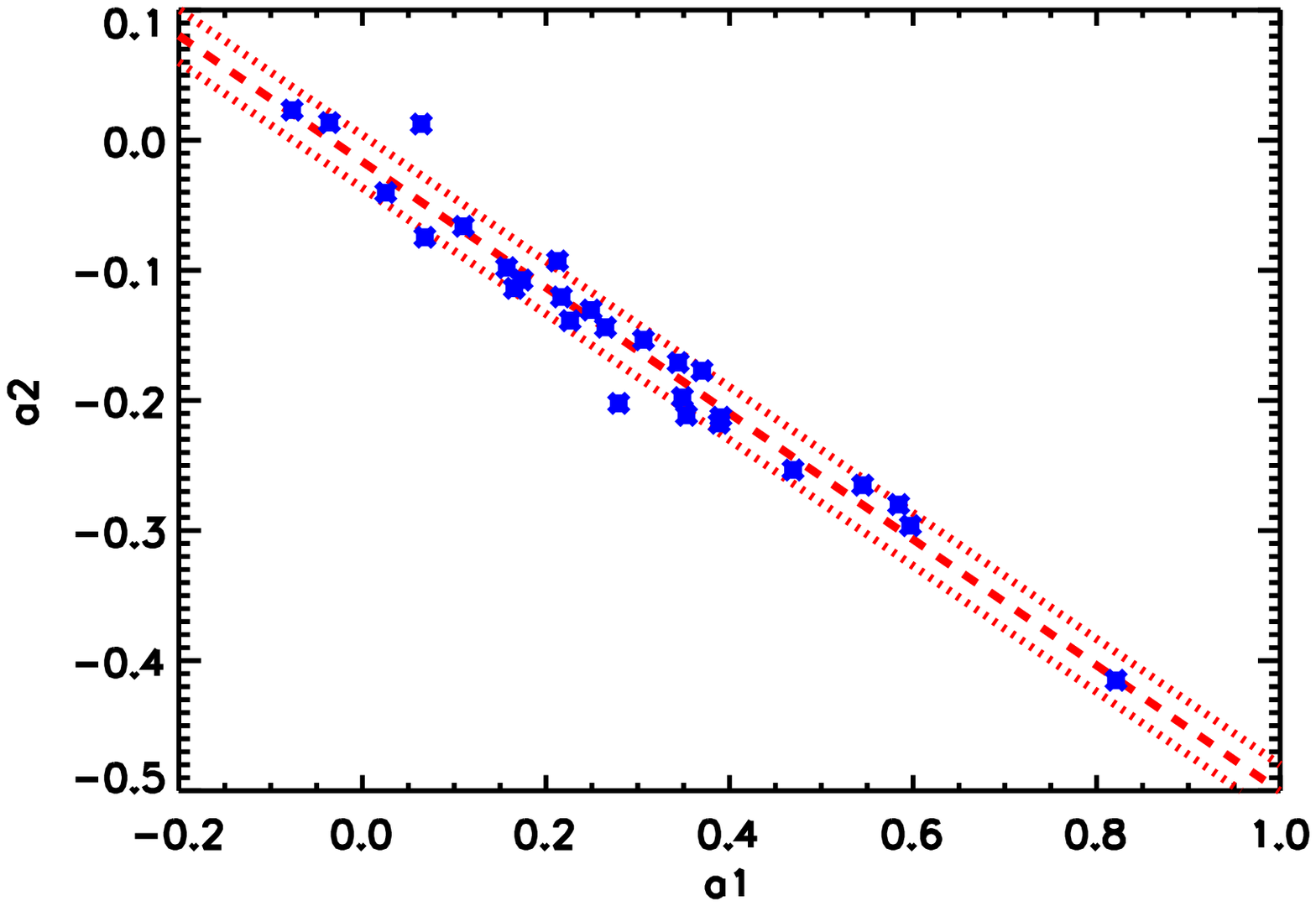}
\includegraphics[width=4.5cm,angle=0]{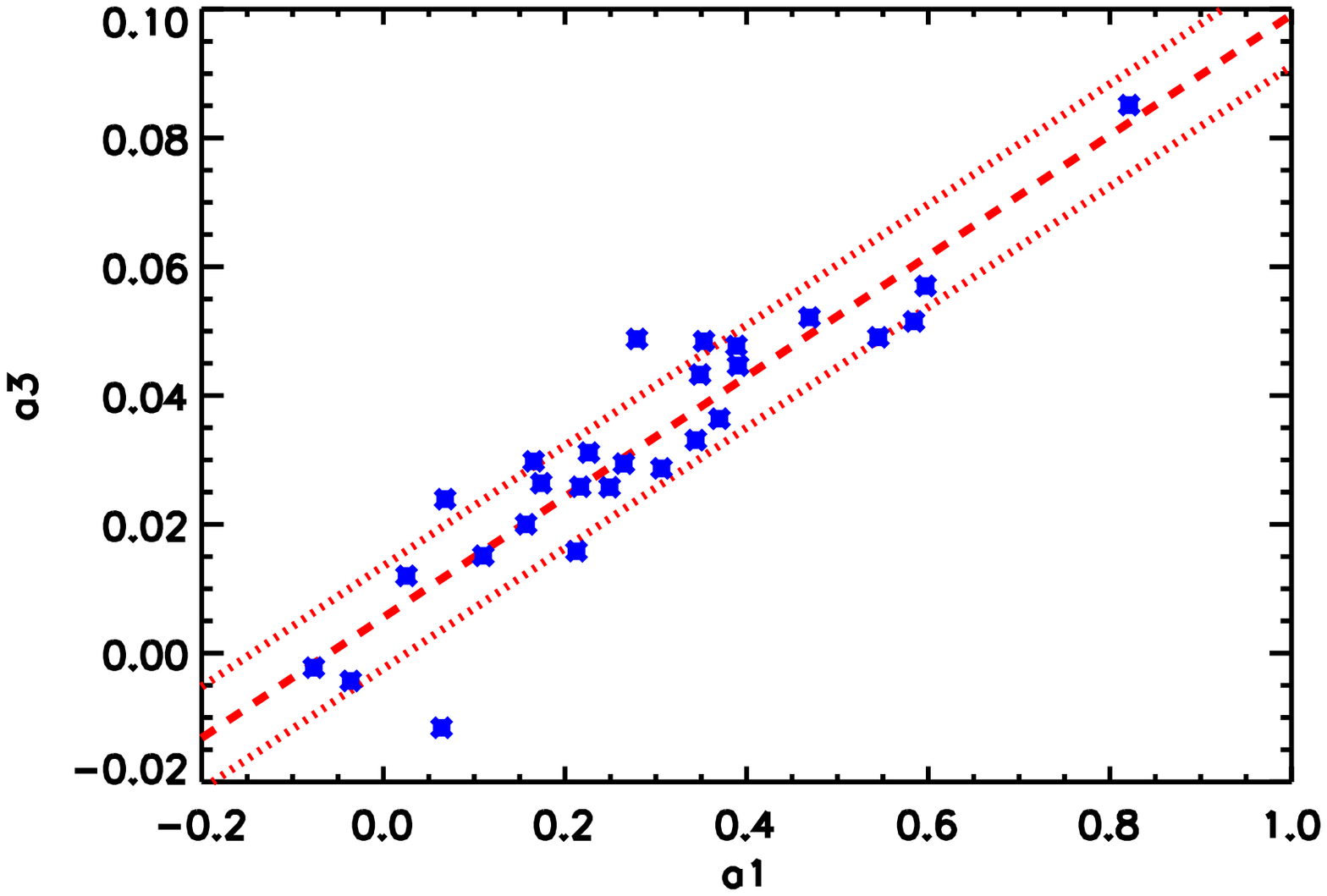}
\includegraphics[width=4.5cm,angle=0]{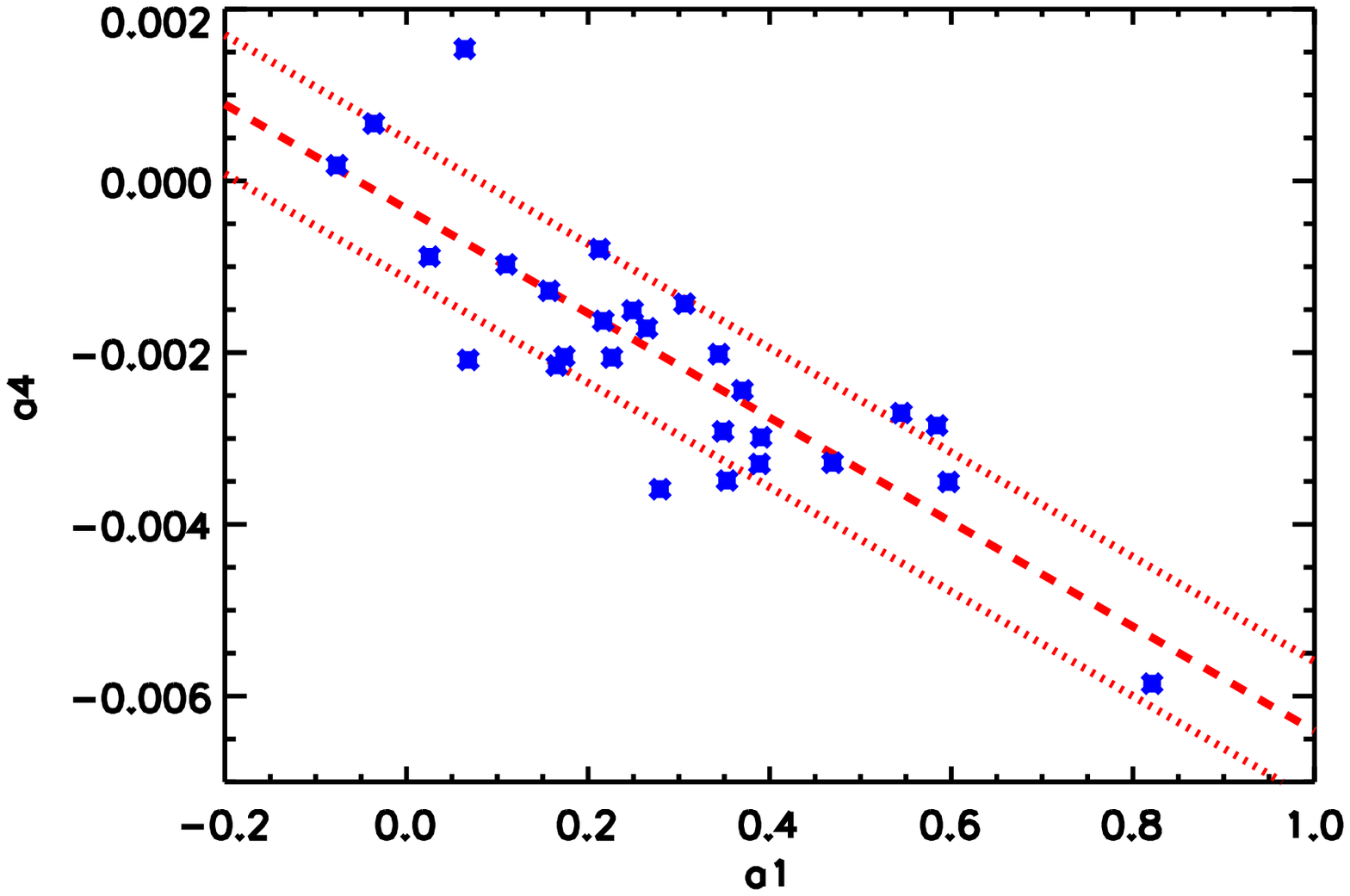}
\includegraphics[width=4.5cm,angle=0]{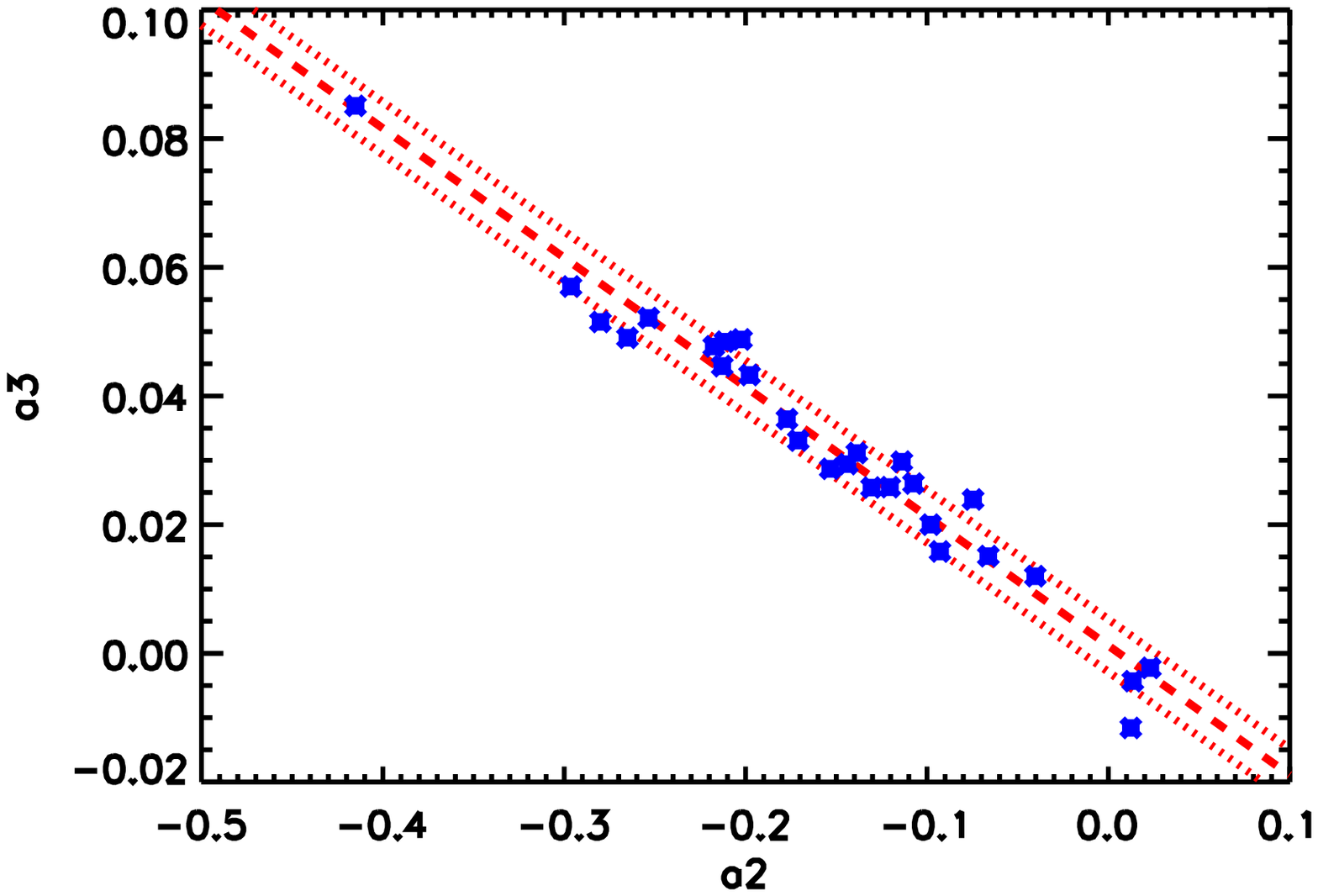}
}
\hbox{
\includegraphics[width=4.5cm,angle=0]{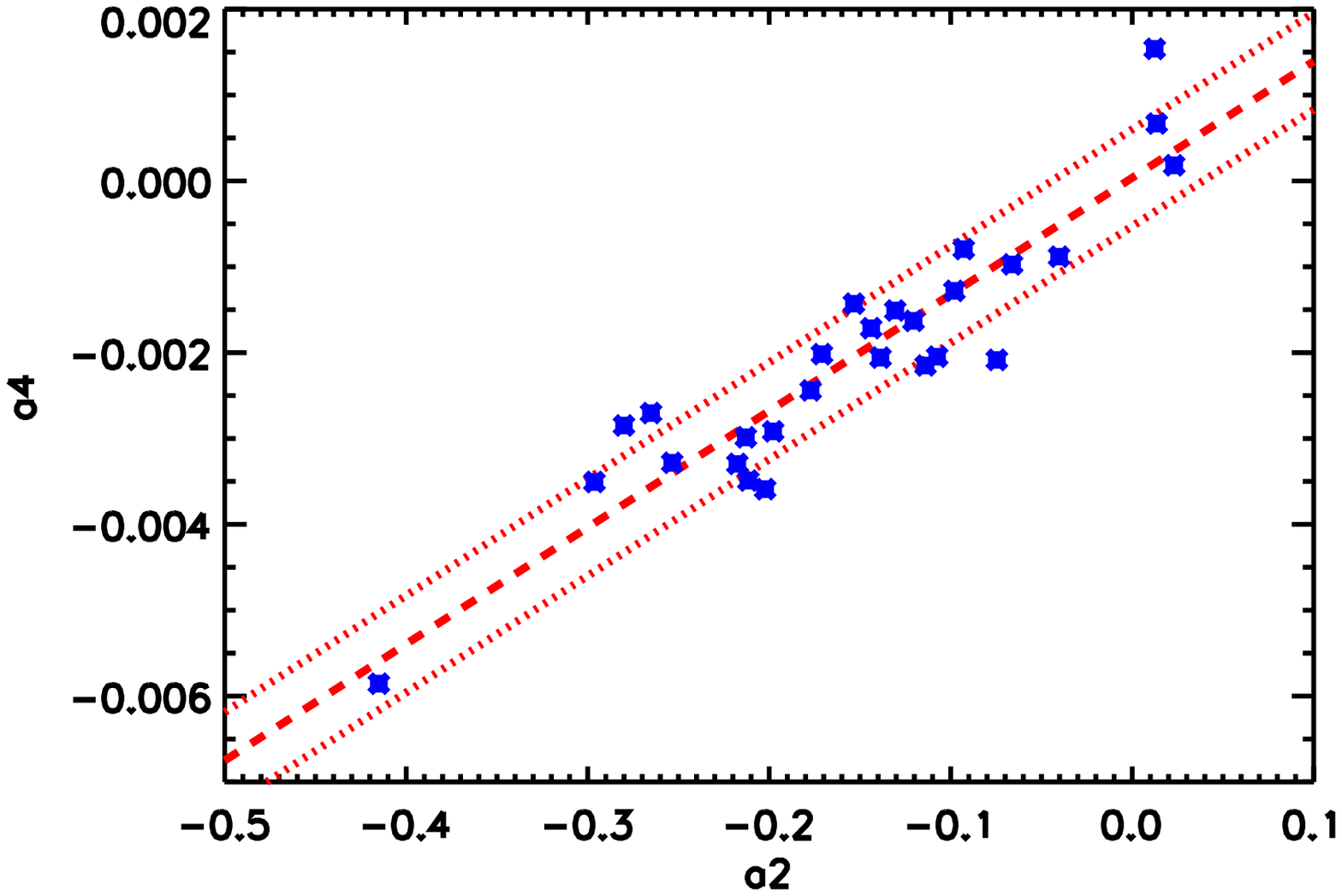}
\includegraphics[width=4.5cm,angle=0]{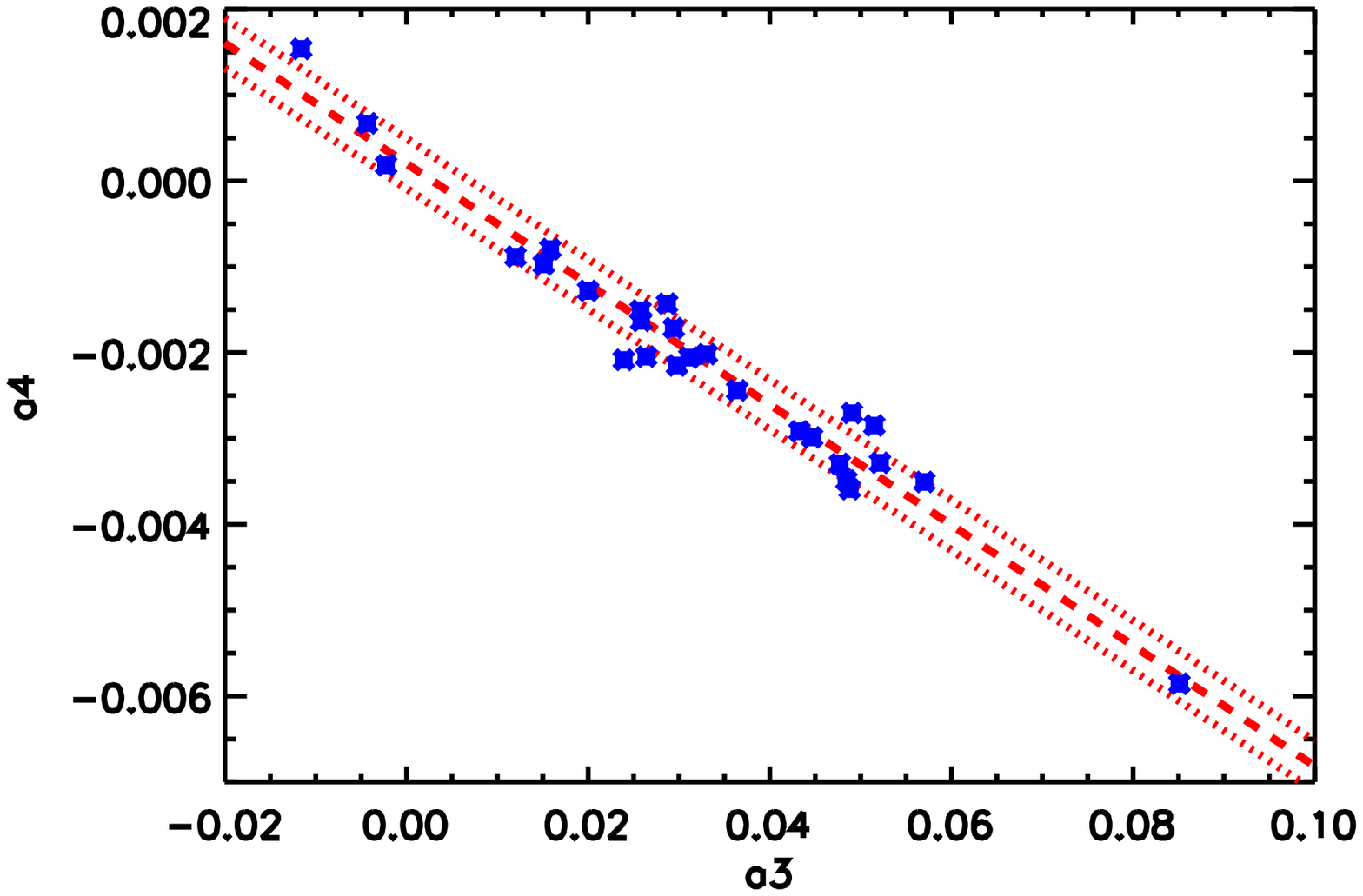}
 }
\caption{The correlations between the coefficients of the best-fit 4th order polynomial models to the MOS1/pn cross-calibration data. The data are shown with blue symbols. The best-fit linear model and the variation characterised by the standard deviation of the differences between the Y-axis data and the corresponding model predictions are shown as red dashed and dotted lines, respectively. 
 }
\label{MOS1_pn_corr.fig}
\end{figure*}

\begin{figure*}
\hbox{
\includegraphics[width=4.5cm,angle=0]{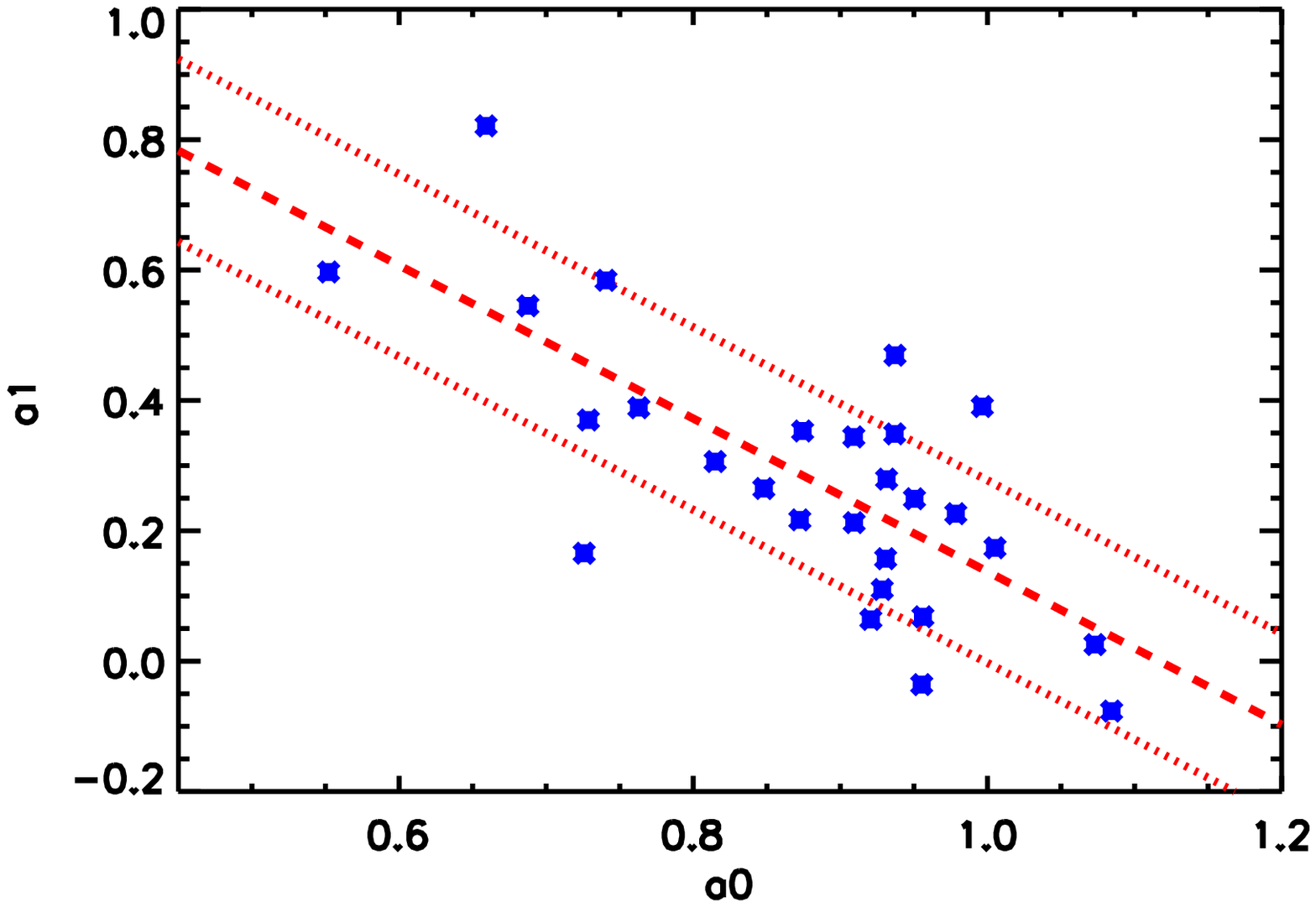}
\includegraphics[width=4.5cm,angle=0]{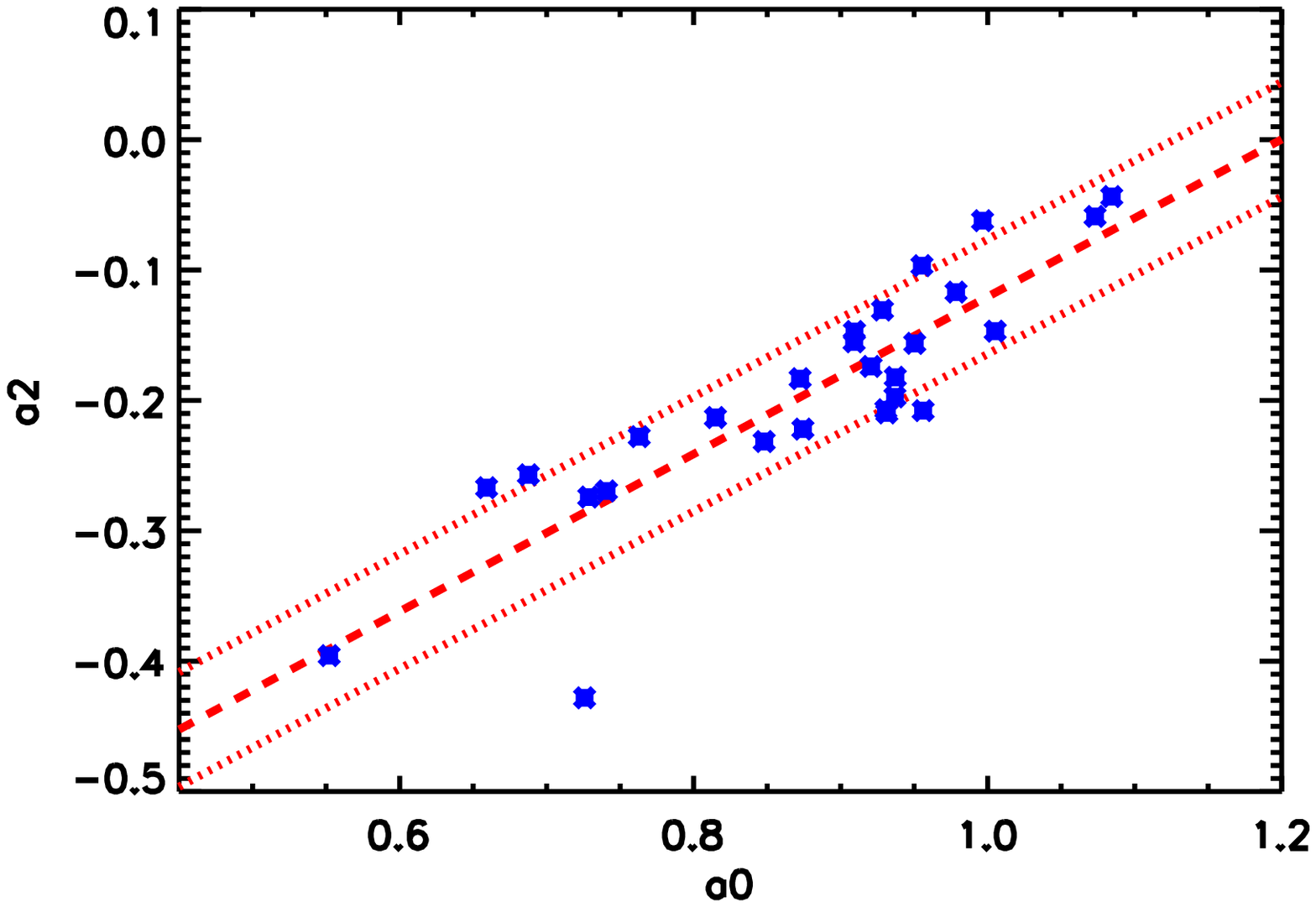}
\includegraphics[width=4.5cm,angle=0]{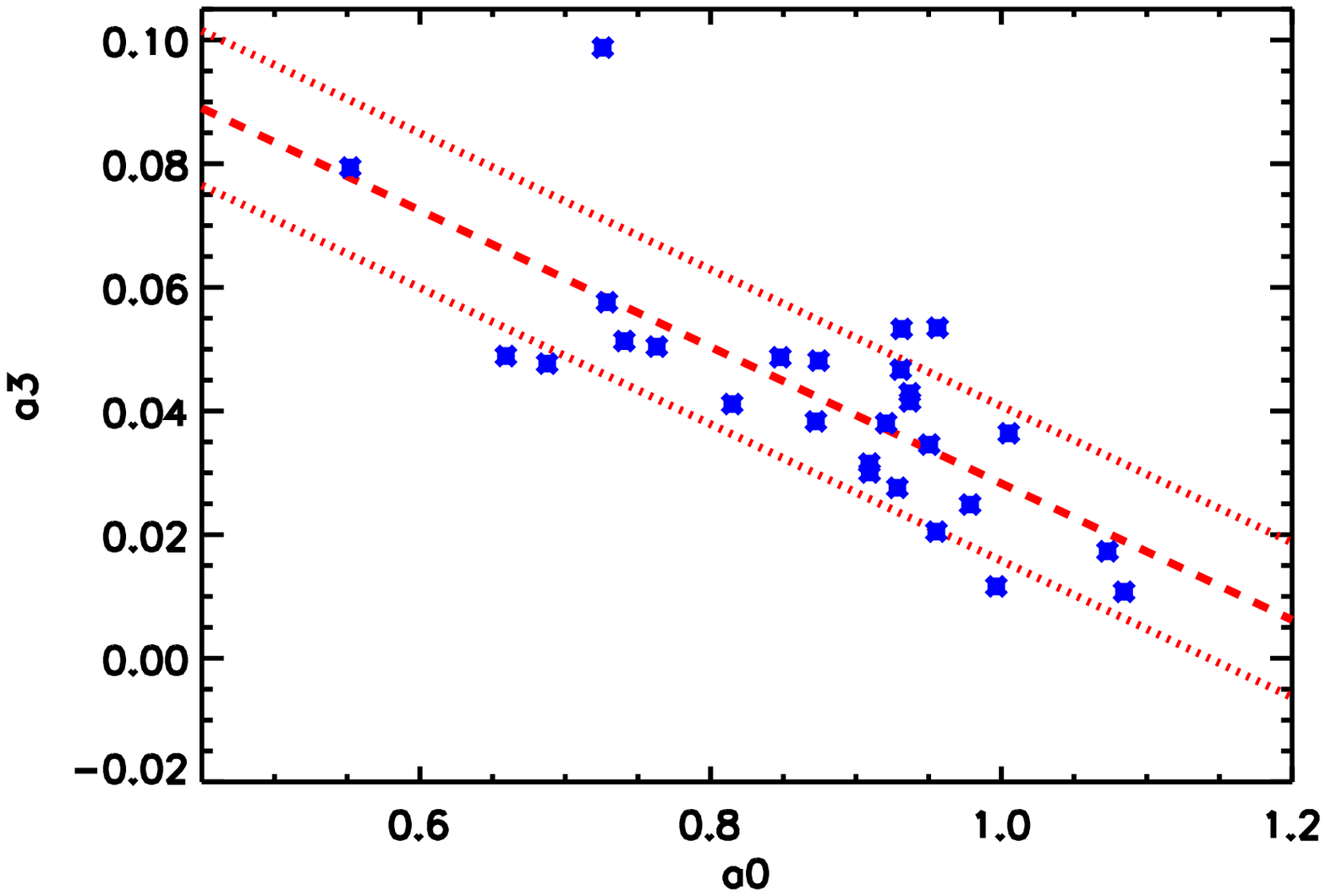}
\includegraphics[width=4.5cm,angle=0]{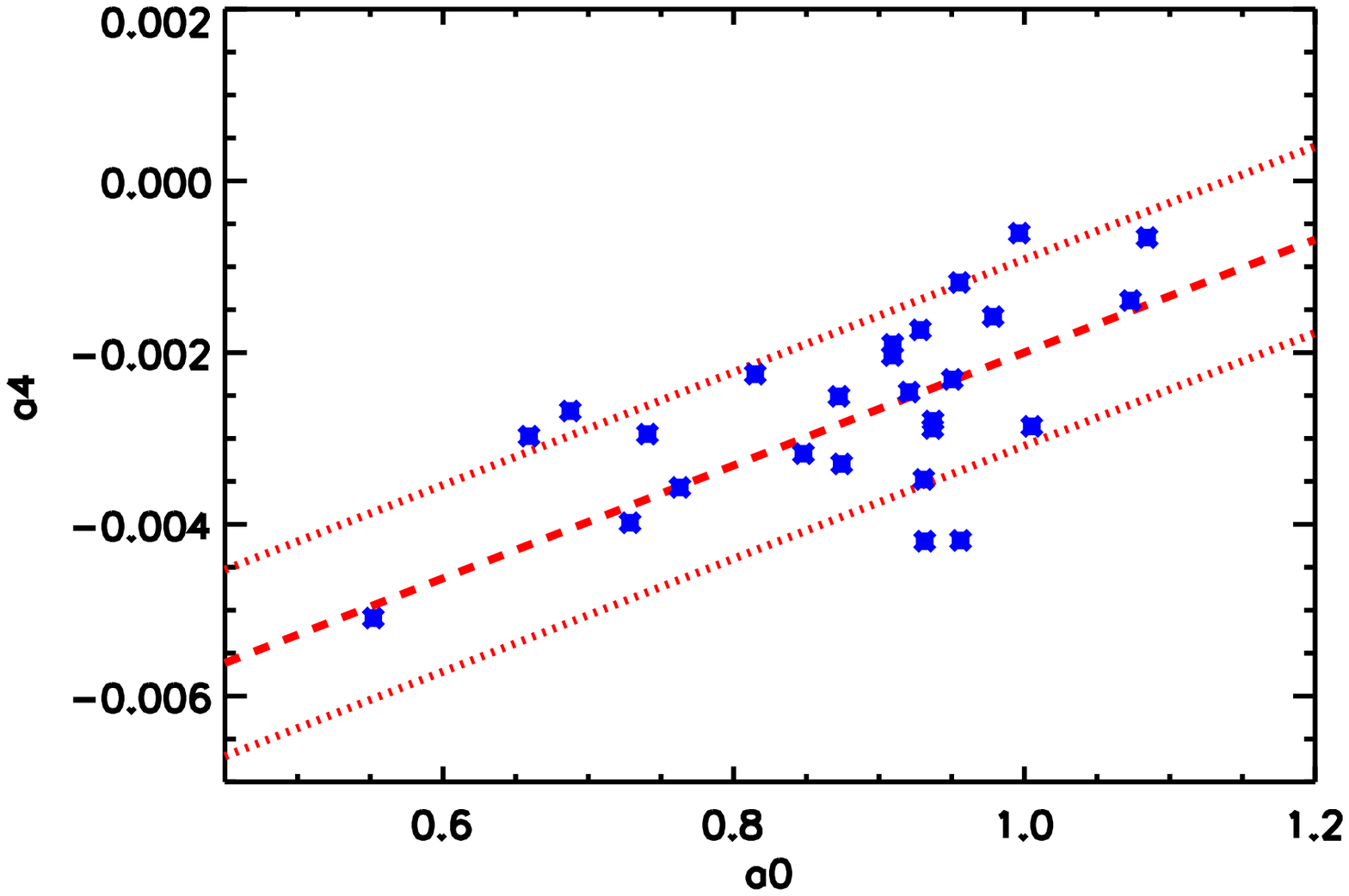}
}
\hbox{
\includegraphics[width=4.5cm,angle=0]{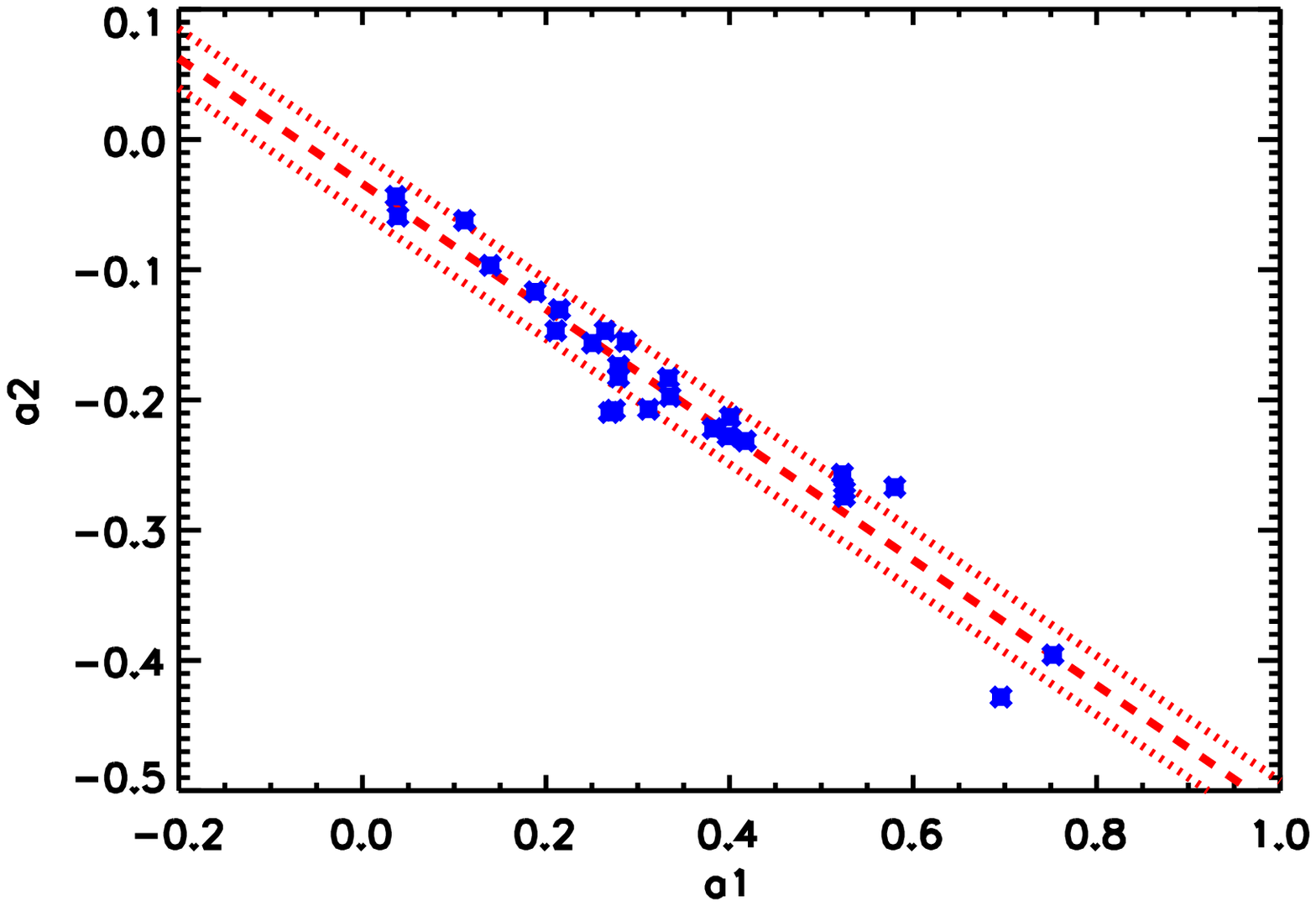}
\includegraphics[width=4.5cm,angle=0]{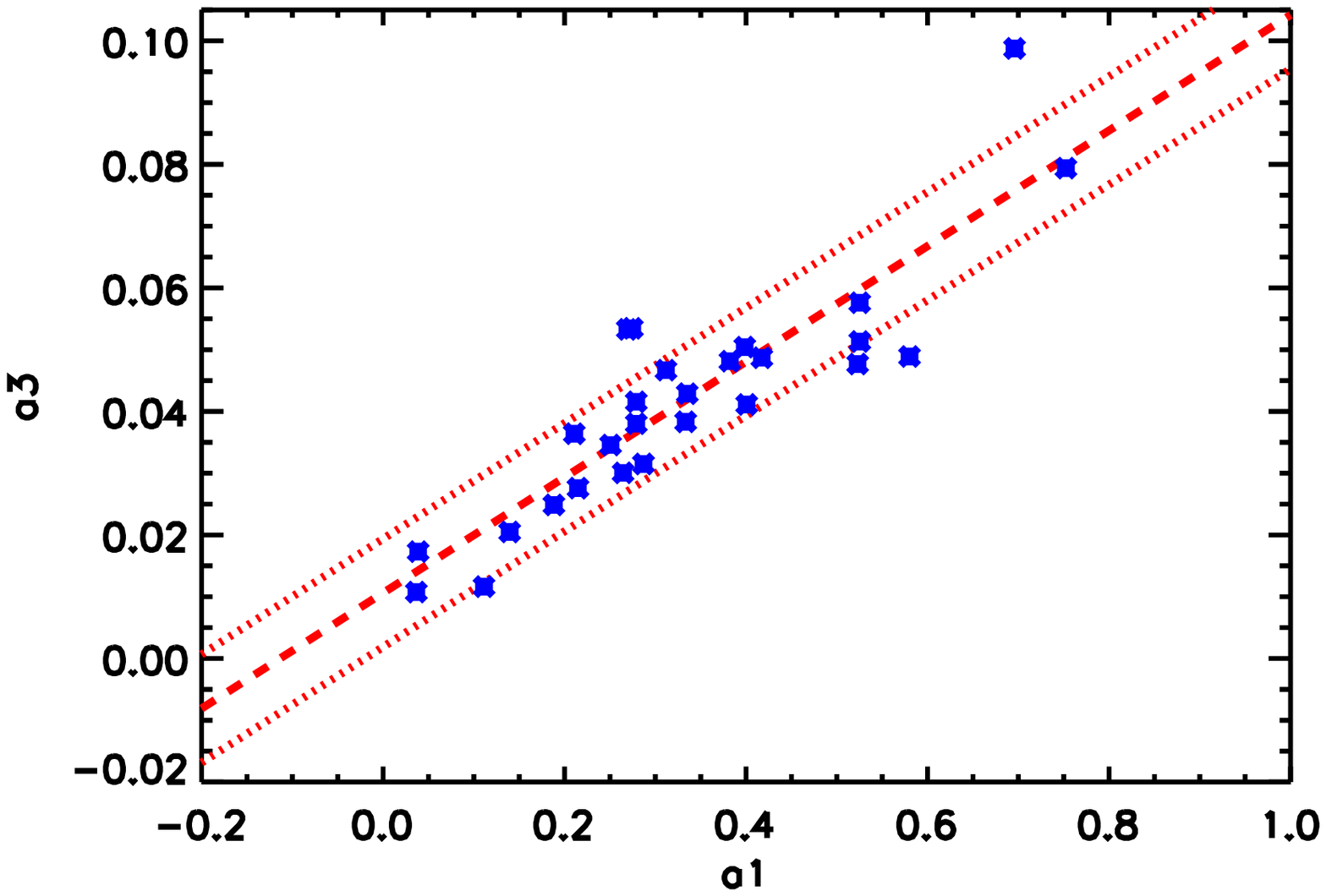}
\includegraphics[width=4.5cm,angle=0]{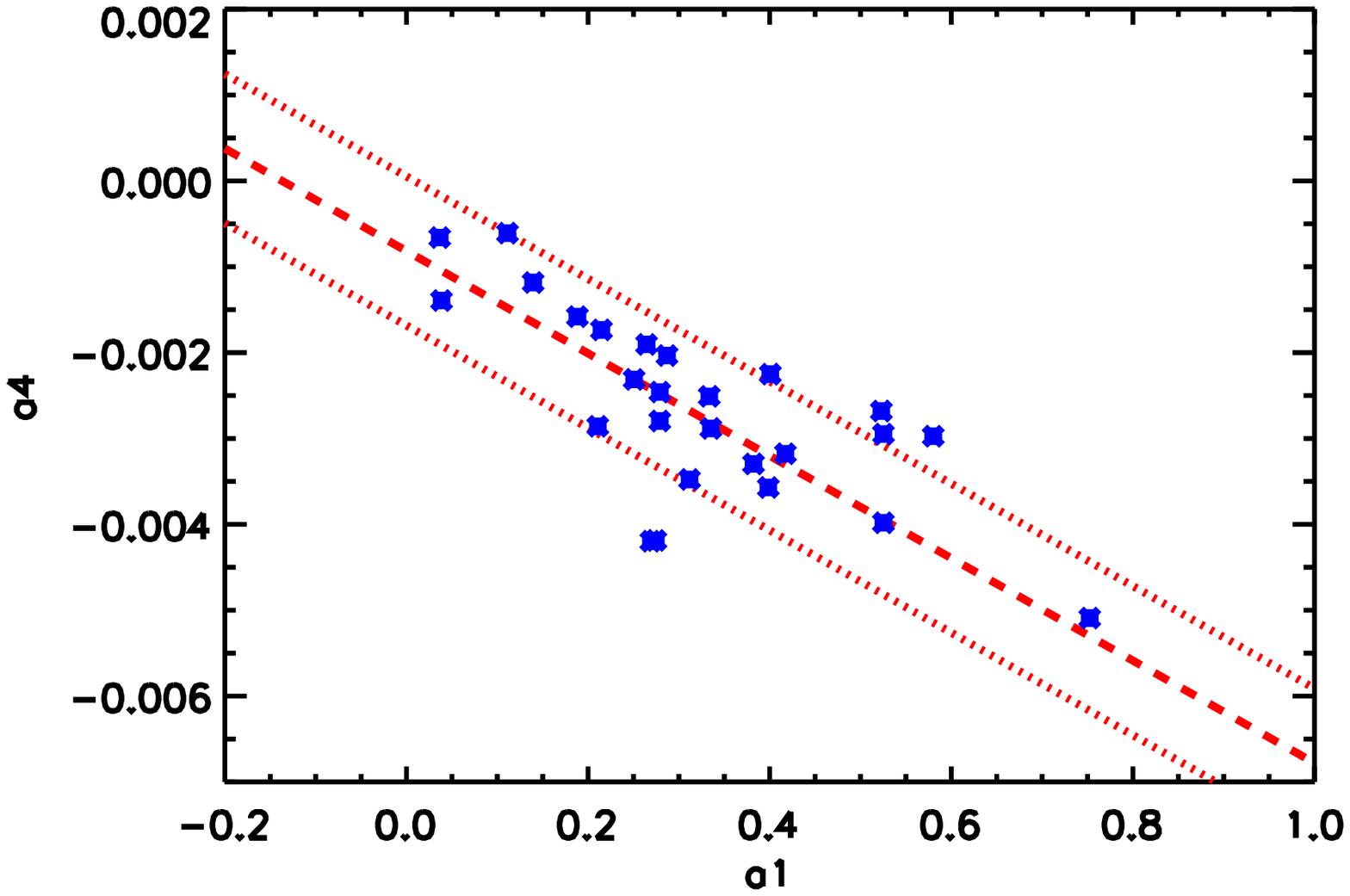}
\includegraphics[width=4.5cm,angle=0]{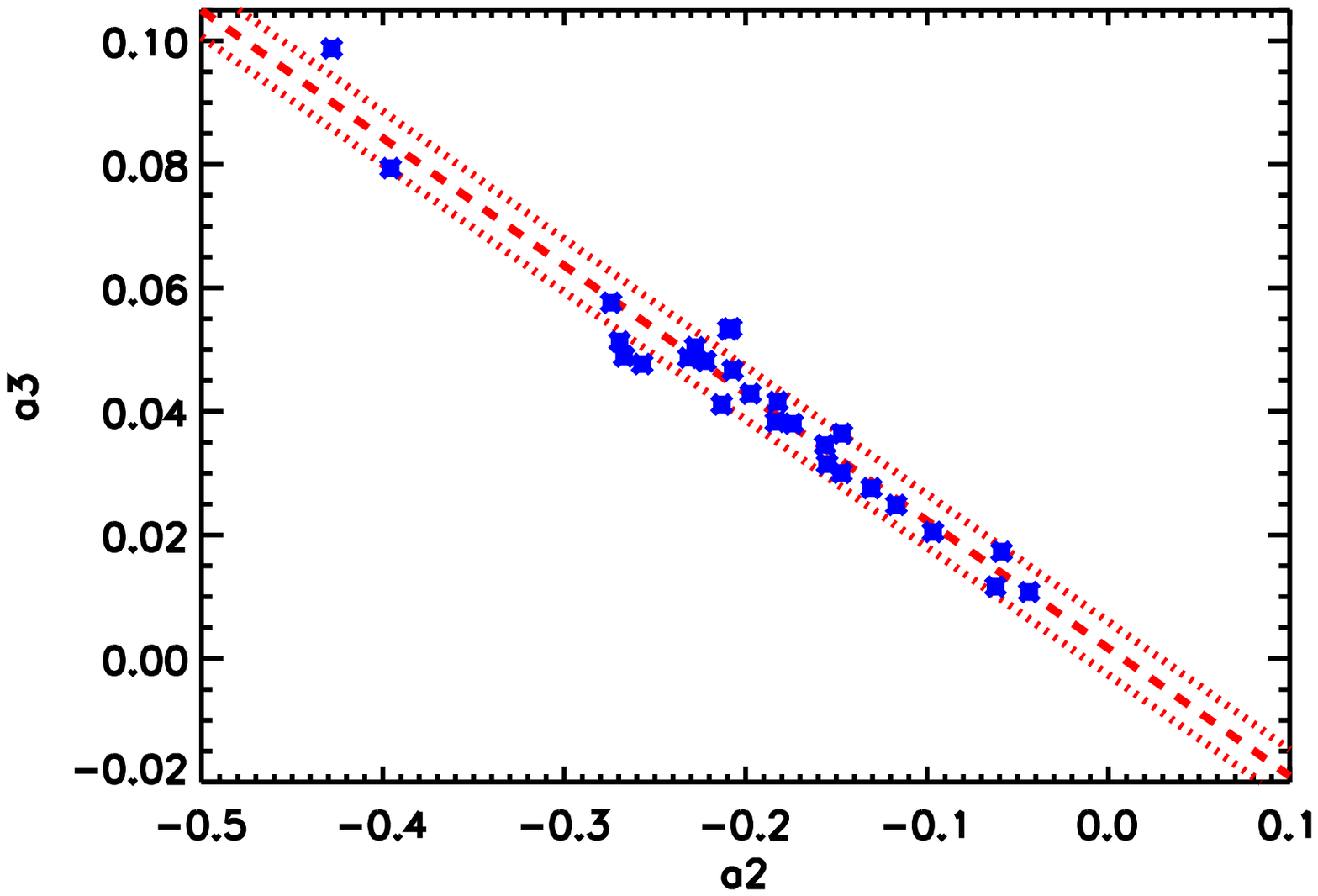}
}
\hbox{
\includegraphics[width=4.5cm,angle=0]{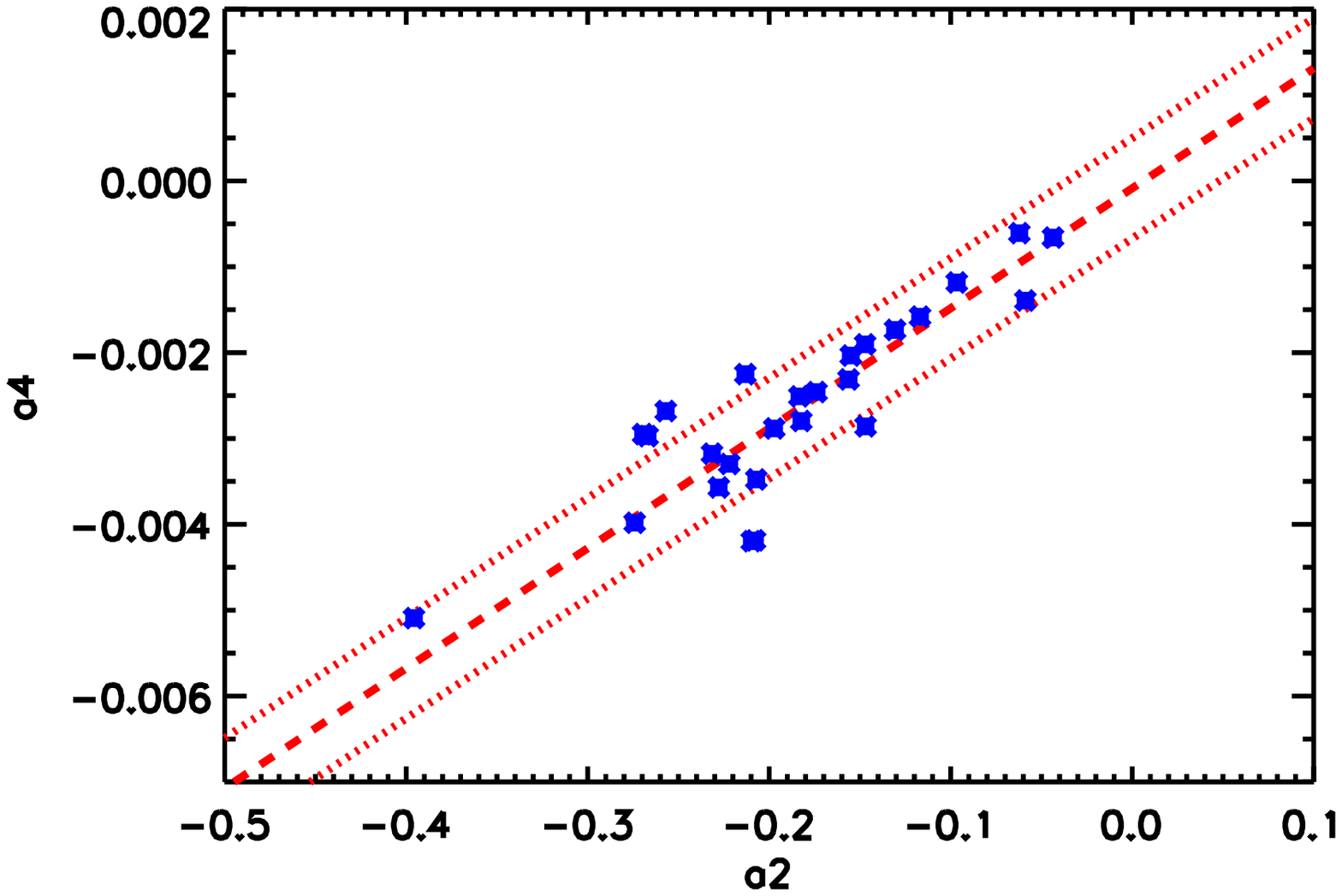}
\includegraphics[width=4.5cm,angle=0]{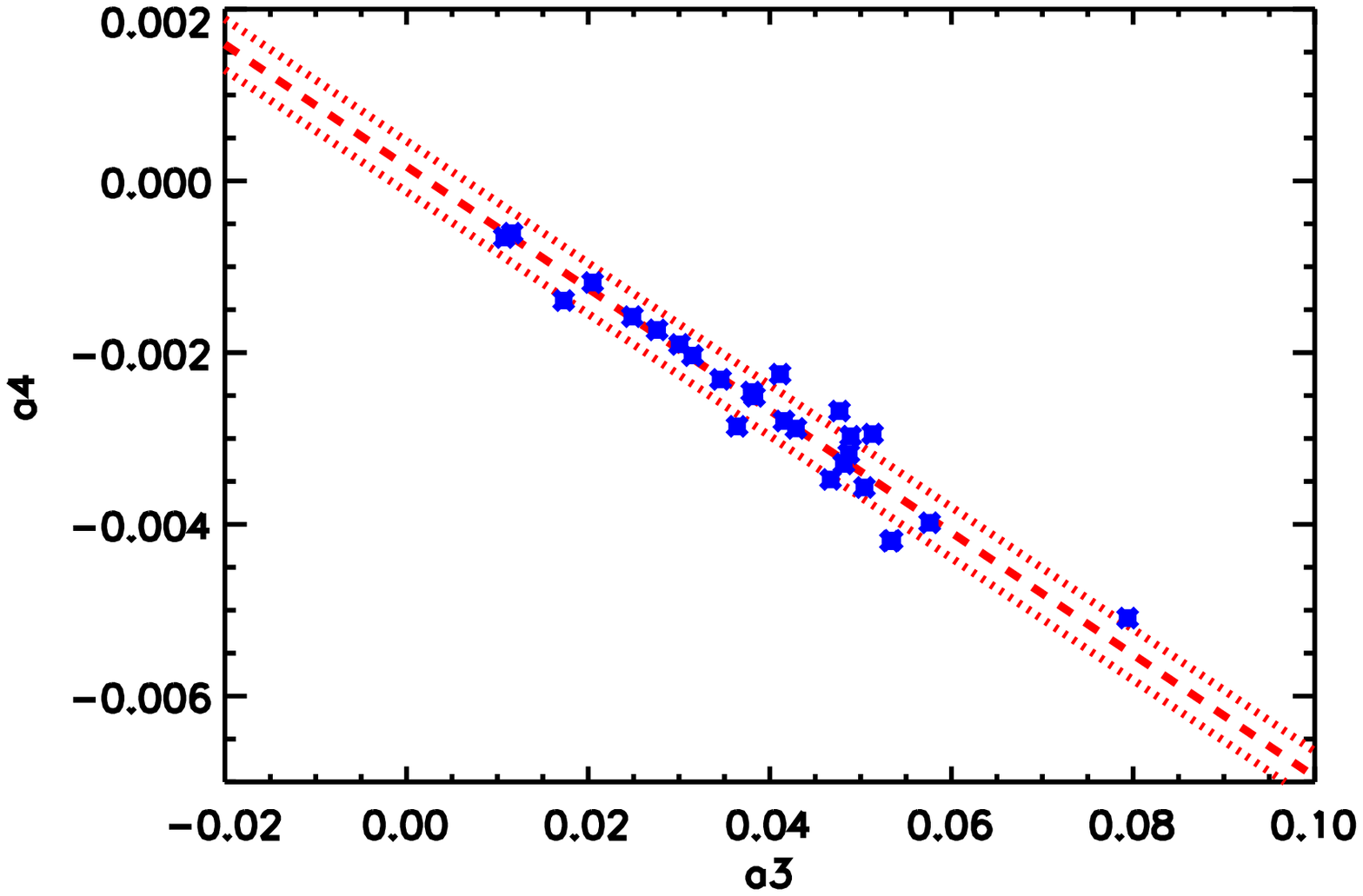}
 }
 \caption{The same as Fig. \ref{MOS1_pn_corr.fig} but for MOS2/pn pair.
 }
\label{MOS2_pn_corr.fig}
\end{figure*}

We visualised the correlations by plotting each parameter as a function of another one for each of the independent pair (see Figs. \ref{MOS1_pn_corr.fig} and \ref{MOS2_pn_corr.fig}).
We quantified these correlations by fitting the best-fit parameters of each pair of the coefficients (a$_i$, a$_j$)
with a linear model a$_j$ = A $\times$ a$_i$ + B (see Fig. \ref{MOS1_pn_corr.fig} and Table \ref{coeff-linear.tab}). Since our sample is not large enough to robustly sample the full parameter ranges, we cannot evaluate the scatter of the a$_j$ data 
at each value of a$_i$. We thus assumed that the scatter is constant in the studied range of a$_i$ values and equal to the standard deviation of the difference of the a$_j$ data and the corresponding model predictions. 

Consistently with the high degree of the correlations between the parameters, the scatter of the parameter correlations ($\sigma_B$ in Table \ref{coeff-linear.tab}) is small compared to the spread of the best-fit parameters (Table \ref{results.tab}). Yet, ignoring the correlation scatter leads to underestimate of the final scatter of the modified effective areas when compared to data. A simple approach of allowing the parameters to vary within the correlation scatter leads to parameter combinations which are not consistent with the data. Due to the nature of the polynomial functions, many of these inconsistent parameter combinations produce models which deviate strongly from the data.

Thus, one would have to develop a more sophisticated method to properly incorporate the parameter correlations into the analysis.
Instead, we devised an another method whereby we rely more on the data rather than the model. Namely, we use the statistical results for the cross-calibration bias in the cluster data sample (the medians and the standard deviations as reported in Table \ref{statsys.tab})
to draw random cross-calibration bias data curves. This procedure ensures that we include the scatter of the data in the analysis.
We then fit these data sets with 4th order polynomials.
In this manner the parameter correlations and their uncertainties are naturally propagated while the models remain consistent with the data. 

\begin{table*}
\small
 \centering
  \caption{The parameters describing the correlations 
  \label{coeff-linear.tab}}
    \begin{tabular}{lcccccccccc}
  \hline\hline
      & \multicolumn{2}{c}{}  & \multicolumn{2}{c}{}  & \multicolumn{2}{c}{}  & \multicolumn{2}{c}{}  & \multicolumn{2}{c}{} \\                                                       \\ 
     & \multicolumn{2}{c}{a0}  & \multicolumn{2}{c}{a1}  & \multicolumn{2}{c}{a2}  & \multicolumn{2}{c}{a3}  & \multicolumn{2}{c}{a4} \\
     & A & B$\pm\sigma_B$    & A & B$\pm\sigma_B$  & A & B$\pm\sigma_B$  & A & B$\pm\sigma_B$  & A & B$\pm\sigma_B$ \\
\hline
\multicolumn{11}{c}{ }  \\
\multicolumn{11}{c}{\bf MOS1/pn}  \\
\multicolumn{11}{c}{ }  \\
a0            & --    & --            & -1.40  &  1.53$\pm$0.04   &  0.65 &  -0.74$\pm$0.04   &  -0.12 &  0.141$\pm$0.011   &  0.0077 & -0.0089$\pm$0.0010 \\   
a1            & -0.68 & 1.09$\pm$0.03 &  --    & --               & -0.48  & -0.016$\pm$0.020 &  0.094  & 0.0056$\pm$0.0080 &  -0.0061 & -0.00032$\pm$0.00081 \\   
a2            &  1.31 & 1.09$\pm$0.05 &  -1.98  & -0.021$\pm$0.041 & -- & -- & -0.20 &  0.0012$\pm$0.0041 &  0.014   &  0.000036$\pm$0.000565 \\   
a3            & -5.85 & 1.08$\pm$0.08 &  9.13  & -0.010$\pm$0.079 & -4.78 & -0.0001$\pm$0.0199   & -- & --  & -0.070 & 0.00020$\pm$0.00029  \\   
a4            & 71.9  & 1.04$\pm$0.10 & -115.9 &  0.046$\pm$0.112 &  63.1  &  -0.024$\pm$0.039  &  -13.7  &  0.0040$\pm$0.0041   & --  & --  \\   
\multicolumn{11}{c}{ }  \\
\multicolumn{11}{c}{\bf MOS2/pn}  \\
\multicolumn{11}{c}{ }  \\
a0            & --    & --    &  -1.17 &  1.31$\pm$0.14 & 0.60 & -0.72$\pm$0.04   & -0.11 & 0.14$\pm$0.01  & 0.0066 & -0.0086$\pm$0.0011 \\   
a1            & -0.68 & 1.11$\pm$0.04 &  --    & --     & -0.48  & -0.03$\pm$0.02 & 0.09 & 0.011$\pm$0.009  &  -0.0060 & -0.00082$\pm$0.00087 \\   
a2            & 1.25   & 1.12$\pm$0.06 &  -1.94 & -0.044$\pm$0.046 & -- & -- & -0.21 &  0.0016$\pm$0.0043  & 0.01 & -0.000086$\pm$0.000584 \\   
a3            & -5.1  & 1.09$\pm$0.09 & 8.4   & -0.018$\pm$0.083 & -4.6   & -0.0025$\pm$0.0202 & -- &  -- & -0.071   & 0.00017$\pm$0.00030  \\   
a4            &  57.4 & 1.04$\pm$0.10 & -101.1 & 0.050$\pm$0.113 & 58.7 &  -0.030$\pm$0.038  & -13.4 &  0.0042$\pm$0.0041  & -- & --  \\   
\hline 
\end{tabular}
\tablefoot{\\
The parameters of the best-fit linear models a$_j$ = A x a$_i$ + B to the coefficient pairs. $\sigma_B$ is evaluated as a standard deviation of the difference between the Y-axis data and the corresponding model prediction. 
 }
\end{table*}

\section{Implementation}
\label{Implementation}
We describe here in detail the procedure for the general user to implement the above information for X-ray spectral analyses of the XMM-Newton/EPIC data (these procedures are implemented in a public on-line tool).
1) The work flow begins with the usual fitting of the X-ray spectrum with the standard SAS-produced arf. 
If the data being analysed are obtained with the pn instrument, the current implementation allows the user to choose whether to estimate the effect of 
the cross-calibration bias between pn and MOS1 or between pn and MOS2. If the user wishes to analyse the data obtained with MOS1 or MOS2, the procedure allows the 
evaluation of the effect of the cross-calibration bias between that instrument and the pn instrument. 
2) Using the statistical results for the measure of the cross-calibration in the cluster sample corresponding to the choice of the instrument pair made in step 1 (Table \ref{statsys.tab}) the user draws a random cross-calibration bias data curve.   
3) The user fits the data set obtained in step 2 with a 4th order polynomial.
4) Depending on the choice of the instrument pair made in step 1, the user multiplies or divides the effective area column in the standard SAS-produced arf used in the beginning of the analysis with the best-fit 4th order polynomial obtained in step 3. If, for example, the user wants to estimate the effect of the MOS1/pn cross-calibration bias on the spectral analysis of the MOS1 (pn) data, the 4th order polynomial obtained in step 3 by fitting the randomised MOS1/pn cross-calibration bias data should be used to multiply (divide) the standard MOS1 effective area.
5) Steps 2-4 are repeated by a large number of times in order to produce a statistically meaningful sample of modified arf:s.
6) The spectral fit of step 1 is repeated, replacing each time the original arf with one of the arf:s generated by the above procedure.
7) The distribution of the best-fit parameters obtained with the modified arf:s can be analysed in order to evaluate the systematic uncertainty related to the cross-calibration bias.

\begin{figure*}
\hbox{
\hspace{-0.5cm}
\includegraphics[width=10cm,angle=0]{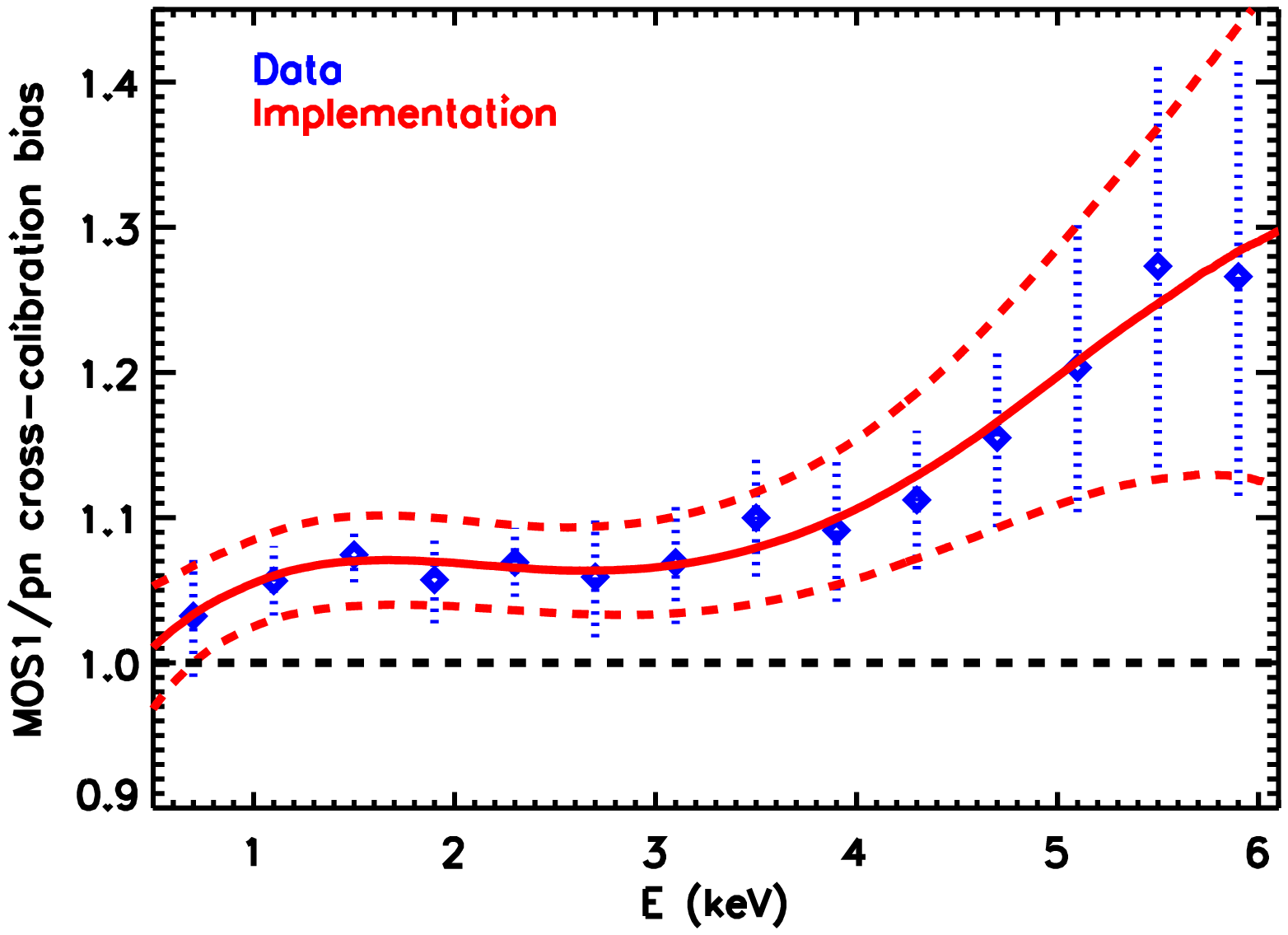}
\hspace{-1cm}
\includegraphics[width=10cm,angle=0]{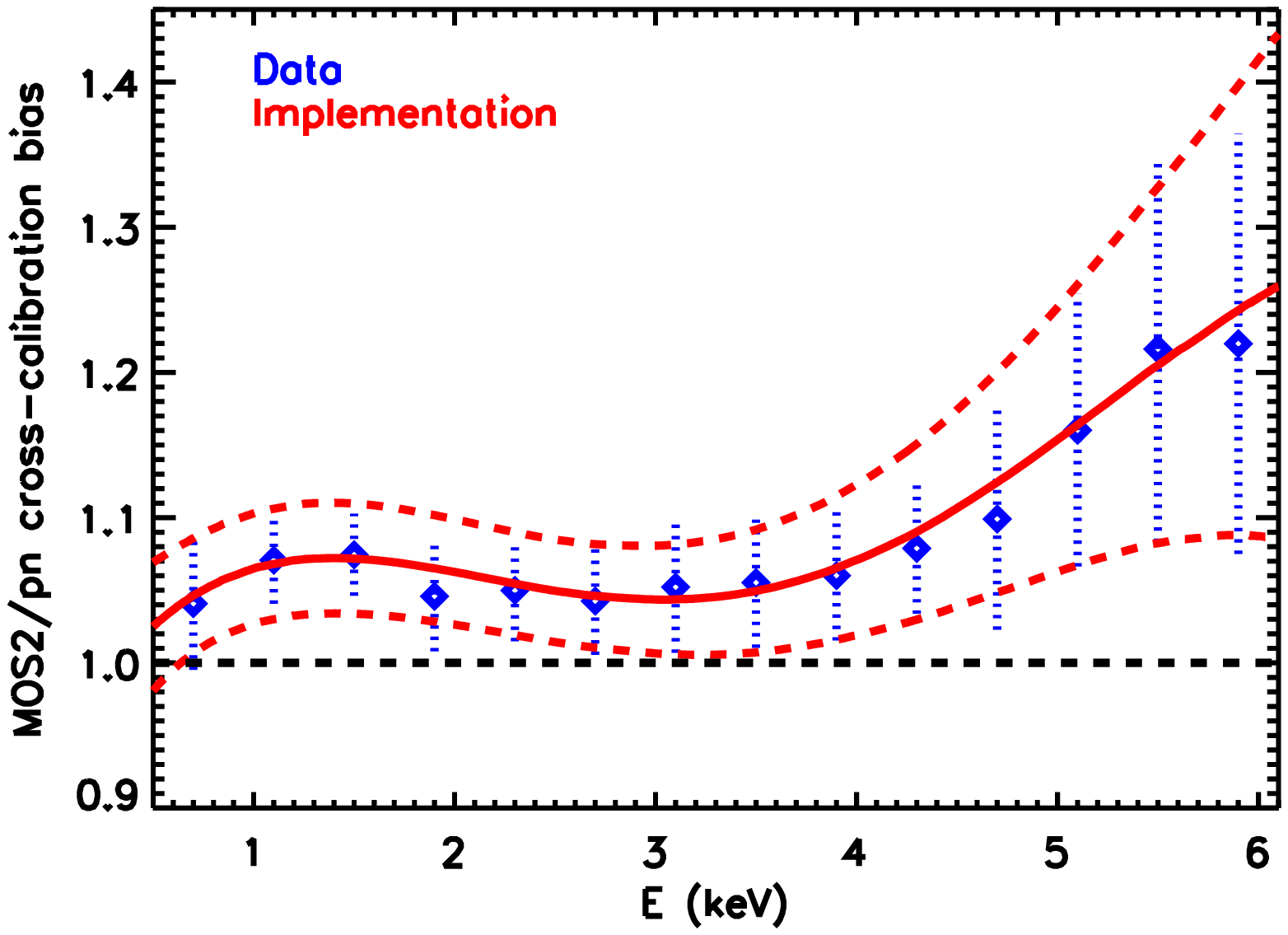}
}
 \caption{The median (solid red line) and the variation as measured by the standard deviation 
(dashed red lines) of the cross-calibration bias parameter J$_2$ of the sample obtained by the procedure laid out in Appendix \ref{Implementation}. The blue symbols indicate the corresponding values obtained from the cluster data (repeated from Fig. \ref{Jmeas.fig}). The results for the MOS1/pn and MOS2/pn pairs are shown in the left and right panels, respectively.  The horizontal dashed line indicates the expectation (unity) in case of no cross-calibration bias.}
\label{implemented.fig}
\end{figure*}

We tested the above procedure by generating modified cross-calibration bias curves for MOS1/pn and MOS2/pn pairs.
Using the distribution of the values at each energy we computed the sample median and the standard deviation.
We found that using 1000 realisations, the median of the randomised sample agrees with that of the cluster data sample within 2\% and the scatter, as measured by the standard deviation, is very similar in the two cases (see Fig \ref{implemented.fig}).

\end{document}